# Conversion rate prediction in online advertising: modeling techniques, performance evaluation and future directions

Tao Xue[1], Yanwu Yang[1], Panyu Zhai[2]

[1]School of Management, Huazhong University of Science and Technology, Wuhan, China

[2]School of Information Management, Zhengzhou University, Science Avenue, Zhengzhou, China

{xuetao.isec, yangyanwu.isec, zhaipanyu.isec}@gmail.com

**Abstract:** Conversion and conversion rate (CVR) prediction play a critical role in efficient advertising decision-making. In past decades, although researchers have developed plenty of models for CVR prediction, the methodological evolution and relationships between different techniques have been precluded. In this paper, we conduct a comprehensive literature review on CVR prediction in online advertising, and classify state-of-the-art CVR prediction models into six categories with respect to the underlying techniques and elaborate on connections between these techniques. For each category of models, we present the framework of underlying techniques, their advantages and disadvantages, and discuss how they are utilized for CVR prediction. Moreover, we summarize the performance of various CVR prediction models on public and proprietary datasets. Finally, we identify research trends, major challenges, and promising future directions. We observe that results of performance evaluation reported in prior studies are not unanimous; semantics-enriched, attribution-enhanced, debiased CVR prediction and jointly modeling CTR and CVR prediction would be promising directions to explore in the future. This review is expected to provide valuable references and insights for future researchers and practitioners in this area.

# 1. Introduction

The rapid development of the Internet has fostered a diverse range of online advertising formats, offering firms a bunch of outlets to deliver commercial messages and engage with potential consumers (Yang et al., 2022). Online advertising has become the mainstream of commercial promotion and the dominant sector of the advertising industry. According to Statista, U.S. online advertising revenue rose from $225 billion in 2023 to $259 billion in 2024, experiencing a 15% annual growth (Statista, 2025). In online advertising, after a user browses and clicks on an advertisement[1], she/he is redirected to a certain page where some conversion actions (e.g., purchasing, registering and downloading) may be performed. Usually, conversion is the ultimate objective for advertisers, which is defined as the last stage in the user journey (Xu et al., 2022). Conversion rate (CVR) is referred to the probability that a given click will lead to a conversion, reflecting the degree to which user engagement with ads is translated into economic returns (Li et al., 2024). As reported in Shopify's annual report, the average CVR of e-commerce advertising increased at an annual rate of 0.4% in 2025, which drove the revenue of the first two quarters of 2025 to $2.68 billion, representing an increase of 31.05% compared with the same period in 2024[2][3]. CVR is a crucial measure to evaluate the effectiveness of advertising campaigns (Zhang et al., 2024b). With the widespread adoption of cost-per-action pricing models, CVR prediction has become a critical demand for efficient advertising decision-making (Li et al., 2021a; Chen et al., 2022a). Inaccurate CVR prediction, either overestimating or underestimating, will result in wasting advertising budgets and missing promotional opportunities (Guo et al., 2023).

In the past decades, CVR prediction has attracted extensive efforts and become a hot research frontier in the field of online advertising (e.g., Ma et al., 2018; Yasui et al., 2020; Huang et al., 2024; Yuan et al., 2025). Existing research has developed plentiful CVR prediction models that employ techniques ranging from multivariate statistics to deep learning. However, it is unclear about the methodological evolution and the relationship between different techniques, which hinders participants in this area from understanding and implementing these models in real practices. To the best of our knowledge, this is the first survey on CVR prediction in the context of online advertising. The objectives of our review are threefold. First, we aim to provide a comprehensive literature review on advertising CVR prediction models, focusing on the underlying techniques and connections among them. Second, we intend to summarize the performance of various CVR prediction models on public and proprietary datasets. Third, we attempt to identify current research trends, main challenges, and potential future directions in this area.

[1] In the following we use "ad" as the abbreviation for advertisement.

[2] https://www.macrotrends.net/stocks/charts/SHOP/shopify/revenue#google_vignette

[3] https://redstagfulfillment.com/average-conversion-rate-for-shopify-stores/

There are several literature reviews recently published in the fields of online advertising and electronic commerce, on related topics including click-through rate (CTR) prediction (Yang & Zhai, 2022), user response prediction (Gharibshah & Zhu, 2021), and purchase prediction (Chen et al., 2024). More specifically, Yang & Zhai (2022) conducted a literature review on advertising CTR prediction, which centers on the evolution between various modeling frameworks; Gharibshah & Zhu (2021) concentrated on approaches for predicting a set of advertising responses including CTR, CVR, engagement, and rating scores, which covers 9 studies on CVR prediction; and Chen et al. (2024) made a bibliometric analysis on the literature related to purchase prediction in B2C e-business, discussed the intellectual structure of major topical clusters, and summarized research focuses and future research directions.

This review differs from previous ones in the following aspects. First, we conduct a comprehensive literature review on CVR prediction in the context of online advertising, covering 99 articles searched from six major academic databases (i.e., Web of Science, ACM Digital Library, IEEE Xplore, EBSCOhost, ScienceDirect, and ABI/INFORM Global). Second, we classify state-of-the-art CVR prediction models into six categories with respect to the underlying techniques and elaborate on connections between these techniques. For each category of CVR prediction models, we present the framework of underlying techniques, their advantages and disadvantages, and discuss how they are utilized for CVR prediction. Third, we outline datasets used in the extant literature and summarize the performance of various models on these datasets. Lastly, we identify current research trends, main challenges, and future directions in this area.

The remainder of this review is structured as follows. Section 2 gives the search and identification process of articles. Section 3 introduces the preliminaries of advertising CVR prediction, including problem definition, features and evaluation metrics. Section 4 reviews state-of-the-art CVR prediction models in the extant literature. Section 5 discusses the model performance on public and proprietary datasets. Section 6 presents the current research trends, main challenges, and future directions. Section 7 concludes this review.

## 2. Literature search and study identification

This review searched publications from six major academic databases: Web of Science, ACM Digital Library, IEEE Xplore, EBSCOhost, ScienceDirect, and ABI/INFORM Global through the full-text search using keywords ("conversion" OR "conversion rate" OR "CVR" OR "purchase") AND ("prediction" OR "forecasting" OR "estimation") AND ("online advertising" OR "Internet advertising"). The search process yielded 164 results from Web of Science, 765 results from ACM, 117 results from IEEE, 26 results from EBSCOhost, 1081 results from ScienceDirect, and 1231 results from ABI/INFORM Global. We eliminated duplicate articles and excluded patents, review articles, news, magazines, reports, posters, and abstracts, resulting in a collection of 2684 articles. Moreover, to get more comprehensive coverage of the related literature, we extended the search by retrieving citations and related works of articles obtained in the previous step in Google Scholar, resulting in 55 additional articles. We manually screened each article to identify whether it addressed advertising CVR prediction by going through the title, abstract, full-text, and datasets. The criteria for manual screening are given as follows. First, we checked whether the title and abstract explicitly referred to advertising CVR prediction. Second, we examined the research problem and methods to identify whether it addressed CVR prediction. Third, we screened the experimental evaluations to check whether the advertising datasets were used for the CVR prediction experiments. As a result, a total of 2640 articles were excluded due to the lack of direct relevance to CVR prediction. Finally, this process resulted in a collection of 99 articles, comprising 4 journal articles, 79 conference proceedings, and 16 pre-prints. The search and selection procedures are summarized in Figure 1. Detailed bibliographic information (e.g., publication year, authorship, and publication venue) of included articles is provided in Table A1 in Appendix A.

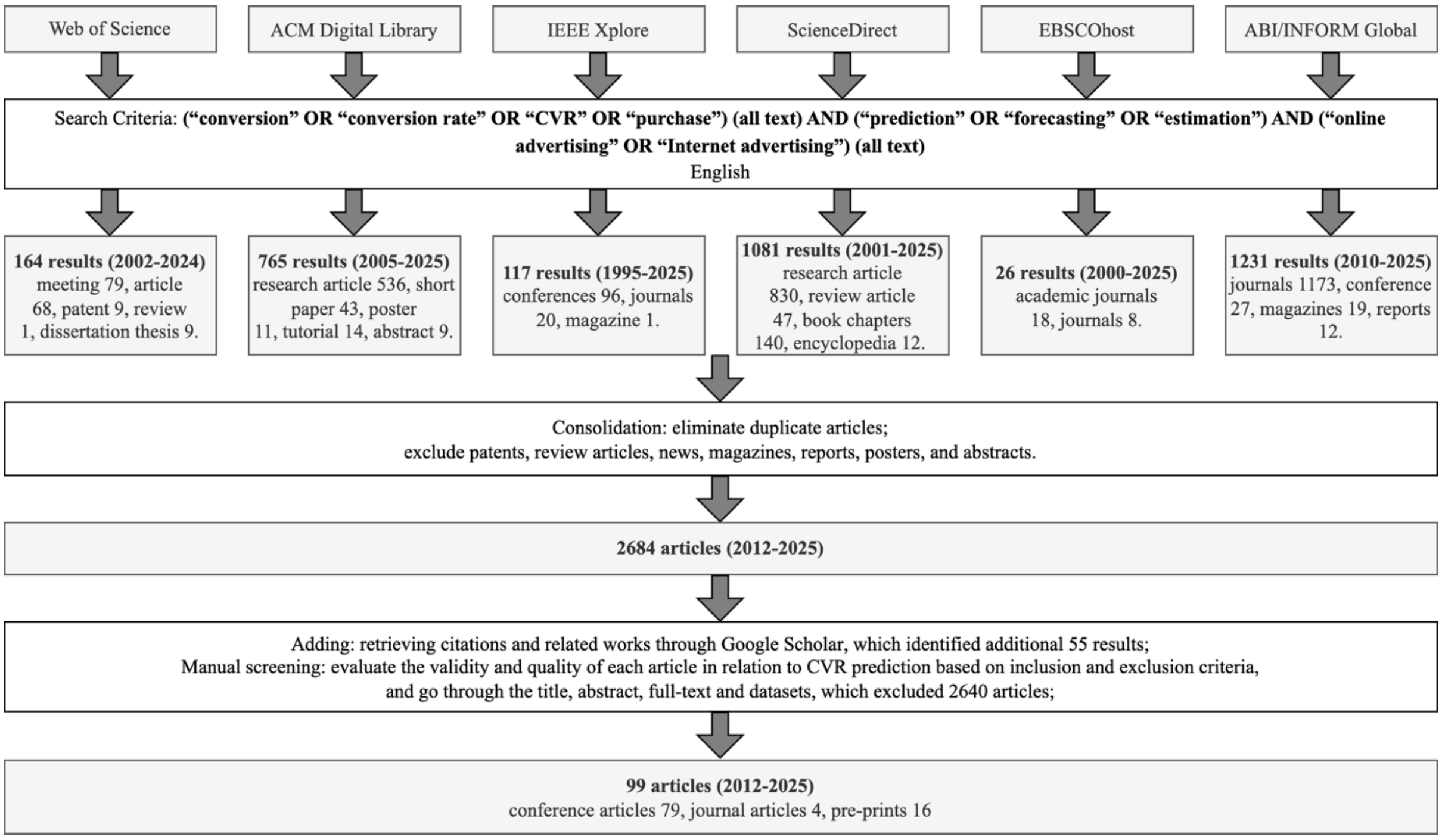

Figure 1. Study search and selection.

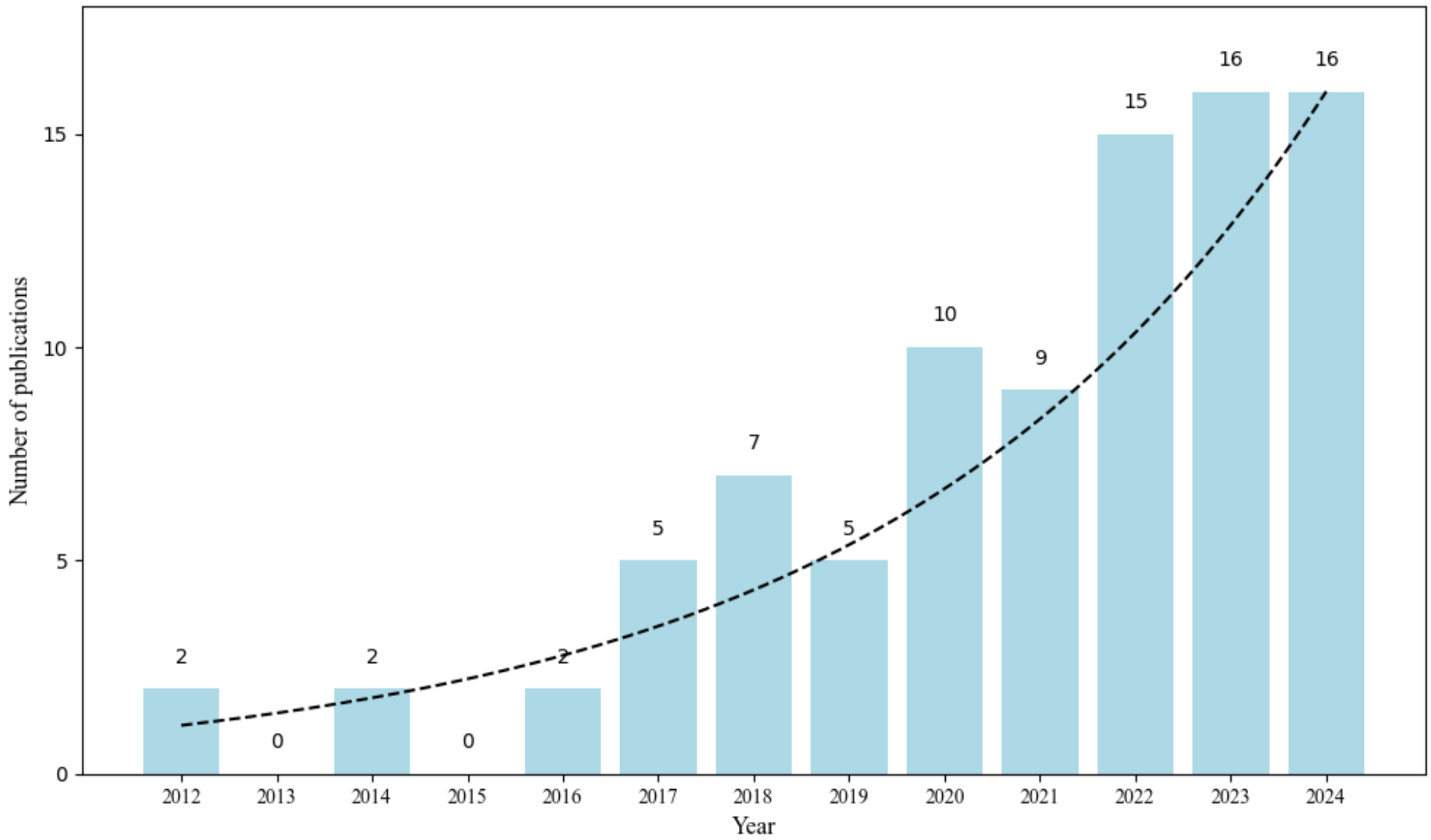

Figure 2. The trend in the number of publications on CVR prediction over years.

Figure 2 depicts publications on CVR prediction research from 2012 to 2024. As shown in Figure 2, the research development in the area of CVR prediction has experienced three periods. In the first period (2012-2016), only a few studies have been published primarily relying on statistical modeling, probably due to the fact that data are scarce. The second period (2017-2021) has witnessed a rapid increase in the publication volume, which is probably spurred by the breakthrough of deep learning and the spreading demand arising from the development of digital business. In the third period (2022-2025), publications have continued to grow steadily, indicating that CVR prediction research has kept being fueled by industrial needs and remained active. Overall, the number of publications shows a significant upward growth over the past decade, and importantly, the increasing trend suggests that research interests on CVR prediction will continue to grow in the near future.

# 3. Preliminaries

In this section, we introduce preliminary knowledge for advertising CVR prediction, including problem definition, features and evaluation metrics.

## 3.1. The definition of CVR prediction problem

In online advertising, the CVR prediction problem can be formulated as follows. Let $\mathcal{D} = \{(\boldsymbol{x}_i, y_i)\}_{i=1}^N$ represents the collected dataset of advertising behaviors such as impressions and clicks. For each sample $i$, $\boldsymbol{x}_i$ represents the input features, including information about user $u_i$, ad $a_i$, and context $c_i$; and $y_i \in \{0,1\}$ is the conversion label. Specifically, if $y_i = 1$ means that $u_i$ performs a specified conversion action (e.g., purchasing, registering and downloading), otherwise $y_i = 0$. CVR prediction is referred to inferring the probability that a user $u_i$ will complete a specific action after clicking on an ad, i.e., $p_{CVR} = p(conversion = 1|click = 1, \boldsymbol{x}_i) = f(\boldsymbol{x}_i)$, where $p_{CVR} \in [0,1]$ is the predicted CVR. Notations used in this paper are presented in Table 1.

Table 1. Notations

| Terms | Definition |
|---|---|
| $y_i$ | Target label: $y_i = 1$ means conversion, otherwise $y_i = 0$. |
| $p_i$ | $p_i \in [0,1]$ is the predicted conversion probability of sample $i$. |
| $\boldsymbol{x}$ | The input feature vector. |
| $\mathcal{L}$ | The loss function. |
| $f_m$ | The weak learner of $m-th$ tree. |
| $F_m$ | The boosting approximation to the predicted CVR. |
| $\Omega(f_m)$ | The regularized term. |
| $\sigma$ | Activation function. |
| $\boldsymbol{w}$ | Weight. |
| $b$ | Bias term. |
| $\boldsymbol{h}$ | The hidden state. |
| $h(t)$ | The hazard rate. |
| $S(t)$ | The survival function. |
| $\boldsymbol{P}$ | Positional embeddings. |
| $\boldsymbol{Q}$ | The query matrix in Transformer. |
| $\boldsymbol{K}$ | The key matrix in Transformer. |
| $\boldsymbol{V}$ | The value matrix in Transformer. |
| $\boldsymbol{f}_i$ | The filter in TCNs. |
| $\boldsymbol{s}_t^n$ | Latent state representations up to touchpoint $t$. |
| $\boldsymbol{r}_t^n$ | Unbiased representations in CRNs. |
| $\boldsymbol{c}_t$ | Advertising channel. |
| $T$ | The distillation temperature. |
| TP | True positive. |
| FP | False positive. |
| TN | True negative. |
| FN | False negative. |

## 3.2. Features for CVR prediction

In the literature (e.g., Zhang et al., 2014; Shan et al., 2018; Liu & Liu, 2021), features used in advertising CVR prediction can be divided into four groups: (1) advertising features, (2) user features, (3) context features, and (4) publisher features, as shown in Table 2. Specifically, advertising features describe creative content and promoted products; user features reflect user demographics and behaviors that can be used to extract user interests; context features provide environmental and temporal information for user-ad interactions; publisher features characterize properties of platforms and pages where ads are displayed, reflecting the channels and mechanisms of advertising delivery.

Table 2. Feature types for CVR prediction

| **Type** | **Feature Examples** |
|---|---|
| Advertising | advertising ID, title, body, website, position, advertiser ID, campaign ID, creative ID, size, category, etc. |
| User | user ID, age, gender, education, marriage, hometown, interest, behavior, etc. |
| Context | device, operation system, browser, timestamp, etc. |
| Publisher | publisher ID, type, location, website, etc. |

Feature engineering is applied after data preprocessing (e.g., cleaning, normalization, deduplication and anomaly detection) to produce meaningful, model-ready features (Kuhn & Johnson, 2019). In multivariate statistical and tree models, extensive feature engineering is usually performed, e.g., feature filtering, feature decomposition, and feature crossing, which incurs substantial time and labor costs (Lee et al., 2012; Zhang et al., 2023). In contrast, deep learning models and their extensions facilitate automated feature engineering and high-order representation learning.

In advertising CVR prediction, certain forms of features are suitable for distinct CVR modeling techniques. For instance, user behaviors exhibit temporal dependencies and dynamic interests. Sequential models (e.g., RNNs, LSTM, and Transformer) have been employed to capture long-term dependencies in user behaviors and characterize the evolution of user interests (Li et al., 2021a). Moreover, user-ad interactions can be constructed as bipartite graphs and further transformed into user-user and ad-ad graphs by leveraging shared neighbor relationships, which are used by graph neural networks (GNNs) to learn connectivity structures and high-order relationships of user-ad interactions (Zheng et al., 2022; Zhai et al., 2023; Jin & Yang, 2025). Furthermore, multimodal features (e.g., advertising titles and copies) provide rich semantic information, which are usually processed through BERT and large language models (LLMs) to learn deep semantic representations (Kitada et al., 2019; Wang et al., 2023a).

## 3.3. Evaluation metrics

In Table 3, we present definitions of evaluation metrics used to assess the performance of advertising CVR prediction models, which are summarized as follows. (a) Accuracy, precision, recall and F1-score have been commonly used in CVR prediction, assessing the quality of estimated CVR probabilities (Agrawal et al., 2022; Wang et al., 2022). (b) Cross-entropy loss function (Logloss), negative log likelihood (NLL), debiased-NLL (D-NLL), normalized Logloss and normalized cross-entropy (NE) have been typically used as loss functions to describe the deviation between the true CVR and the predicted one (Terven et al., 2025). (c) Area under the ROC curve (AUC), group AUC (GAUC), area under the curve of precision/recall (PRAUC), multi-AUC, weighted AUC, normalized discounted cumulative gain (NDCG), Kolmogorov-Smirnov (KS) score and Debiased-AUC (D-AUC) have been used to evaluate the ranking ability of CVR prediction models (Pan et al., 2019; Xi et al., 2021; Guo et al., 2023). (d) Mean square error (MSE), mean absolute error (MAE), root means squared error (RMSE), weighted MSE (MSEW), mean absolute percentage error (MAPE), explained variance score (EVS), R squared score ($R^2$), Bias, Brier score (BS), calibration error (i.e., ECE, MVCE, F-ECE, and F-RCE) and predicted CVR over the actual CVR (PCOC) measure the degree to which predicted probabilities align with actual CVR (Chan et al., 2023; Lane, 2025). (e) Relative improvement and MTL gain quantify the performance improvements of CVR prediction models over baselines (Tang et al., 2020; Ding et al., 2025), and compliance rate (CR) and utility are two business-aware metrics: the former evaluates the advertising cost control ability of the platform (Vasile et al., 2017), and the latter reflects the contribution of a CVR prediction model to the profit realization (Cui et al., 2024).

From Table 3, we observe that AUC and Logloss are the most widely adopted metrics in the extant literature. Moreover, calibration metrics (e.g., calibration errors) have attracted increasing attention in recent years. In addition, business-aware metrics (i.e., CR and utility) have rarely been considered, probably because including them into loss functions may lead to the high complexity and low computational efficiency of CVR prediction models.

Table 3. Evaluation metrics for CVR prediction.

| Metrics | Definition | References |
|---|---|---|
| Accuracy | $Accuracy = \frac{TP+TN}{N}$ | Li et al. (2018); Cui et al. (2018); Yoshikawa & Imai (2018); Du et al. (2019b); Gligorijevic et al. (2020); Agrawal et al. (2022); Yasui & Kato (2022); Ban et al. (2024a) |
| Precision | $Precision = \frac{TP}{TP+FP}$ | Du et al. (2019b); Gligorijevic et al. (2020); Guo et al. (2024); Zhang et al. (2024b) |
| Recall | $Recall = \frac{TP}{TP+FN}$ | Du et al. (2019b); Gligorijevic et al. (2020); Wang et al. (2022); Guo et al. (2024) |
| F1 | $F1 = \frac{2*Precision*Recall}{Precision+Recall}$ | Zhang et al. (2014); Wang et al. (2022); Bi ģci (2025) |
| AUC | AUC is the area under the ROC curve, reflecting the trade-off of the true positive rate (TPR) and the false positive rate (FPR).<br>$TPR = \frac{TP}{TP+FN}$<br>$FPR = \frac{FP}{FP+TN}$ | Lee et al. (2012); Rosales et al. (2012); Ji et al. (2016); Yang et al. (2016); Bhamidipati et al. (2017); Ji & Wang (2017); Li et al. (2018); Ma et al. (2018); Ren et al. (2018); Shan et al. (2018); Yoshikawa & Imai (2018); Du et al. (2019b); Pan et al. (2019); Zhou et al. (2019); Gligorijevic et al. (2020); Kumar et al. (2020); Qiu et al. (2020); Tang et al. (2020); Wen et al. (2020); Yang et al. (2020); Zhang et al. (2020); Gu et al. (2021); Hou et al. (2021); Li et al. (2021a); Li et al. (2021b); Su et al. (2021); Wei et al. (2021); Xi et al. (2021); Yang et al. (2021); Agrawal et al. (2022); Chen et al. (2022a); Chen et al. (2022b); Dai et al. (2022); Ding et al. (2022); Guo et al. (2022); Lin et al. (2022); Wang et al. (2022); Xie et al. (2022); Xu et al. (2022); Yang & Zhan (2022); Yao et al. (2022); Yasui & Kato (2022); Zhang et al. (2022); Chan et al. (2023); Dai et al. (2023); Guo et al. (2023); Liu et al. (2023a); Liu et al. (2023b); Min et al. (2023); Ouyang et al. (2023a); Ouyang et al. (2023b); Tan et al. (2023); Wang et al. (2023c); Yang et al. (2023); Zhang et al. (2023); Zhu et al. (2023); Ban et al. (2024a); Ban et al. (2024b); Feng et al. (2024); Guo et al. (2024); Huang et al. (2024); Li et al. (2024); Liu et al. (2024a); Liu et al. (2024b); Su et al. (2024); Tang et al. (2024); Yang et al. (2024); Yu et al. (2024); Zhang et al. (2024a); Zhang et al. (2024b); Zhao et al. (2024); Bi ģci (2025); Cheng et al. (2025); Dai et al. (2025); Ding et al. (2025); Dishi et al. (2025); Fei et al. (2025); Yi et al. (2025); Yuan et al. (2025); Zhuang et al. (2025) |
| GAUC | $GAUC = \frac{\sum_u w_u \times AUC_u}{\sum_u w_u}$ | Wen et al. (2020); Zhang et al. (2020); Hou et al. (2021); Lin et al. (2022); Jin et al. (2023) |
| PR-AUC | PR-AUC is the area under the precision-recall curve, reflecting the relationship of precision and recall. | Yasui et al. (2020); Gu et al. (2021); Hou et al. (2021); Yang et al. (2021); Chen et al. (2022a); Yang & Zhan (2022); Dai et al. (2023); Wang et al. (2023c); Liu et al. (2024a); Yu et al. (2024); Ding et al. (2025); Zhuang et al. (2025) |
| Multi-AUC | $MAUC = \frac{2}{\lvert C\rvert(\lvert C\rvert-1)} \sum_{\{c_i,c_j\}\in C} AUC(c_i, c_j)$ | Shan et al. (2018) |
| Weighted AUC | $WAUC = \frac{\sum_{t\in\mathcal{T}} AUC_t \cdot N_t}{\sum_{t\in\mathcal{T}} N_t}$ | Pan et al. (2019) |
| D-AUC | $D - AUC = \frac{\sum_{i\in\mathcal{D}^-_{click}, j\in\mathcal{D}^+_{click}} \frac{1}{\hat{p}_{CTR}(i)} \cdot \frac{1}{\hat{p}_{CTR}(j)} \cdot \mathbb{I}(\hat{p}_{CVR}(i) > \hat{p}_{CVR}(j))}{(\sum_{i\in\mathcal{D}^-_{click}} \frac{1}{\hat{p}_{CTR}(i)}) \cdot (\sum_{j\in\mathcal{D}^+_{click}} \frac{1}{\hat{p}_{CTR}(j)})}$ | Xu et al. (2022) |
| Logloss | $Logloss = -\frac{1}{N}\sum_i^N y_i \, log\widehat{y_i} + (1 - y_i)\log(1 - \widehat{y_i})$ | Cui et al. (2018); Ren et al. (2018); Yoshikawa & Imai (2018); Du et al. (2019a); Kumar et al. (2020); Yasui et al. (2020); Li et al. (2021a); Li et al. (2021b); Liu & Liu (2021); Ding et al. (2022); Guo et al. (2022); Xie et al. (2022); Yao et al. (2022); Chan et al. (2023); Guo et al. (2023); Shtoff et al. (2023); Wang et al. (2023c); Zhang et al. (2023); Huang et al. (2024); Tang et al. (2024); Yang et al. (2024); Zhang et al. (2024a); Zhao et al. (2024); Bi ģci (2025); Ding et al. (2025) |
| Normalized Logloss | Normalized Logloss is the Logloss normalized by that of the naive predictor that predicts the average CVR of the training set. | Yasui et al. (2020) |
| NLL | $NLL = -\sum_{i=1}^{n} \log( y_i \log(p_i)$ | Chapelle (2014); Gu et al. (2021); Wei et al. (2021); Yang et al. (2021); Chen et al. (2022a); Gao & Yang (2022); Xu et al. (2022); Yang & Zhan (2022); Yasui & Kato (2022); Dai et al. (2023); Liu et al. (2023a); Liu et al. (2024a); Yu et al. (2024); Fei et al. (2025) |
| D-NLL | $D - NLL = \frac{1}{\sum_{i \in \mathcal{D}_{click}} \frac{1}{\hat{p}_{CTR}(i)}} \sum_{i \in \mathcal{D}_{click}} \frac{1}{\hat{p}_{CTR}(i)} \cdot NLL_{CVR}(i)$ | Xu et al. (2022) |
| MAPE | MAPE is employed to evaluate the percentage error. | Cui et al. (2024) |
| NE | $NE = \frac{-\frac{1}{N}\sum_{i=1}^{n}(\frac{1+y_i}{2}\log(p_i)+\frac{1-y_i}{2}\log(1-p_i))}{-(plop(p)+(1-p)\log(1-p))}$ | Zhang et al. (2025a) |
| MSE | $MSE = -\frac{1}{N}\sum_{i}^{N}(\hat{y_i} - y_i)^2$ | Kitada et al. (2019); Yao et al. (2023) |
| MSEW | $MSEW = \frac{1}{N}\sum_{i}^{N}((y_i - p(x_i)) \cdot v_i)^2$ | Vasile et al. (2017) |
| MAE | $MAE = -\frac{1}{N}\sum_{i}^{N} \lvert \hat{y_i} - y_i \rvert$ | Yao et al. (2023) |
| RMSE | $RMSE = \sqrt{\frac{1}{N}\sum_{i}^{N}(\hat{y_i} - y_i)^2}$ | Jiang & Jiang (2017); Shan et al. (2018) |
| $R^2$ | $R^2 = 1 - \frac{\sum_{i}^{N}(\hat{y_i}-y_i)^2}{\sum_{i}^{N}(\bar{y_i}-y_i)^2}$ | Yao et al. (2023) |
| PCOC | $PCOC = \lvert log \frac{\sum_{i=1}^{n}\hat{y_i}}{\sum_{i=1}^{n} y_i} \rvert$ | Chan et al. (2023); Yang et al. (2023); Dai et al. (2025) |
| EVS | $EVS = 1 - \frac{Var(y-\hat{y})}{Var(y)}$ | Yao et al. (2023) |
| KS | $KS = \max \lvert F_{positive}(x) - F_{negative}(x) \rvert$ | Wang et al. (2022); Dai et al. (2023); Huang et al. (2024) |
| NDCG | $NDCG = \frac{1}{G_n} \cdot \sum_{i=1}^{n} \frac{2^{g_i}-1}{\log(1+i)}$ | Shan et al. (2018); Kitada et al. (2019) |
| BS | $BS = \frac{1}{\lvert\mathcal{D}\rvert}\sum_{i}^{\lvert\mathcal{D}\rvert}(y_i - p_i)^2$ | Guo et al. (2022); Guo et al. (2023); Zhang et al. (2024a) |
| Bias | $Bias = \frac{1}{\lvert\mathcal{D}\rvert}\sum_{j=1}^{\lvert Z\rvert} N_{Z_j} \frac{\lvert \sum_{z_i=z_j, i\in\mathcal{D}}(y_i-p_i)\rvert}{\sum_{z_i=z_j, i\in\mathcal{D}} y_i}$ | Liu et al. (2023a) |
| Calibration errors | Expected calibration error: $ECE = \frac{1}{\lvert\mathcal{D}\rvert}\sum_{t=1}^{T} \lvert \sum_{(x_i,y_i)\in\mathcal{D}_t}(y_i, \hat{p}_i) \rvert$ Multi-view calibration error: $MVCE = \frac{1}{RT}\sum_{r=1}^{R}\sum_{t=1}^{T}\frac{\lvert\sum_{(x_i,y_i)\in\mathcal{D}_{r,t}}(y_i,\hat{p}_i)\rvert}{\lvert\mathcal{D}_{r,t}\rvert}$ Field-level expected calibration error: $F - ECE = \frac{1}{\lvert\mathcal{D}\rvert}\sum_{z=1}^{\lvert Z\rvert} \lvert \sum_{i=1}^{\lvert D\rvert}(y_i - \hat{p}_i)\mathbf{1}_{[z_i=z]} \rvert$ | Guo et al. (2022); Chan et al. (2023); Guo et al. (2023); Yang et al. (2024); Zhang et al. (2024a); Zhao et al. (2024); Dai et al. (2025); Yuan et al. (2025) |
| | Field-level relative calibration error: $F - RCE \frac{1}{\lvert\mathcal{D}\rvert}\sum_{z=1}^{\lvert Z\rvert} N_z \frac{\lvert\sum_{i=1}^{\mathcal{D}}(y_i - \hat{p}_i)\mathbf{1}_{[z_i=z]}\rvert}{\sum_{i=1}^{\mathcal{D}}(y_i+\epsilon)\mathbf{1}_{[z_i=z]}}$ | |
| Relative improvement | $RI_{metric} = \frac{Metric(Model)-Metric(Vanilla)}{Metric(Oracle)-Metric(Vanilla)} \times 100\%$ <br> $RelaImpr = \frac{Metric(model)-0.5}{Metric(base)-0.5} \times 100\%$ | Jiang & Jiang (2017); Saito et al. (2020); Gu et al. (2021); Su et al. (2021); Yang et al. (2021); Chen et al. (2022a); Gao & Yang (2022); Dai et al. (2023); Liu et al. (2023a); Wang et al. (2023c); Ban et al. (2024b); Ding et al. (2025); Dishi et al. (2025) |
| MTL Gain | $MTL\ gain = \begin{cases} M_{MTL} - M_{single}, M > 0 \\ M_{single} - M_{MTL}, M < 0 \end{cases}$ | Tang et al. (2020) |
| Utility | $Utility = \sum_i \int_0^{p(x_i)v_i} (y_i \cdot v_i - \tilde{c}) \Pr(\tilde{c} \mid c_i) d\tilde{c}$ | Vasile et al. (2017) |
| CR | $CR = \frac{1}{N}\sum_y^{\mathcal{Y}} \eta(0.8 \leq \frac{\hat{y}}{y} \leq 1.2)$ | Cui et al. (2024) |

# 4. State-of-the-art CVR prediction models

In the advertising literature, CVR prediction models can be divided into six categories according to the underlying techniques: (1) multivariate statistical models, (2) tree models, (3) deep learning models, (4) causal learning models, (5) knowledge distillation models, and (6) multi-task learning models. The categories of state-of-the-art CVR prediction models and the distribution of publications covered in this review are presented in Table 4. In the following, for each category of CVR prediction models, we present the framework of underlying techniques, advantages and disadvantages, and related studies.

Table 4. Categories of CVR prediction models in the literature.

| Category | Underlying Techniques | CVR Prediction Models | References |
|---|---|---|---|
| Multivariate statistical models | LR | Delayed feedback model (DFM); Dynamic transfer learning with reinforced word model (Trans-RWM); Combined regression and triplet-wise ranking model (CRT) | Lee et al. (2012); Rosales et al. (2012); Chapelle (2014); Yang et al. (2016); Bhamidipati et al. (2017); Jiang & Jiang (2017); Vasile et al. (2017); Tallis & Yadav (2018); Shan et al. (2018) |
| | Survival analysis | Additive hazard model (AH); Probabilistic multi-touch attribution model (PMTA); Additional multi-touch attribution model (AMTA); Nonparametric delayed feedback model (NoDeF) | Zhang et al. (2014); Ji et al. (2016); Ji & Wang (2017); Yoshikawa & Imai (2018) |
| Tree models | GBDT | Boosted trees model with feature engineering (BTMFE) | Lu et al. (2017); Du et al. (2019a) |
| | XGBoost | Hierarchical history modeling model | Liu & Liu (2021); Zhang et al. (2023) |
| Deep learning models | MLP | Generalized delayed feedback model (GDFM); Historical data reuse model (HDR); Masked multi-domain model (MMN); Self-supervised pre-training model; Universal and segmentation-specific representation model (USSR); Deep ensemble shape calibration model (DESC); DelayAdapter | Yang & Zhan (2022); Chan et al. (2023); Min et al. (2023); Ouyang et al. (2023b); Shtoff et al. (2023); Tan et al. (2023); Yang et al. (2024); Yu et al. (2024); Dai et al. (2025) |
| | DNNs | Gradient-flow based fine-tuning (GFlow-FT); Behavioral auxiliary model (Aux); Automatic fusion model (AutoFuse); Multi-interval screening and synthesizing model (MISS); Delayed feedback modeling via neural satellite networks (DFSN); Confidence-aware multi-field calibration model (ConfCalib) | Ding et al. (2022); Guo et al. (2023); Jin et al. (2023); Liu et al. (2023a); Liu et al. (2024a); Zhao et al. (2024); Bicici (2025) |
| | RNNs | Deep neural net with attention multi-touch attribution model (DNAMTA); Ensemble CNN + LSTM; Dual-attention recurrent neural model (DARNN); Local attention in GRU-based RNNs (LATT); Deep time aware conversion model (DTAIN); Deep learning model for multi-touch attribution model (DeepMTA); Two-stage deep learning model (TS-DL); Action attention bidirectional GRU (AAGRU); Deep journey hierarchical attention model (DJHAN) | Li et al. (2018); Cui et al. (2018); Ren et al. (2018); Du et al. (2019b); Zhou et al. (2019); Gligorijevic et al. (2020); Qiu et al. (2020); Yang et al. (2020); Su et al. (2021); Ban et al. (2024a); Ban et al. (2024b) |
| | Transformer | Attentive capsule model (ACN); Distribution-constrained batch Transformer (DCBT); Adapter-based CVR model | Li et al. (2021a); Yang et al. (2023); Li et al. (2024) |
| | TCNs | Stage-TCN; TCN with an attention mechanism (Att-TCN) | Agrawal et al. (2022); Xie et al. (2022) |
| Causal learning models | Counterfactual learning | Dual learning algorithm for delayed feedback (DLA-DF); Feedback shift importance weight model (FSIW); Delayed feedback modeling with real negatives (DEFER); Elapsed-time sampling delayed feedback model (ES-DFM); Non-negative delayed feedback learning model (nnDF); Unbiased delayed feedback label correction model (ULC); Influence function-empowered framework for delayed feedback modeling (IF-DFM) | Saito et al. (2020); Yasui et al. (2020); Gu et al. (2021); Yang et al. (2021); Yasui & Kato (2022); Wang et al. (2023c); Ding et al. (2025) |
| | CRNs | Causal attribution mechanism for multi-touch attribution (CAMTA); Causal multi-touch attribution model (CausalMTA); Deep causal representation for multi-touch attribution (DCRMTA) | Kumar et al. (2020); Yao et al. (2022); Tang et al. (2024) |
| Knowledge distillation models | - | Knowledge distillation calibration model (KD calibration); Uncertainty-regularized knowledge distillation model (UKD); Entire-space variational information exploitation model (EVI); Hardness-aware privileged features distillation model (HA-PFD) | Guo et al. (2022); Xu et al. (2022); Fei et al. (2025); Yuan et al. (2025) |
| Multi-task learning models | Hard parameter sharing | Entire space multi-task model (ESMM); GRU + conditional attention model; Multi-task field-weighted factorization machine model (MT-FwFM); Entire space delayed feedback model (ESDF); Elaborated entire space supervised multi-task model ($ESM^2$); Multi-task inverse propensity weighting estimator (Multi-IPW) and Multi-task doubly robust estimator (Multi-DR); Following the prophet model (FTP); Causal feature selection for multi-task learning model (CFS-MTL); Doubly robust estimator to control bias (DR-bias) and doubly robust estimator to control mean squared error (DR-MSE); Entire space counterfactual multi-task model ($ESCM^2$); Multi-head online learning model (MHOL); Dually enhanced delayed feedback model (DDFM); Contrastive learning for CVR prediction model (CL4CVR); Exposure guided embedding alignment model (EGEAN); Non-click samples improved semi-supervised model (NISE); Entire-space weighted AUC model (EWAUC); Direct dual propensity optimization model (DDPO); Entire-space dual multi-task learning model (ChorusCVR) | Ma et al. (2018); Kitada et al. (2019); Pan et al. (2019); Wang et al. (2020); Wen et al. (2020); Zhang et al. (2020); Li et al. (2021b); Chen et al. (2022b); Dai et al. (2022); Wang et al. (2022); Gao & Yang (2022); Dai et al. (2023); Ouyang et al. (2023a); Feng et al. (2024); Huang et al. (2024); Liu et al. (2024b); Su et al. (2024); Cheng et al. (2025); Dishi et al. (2025); Zhang et al. (2025a) |
| | Soft parameter haring | Entire space CVR model with automated hierarchical representation integration (AutoHERI); Adaptive information transfer multi-task model (AITM); Multi-view multi-task embedding with hierarchical attention model (MVTA); Direct entire-space causal multi-task model (DCMT); Multi-task conditional attention network-based logistics advertising conversion prediction model (MCAC); Deep hierarchical ensemble network (DHEN) | Wei et al. (2021); Xi et al. (2021); Yao et al. (2023); Zhu et al. (2023); Guo et al. (2024); Zhuang et al. (2025) |
| | Expert sharing | Progressive layered extraction model (PLE); Delayed feedback modeling with unbiased estimation (DEFUSE); Adaptive fine-grained task relatedness modeling approach (AdaFTR); Task adaptive multi-learner model (TAML); Multi-task mixture-of-experts calibration model (MTMEC); Adversarial-enhanced causal multitask model (AECM); Adaptive entire-space multi-scenario multi-task transfer learning model ($AEM^2TL$) | Tang et al. (2020); Hou et al. (2021); Chen et al. (2022a); Lin et al. (2022); Zhang et al. (2022); Liu et al. (2023b); Cui et al. (2024); Zhang et al. (2024a); Zhang et al. (2024b); Yi et al. (2025) |

## 4.1. Multivariate statistical models

In the literature (e.g., Ji et al., 2016; Yang et al., 2016; Ji & Wang, 2017; Jiang & Jiang, 2017), two multivariate statistical models (i.e., logistic regression and survival analysis) have been used to describe relationships between various features and CVR.

### 4.1.1. Logistic regression (LR)

Logistic regression (LR) is one of the most widely used models for CVR prediction in online advertising (Lee et al., 2012), which learns a linear combination function of multiple features and applies a sigmoid function to transform it into a conversion probability. Mathematically, LR can be formulated as $p = \sigma\left(f_{LR}(\boldsymbol{x})\right) = \sigma(\boldsymbol{w}^T\boldsymbol{x}) = \frac{1}{1+e^{-w^T x}}$, where $p$ is the predicted CVR, $\sigma$ is the activation function, $f_{LR}(\boldsymbol{x})$ is the linear combination of the feature vector $\boldsymbol{x}$, and $\boldsymbol{w}$ is the learnable weight vector. Figure 3 presents the LR modeling framework for CVR prediction.

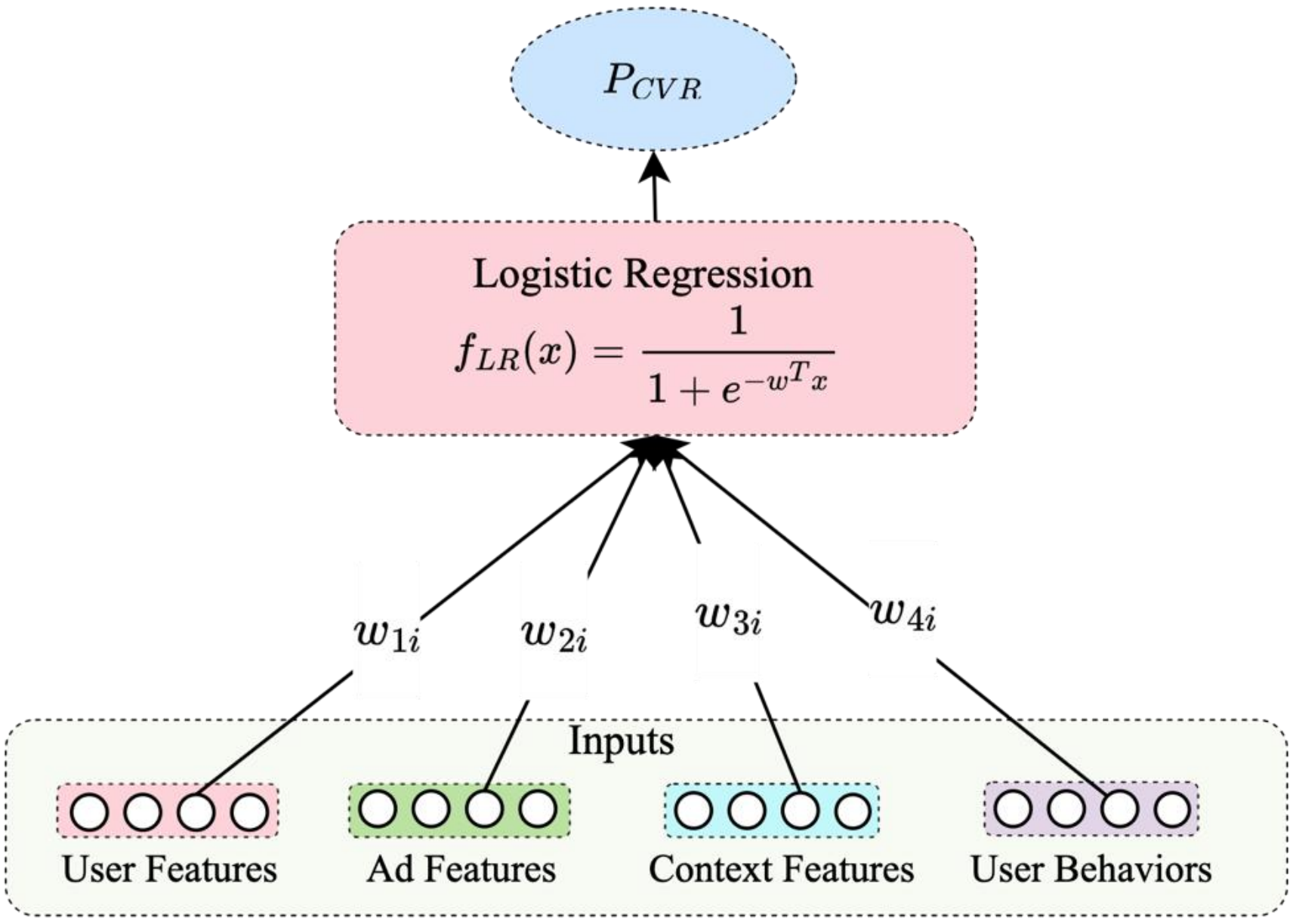

Figure 3. The LR modeling framework for CVR prediction.

The LR-based CVR prediction model offers several advantages: (1) it is simple and efficient; (2) its weight coefficients provide an intuitive indication of how each feature influences the probability of conversion. However, LR fails to capture non-linear relationships and interactions among features (Ling et al., 2017), and is prone to outliers.

Assuming the time delay between advertising impressions and conversions follows an exponential distribution, Chapelle (2014) employed the LR to predict conversion probability; Tallis & Yadav (2018) employed LR as the base to develop three CVR prediction models (i.e., the historic CVR feature model, the time decay weighting model and the mixture of long-term and short-term model); Shan et al. (2018) integrated the triplet-wise ranking into the LR to estimate CVR, jointly optimizing the regression loss and the triplet-wise ranking loss. Lee et al. (2012) estimated the distribution parameters of conversion at different levels (i.e., user, publisher and advertiser) with separate binomial distributions, and fed the estimates into an LR framework to obtain the CVR prediction scores; Jiang & Jiang (2017) proposed an LR model incorporating the impact of advertising creatives to predict CVR; Vasile et al. (2017) proposed a LR model with the cost-per-conversion weight for CVR prediction, where the Logloss of each sample is weighted with its corresponding cost to favor high-value conversions; Bhamidipati et al. (2017) proposed a two-stage LR model for conversion prediction, where a vanilla LR model is trained with a low-noise subset of features, and then its weights are frozen together with a set of cross features are used to further fine-tune the model.

#### 4.1.2. Survival analysis (SA)

In online advertising, the process from exposures to conversions is not instantaneous, and the effects of exposures fade over time (Ji et al., 2016). Survival analysis (SA) employs the hazard rate and survival function to model the impacts of exposures and estimate the conversion probability at a specific time.

Specifically, the hazard rate (i.e., $h(t)$) represents the occurrence rate of the conversion at time $t$ on the condition that the user has not converted before a specific time $t$, and the survival function (i.e., $S(t)$) is used to compute the cumulative probability that the conversion time is later than $t$ (Zhang et al., 2014), which are given as

$$\begin{cases} S(t) = exp\left(-\int_0^t h(t)\,dt\right) \\ h(t) = \lim\limits_{\Delta t \to 0} \frac{Pr(t \le T \le t+\Delta t | T > t)}{\Delta t} = \frac{\psi(t)}{S(t)}, \end{cases} \quad (1)$$

where $T$ denotes the conversion delay between an exposure and conversion, and $\psi(t)$ is the probability density function.

SA-based models for CVR prediction are effective in capturing the time-varying nature of exposures and accounting for delays in user conversion. Meanwhile, SA-based CVR prediction models can quantify the conversion contribution of individual ad touchpoints (i.e., user-ad interactions) via the hazard function (Zhang et al., 2014). However, SA models assume that the effect of exposures is additive and parameter distributions of the survival function follow specific forms (e.g., exponential distribution and Weibull distribution), which may not fully align with real-world advertising scenarios (Li et al., 2018; Ren et al., 2018).

In the literature, SA-based CVR prediction models use the hazard function to capture user conversion dynamics. Zhang et al. (2014) formulated the hazard function with the impact of advertising strength and the decaying speed that reflect the influence of an user behavior on the conversion and the decay of advertising effectiveness over time, respectively, and used the survival function to obtain conversion probability; Ji et al. (2016) and Ji & Wang (2017) employed the Weibull distribution to describe the influence of exposures and used LR to model intrinsic CVR (i.e., the conversion probability based on user features), which are combined into a probabilistic framework to estimate the conversion probability; Yoshikawa & Imai (2018) employed SA to model the distribution of the delay between click and conversion without a predefining parametric distribution, which is combined with LR to predict CVR in a specific period.

## 4.2. Tree models

Tree models have attracted considerable attention in online advertising because of their capabilities of handling heterogeneous features and aggregating information from diverse sources (Zhang et al., 2023). In the literature, gradient boosting decision tree (GBDT) and extreme gradient boosting (XGBoost) have been employed to make CVR prediction. Figure 4 presents the tree modeling framework for CVR prediction.

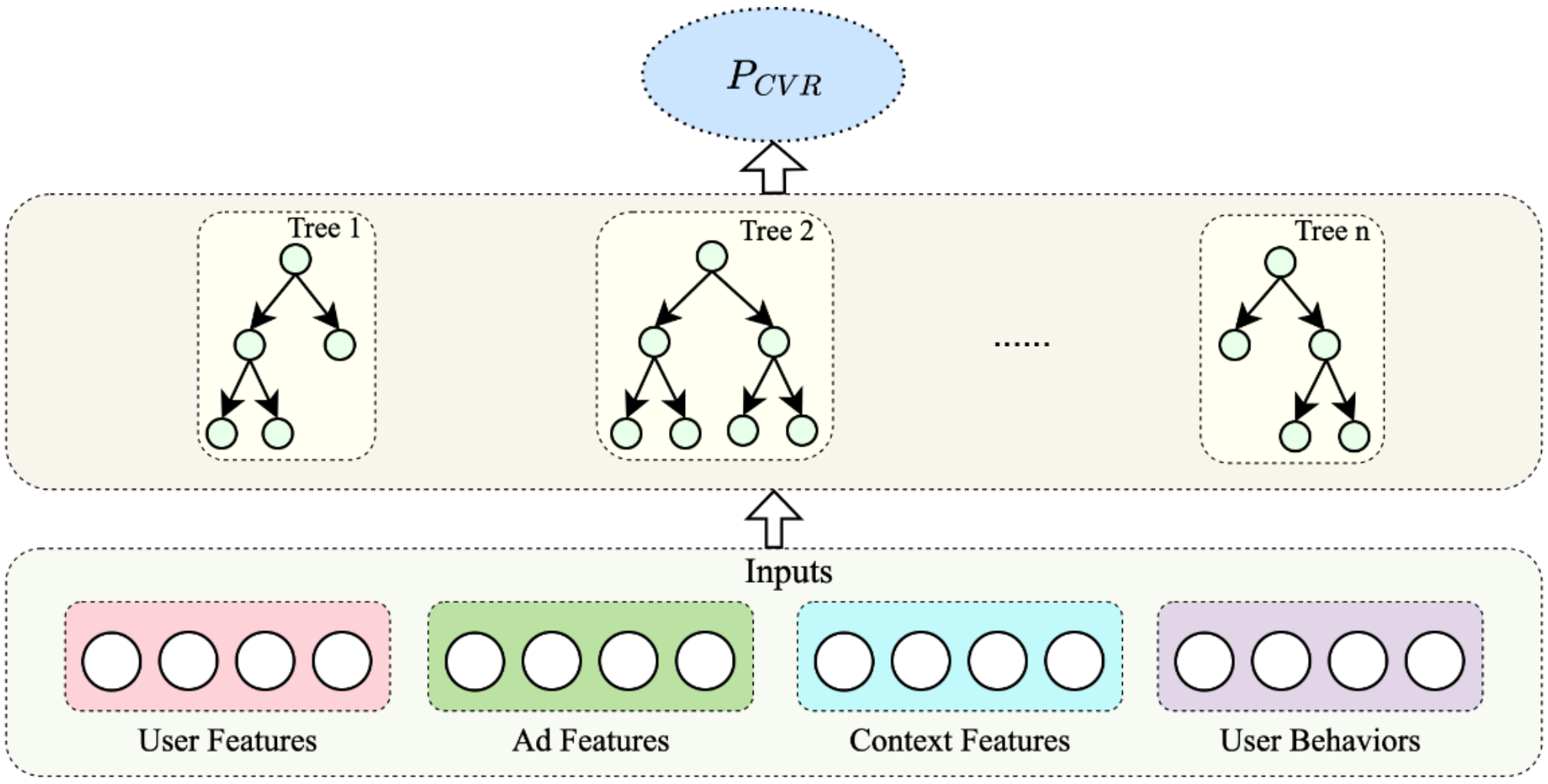

Figure 4. The tree modeling framework for CVR prediction.

### 4.2.1. Gradient boosting decision tree (GBDT)

GBDT is a representative tree model built upon the core principle of boosting. In advertising CVR prediction, GBDT iteratively trains multiple decision trees to approximate the predicted conversion probability.

Specifically, the $m-th$ iteration can be represented as $F_m(\boldsymbol{x}) = F_{m-1}(\boldsymbol{x}) + \eta f_m(\boldsymbol{x})$, where $F_m(\boldsymbol{x})$ is the boosting approximation to the predicted CVR, $\boldsymbol{x}$ is the feature vector, $\eta$ is the learning coefficient, and $f_m(\boldsymbol{x})$ is the regression tree, which partitions the feature space into disjoint regions and assigns a leaf value to each region. $f_m(\boldsymbol{x})$ is learned by fitting the first-order negative gradient. For details about GBDT, refer to Appendix B.1. Through the partitioning mechanism, GBDT is able to model complex non-linear combinations of features, and provides reliable estimates of post-click conversion probability.

GBDT has several advantages (Du et al., 2019a): (1) it is interpretable in that it explicitly reveals how each feature contributes to CVR prediction; (2) it is robust to outliers and missing values; (3) it excels at handling heterogeneous data. However, GBDT is prone to overfitting, resulting in the limited generalization on high-dimensional, sparse advertising datasets.

Lu et al. (2017) used GBDT to construct a data-driven tree that only activates high-confidence leaf nodes, applied the Beta-Binomial model to estimate the conversion probability for each leaf node, and combined the results from all eligible trees using a variance-weighted average to produce the conversion probability; Du et al. (2019a) used GBDT to iteratively train an ensemble of decision trees by minimizing the Logloss, and computed the CVR prediction by summing the tree outputs.

### 4.2.2. Extreme gradient boosting (XGBoost)

XGBoost is an extension of GBDT, which uses the first-order gradient and the second-order gradient of the loss function to estimate the gain at each round of tree construction, in order to facilitate the approximation of conversion probability. Moreover, XGBoost incorporates a regularization term to control the model complexity and prevent the overfitting problem.

Formally, the regularized objective function $\mathcal{L}_m$ of XGBoost is defined as $\mathcal{L}_m = \sum_{i=1}^{n} l(y_i, y_i^{m-1} + f_m(\boldsymbol{x}_i)) + \sum_{m=1}^{M} \Omega(f_m)$, where $l(y_i, y_i^{m-1} + f_m(\boldsymbol{x}_i))$ is the loss of sample $\boldsymbol{x}_i$, $y_i^{m-1}$ is the predicted logits of conversion from previous round, $f_m(\boldsymbol{x}_i)$ is the base learners, and $\Omega(f_m)$ is the regularization term. The second-order Taylor expansion is employed to approximate the above loss function, i.e., $\mathcal{L}_m = \sum_{i=1}^{n}[g_i f_m(\boldsymbol{x}_i) + \frac{1}{2} h_i f_m(\boldsymbol{x}_i)^2] + \Omega(f_m)$, where $g_i = l'(y_i, \hat{y}_i)$ and $h_i = l''(y_i, \hat{y}_i)$ are the first and second gradients of the loss function, respectively.

XGBoost has several advantages: (1) it is sparsity-aware; (2) it controls model complexity and prevents overfitting by using regularization terms; (3) it supports parallel computing (Liu & Liu, 2021). However, the efficiency and scalability of XGBoost are degraded on high-dimensional and large-scale data.

Liu & Liu (2021) and Zhang et al. (2023) extracted statistical and temporal features from large-scale advertising logs, and utilized XGBoost to learn a regularized ensemble of regression trees and sum outputs to obtain the conversion probability.

## 4.3. Deep learning models

Although multivariate statistical models and tree models are simple and easy to implement, they fail to capture non-linear feature interactions and sequential dependencies (Su et al., 2021). Hence, deep learning models have been developed to capture expressive representations for CVR prediction. In the literature, five deep learning models have been employed for CVR prediction, including multilayer perceptron (MLP), deep neural networks (DNNs), recurrent neural networks (RNNs), Transformer and temporal convolution networks (TCNs).

### 4.3.1. Multilayer perceptron (MLP)

MLP transforms a set of features into meaningful representations ($\boldsymbol{H}$) through a hidden layer that consists of multiple neurons and non-linear activation functions, i.e., $\boldsymbol{H} = MLP(\boldsymbol{x}; \boldsymbol{\theta})$, where $\boldsymbol{x}$ is the feature vector and $\boldsymbol{\theta}$ is the set of parameters. The representations are then fed into a prediction layer with a sigmoid function to estimate the conversion probability. For details about MLP, refer to Appendix B.2. Figure 5 presents the MLP modeling framework for

CVR prediction.

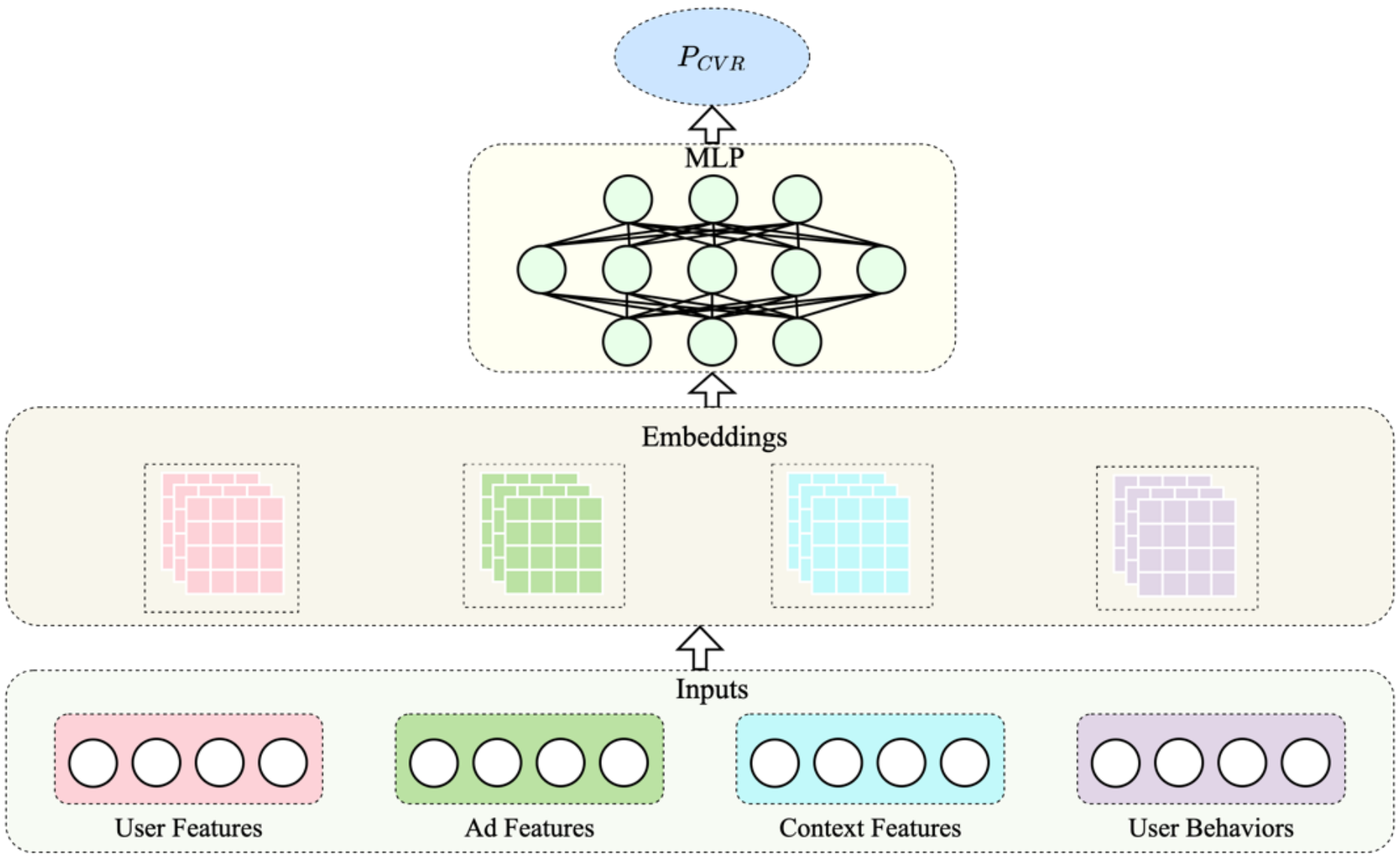

Figure 5. The MLP modeling framework for CVR prediction.

Although MLP can capture complex non-linear relationships among features by learning non-linear functions (Mao et al., 2023), a vanilla MLP is inefficient to model multiplicative feature interactions (Wang et al., 2021).

MLP is usually used as a core building block in CVR prediction models to learn feature representations (Shtoff et al., 2023; Tan et al., 2023). Well-trained CVR prediction models usually perform sub-optimally in sales promotion. To this end, Chan et al. (2023) retrieved historically similar promotions as auxiliary training data to fine-tune the CVR prediction model, and developed an MLP-based transition module to capture the transition from non-promotional to promotional conversions. Online advertising systems may support diverse conversion types (e.g., purchasing, registering and downloading) and various ad formats (e.g., banner ads and email ads). To capture such heterogeneity, Ouyang et al. (2023b) proposed a multi-domain CVR prediction framework where each pair of conversion type and ad format was formulated as a domain and each CVR prediction tower employed an MLP with domain-specific parameters, and obtained the final predicted CVR by aggregating results from multiple domain-specific towers via an auto-masking mechanism. Min et al. (2023) developed an MLP-based pluggable module to generate sequential features for users with sparse behaviors based on their attributes (e.g., gender, occupation and hobbies), which were fed into the prediction model along with the original sequence features to obtain the CVR.

### 4.3.2. Deep neural networks (DNNs)

Compared to MLP, DNNs employ deeper architectures composed of multiple stacked hidden layers to progressively learn complex feature representations, with deeper layers capturing higher-level feature representations from lower-level inputs (Wang et al., 2017; Ling et al., 2017; Yang & Zhai, 2022).

Formally, the updating process of the hidden layer is formulated as $\boldsymbol{h}_l = \sigma_l(\boldsymbol{W}_l \boldsymbol{h}_{l-1} + b_l), l = 1,2, \ldots, L$, where $\boldsymbol{h}_l$ denotes the hidden state of the $l-th$ hidden layer, $\sigma_l$ is activation function, $\boldsymbol{W}_l$ is the learned weight matrix, and $b_l$ is the bias. For more details about DNNs, refer to Appendix B.3. Figure 6 presents the DNNs modeling framework for CVR prediction.

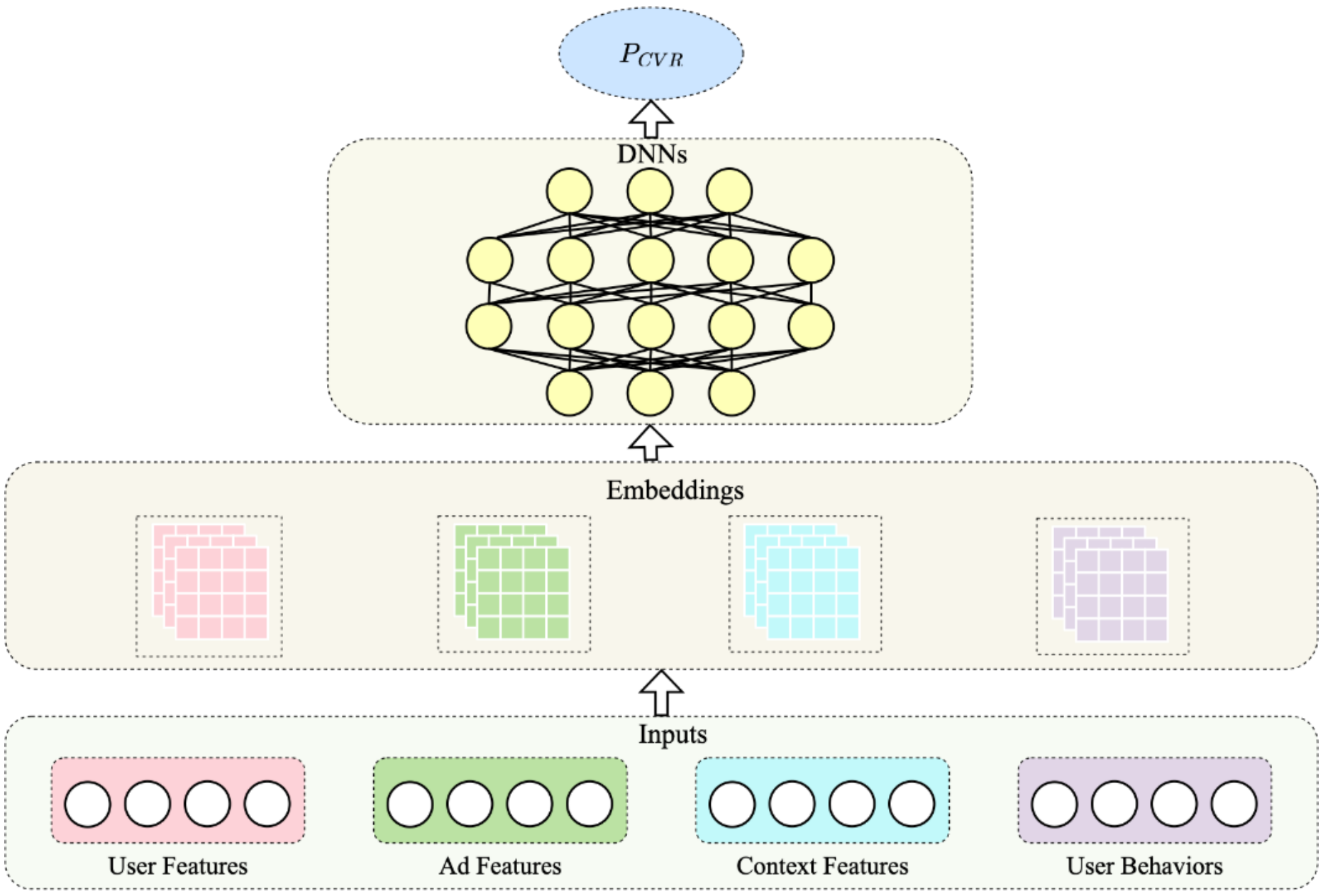

Figure 6. The DNNs modeling framework for CVR prediction.

DNNs are capable of learning higher-level and non-linear feature interactions through stacked hidden layers (Jin et al., 2023). However, DNNs have several limitations: (1) they fail to capture sequential patterns in user behaviors; (2) their computational complexity grows with the depth of hidden layers (Biçici, 2025); (3) they suffer from limited interpretability, which hinders the identification of key features influencing prediction results.

In the literature on online advertising, DNNs have been employed to learn complex feature representations for estimating conversion probabilities (Guo et al., 2023; Jin et al., 2023; Zhao et al., 2024). Considering the trade-off between label accuracy and data freshness in the delayed feedback issue, Liu et al. (2023a) employed a DNNs-based main model trained with a long waiting window to ensure high label accuracy, along with two DNNs-based satellite models trained on the latest data, and calculated the CVR by aggregating the output logits from the main model and satellite models. Liu et al. (2024a) developed a DNNs-based multi-interval screening model with multiple output heads to independently estimate conversion probabilities within different waiting windows, and generated dynamic weights via dense layers to aggregate the head-specific predictions.

Recently, researchers have made efforts to improve the training stability and learning capacity of deeper DNNs. Ding et al. (2022) proposed a transfer learning method to address the overfitting problem of DNNs by using a gradient pruning technique, which was implemented in a three-layer DNN module for CVR prediction. To mitigate the vanishing gradient problem with the increasing depth of deep learning models, Biçici (2025) integrated residual connections into DNNs component to learn more complex and abstract representations for CVR prediction.

#### 4.3.3. Recurrent neural networks (RNNs)

In online advertising, users usually engage in a series of interactions (e.g., impressions and clicks) with ads before conversions (Jin et al., 2025). However, feed-forward neural networks (e.g., MLP and DNNs) are limited in capturing sequential patterns from user behaviors. To address this issue, RNNs have been adopted to capture sequential dependencies for CVR prediction (Ban et al., 2024b). For details about RNNs, refer to Appendix B.4. Figure 7 presents the RNNs modeling framework for CVR prediction.

Although RNNs perform well in processing user-ad interactions, they suffer from gradients vanishing and exploding, which hinders learning long-term dependencies of interaction behaviors. To mitigate these issues, several RNNs variants, e.g., long short-term memory (LSTM) and gated recurrent unit (GRU), have been proposed by introducing gating mechanisms to control the flow of information and update memory states (Su et al., 2021). Specifically, LSTM introduces the memory cell $\boldsymbol{c}_t$ and three gates (i.e., the input gate $\boldsymbol{i}_t$, the forget gate $\boldsymbol{f}_t$, and the output gate $\boldsymbol{o}_t$) to learn sequential representations by selectively retaining and updating information of interaction sequences. The learning process is defined as $\boldsymbol{h}_t = LSTM_t(\boldsymbol{x}_t; \theta(\boldsymbol{c}_t, \boldsymbol{i}_t, \boldsymbol{f}_t, \boldsymbol{o}_t))$, where $\boldsymbol{x}_t$, $\boldsymbol{h}_t$ and $\theta$ are the input feature vector, output representation and learnable parameters of the $t$-th LSTM unit, respectively. As a simplified variant of LSTM, GRU employs the update gate $\boldsymbol{z}_t$ and the reset gate $\boldsymbol{r}_t$ to efficiently capture sequential patterns with fewer parameters. Similarly, the computation process of GRU can be represented as $\boldsymbol{h}_t = GRU_t(\boldsymbol{x}_t; \theta(\boldsymbol{z}_t, \boldsymbol{r}_t))$. However, LSTM and GRU incur high computational costs, particularly in terms of training time and memory consumption.

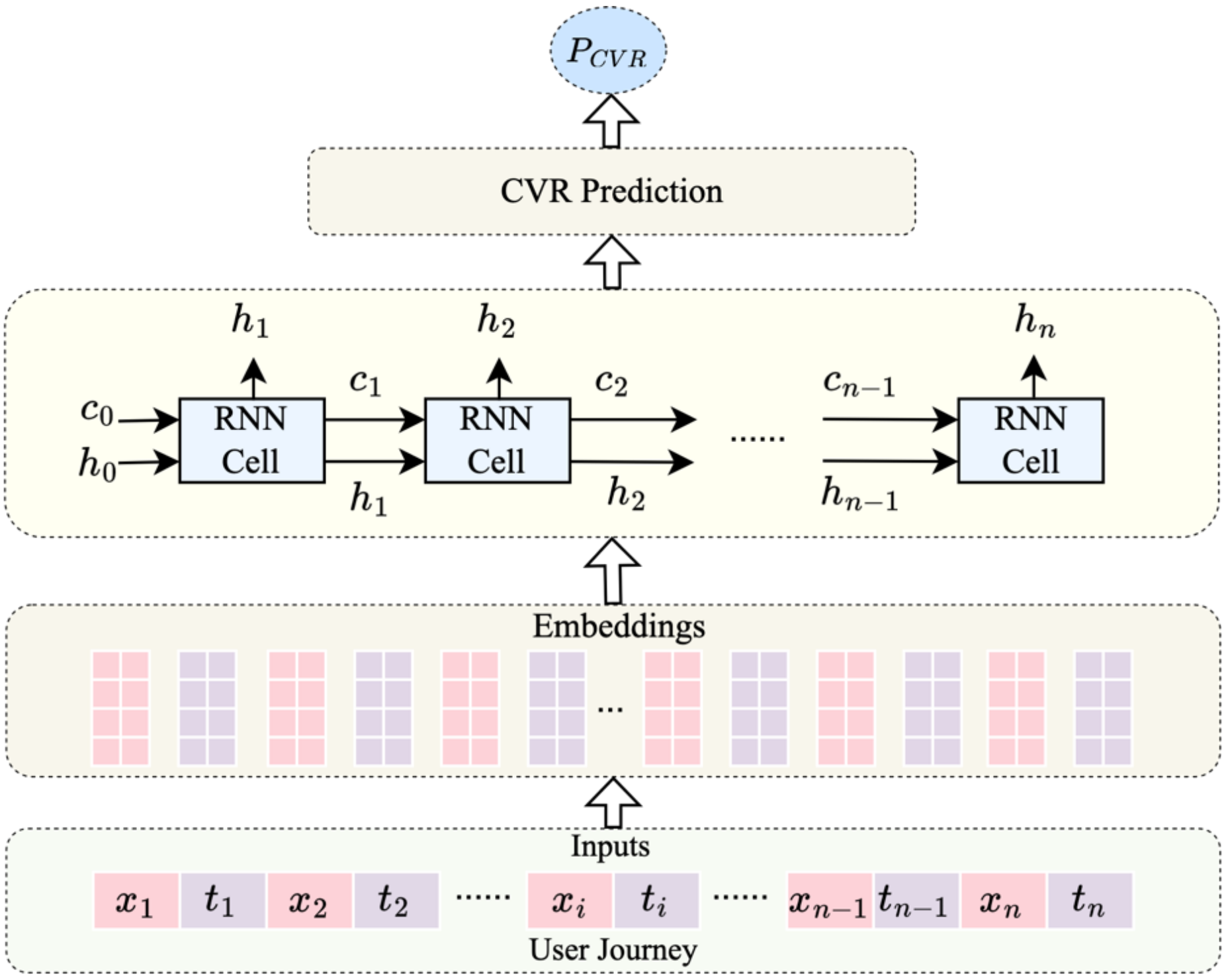

Figure 7. The RNNs modeling framework for CVR prediction.

RNNs and their variants are commonly used to capture the sequential dependencies in behavioral sequences. Qiu et al. (2020) used RNNs to capture sequential representations for calculating conversion probabilities; Cui et al. (2018) used multi-layer LSTM to capture dependencies in consumer journey and the Monte Carlo strategy to simulate temporal evolution of probabilities distribution in user behaviors for CVR prediction; Zhou et al. (2019) developed attention-based Bi-GRU for CVR prediction, which represents behavioral trajectory via Bi-GRU and assigns an attention weight to each activity using a global attention mechanism;

Gligorijevic et al. (2020) employed Bi-GRU to learn users' trajectory representations that were fed into fully connected layers to obtain conversion probabilities. Advertising identifiers can help to track logs of user-ad interactions and characterize user behaviors. Ban et al. (2024a) and (2024b) encapsulated advertising identifiers into structured journeys, and employed GRU with attention mechanisms to learn journey representations for CVR prediction.

In online advertising, the attribution analysis provides valuable information for enhancing CVR prediction (Du et al., 2019b; Yang et al., 2020). Li et al. (2018) utilized LSTM to learn representations of touchpoint sequences, and applied an attention mechanism to compute attribution weights for each touchpoint, which are fused with user-context information to obtain the CVR; Ren et al. (2018) designed a dual-attention RNNs that simultaneously apply attention mechanisms on impression-level and click-level hidden states that were dynamically combined to learn conversion attributions of impression and click behaviors.

#### 4.3.4. Transformer

Although RNNs and their variants can model the dependencies in user behaviors, the hidden state updates are executed sequentially along behavioral sequences, which makes training on longer sequences memory-intensive. In contrast, Transformer utilizes self-attention, which directly computes the dependencies between each behavior and all other behaviors in parallel, to efficiently capture long-range dependencies in behavioral sequences (Zhang et al., 2025b).

For a behavioral sequence, positional information in the sequence is encoded as a set of positional embeddings $\boldsymbol{P} = \{\boldsymbol{p}_1, \boldsymbol{p}_2, \dots, \boldsymbol{p}_N\}$, which are added into the input embeddings $\boldsymbol{E} = \{\boldsymbol{e}_1, \boldsymbol{e}_2, \dots, \boldsymbol{e}_N\}$ to obtain the positional sequence embeddings $\boldsymbol{X} = \boldsymbol{P} + \boldsymbol{E}$. Subsequently, $\boldsymbol{X}$ is mapped into three vectors (i.e., query $\boldsymbol{Q}$, key $\boldsymbol{K}$, and value $\boldsymbol{V}$), and processed through multi-head self-attention mechanism to capture sequential dependencies, i.e., $Multihead(\boldsymbol{Q}, \boldsymbol{K}, \boldsymbol{V}) = [head^{(1)}, head^{(2)}, \dots, head^{(n)}]\boldsymbol{W}_0$, where $n$ is the number of attention heads, $\boldsymbol{W}_0$ is a learnable matrix, and each head is computed as $head^{(h)} = Softmax(\boldsymbol{Q}\boldsymbol{K}^T/\sqrt{d})\boldsymbol{V}$. The output of the multi-head attention is passed through a position-wise feed-forward network with residual connections and layer normalization to produce representations for CVR prediction. For details about Transformer, see Appendix B.5. Figure 8 presents the Transformer modeling framework for CVR prediction.

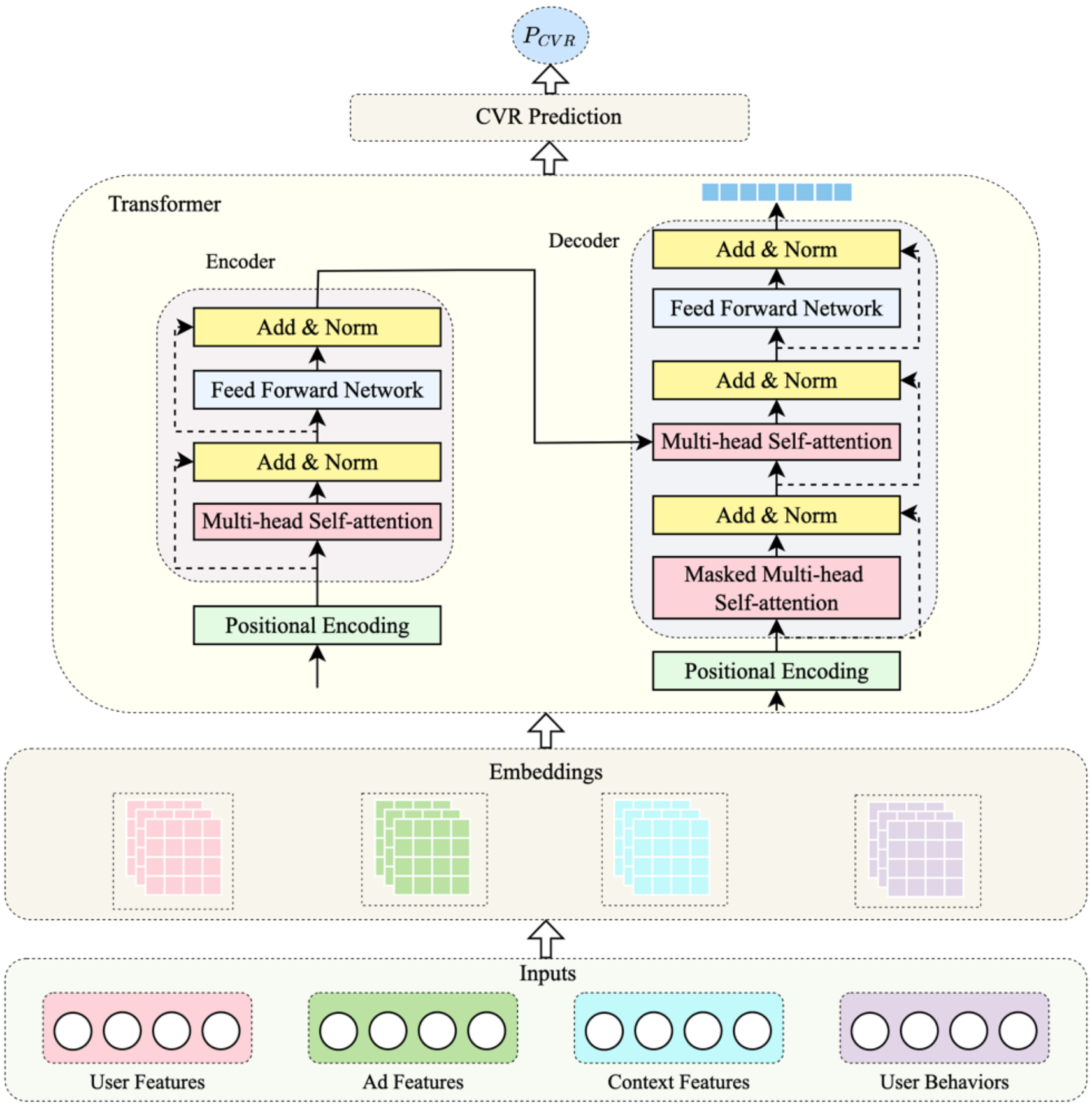

Figure 8. The Transformer modeling framework for CVR prediction.

Transformer processes sequential data in parallel through the self-attention mechanism, substantially improving the training speed. However, the quadratic complexity with respect to the sequence length in self-attention computation leads to parameter inefficiency and considerable computational demands (Zhang et al., 2025b).

Li et al. (2021a) combined a Transformer-based encoder learning low-level features from user behaviors and a capsule network extracting multiple latent interests, and designed a modified dynamic routing algorithm to facilitate CVR prediction; Yang et al. (2023) introduced a Transformer module to enhance cold sample representations by extracting related information from warm samples in the same batch, and used a two-layer MLP classifier to predict conversion probabilities. In order to protect data privacy and reduce communication costs between data environments and ad platforms, Li et al. (2024) proposed a model training framework for privacy-preserving CVR prediction, where batch-level aggregated gradients and label differential privacy techniques were employed to prevent information leakages, and LoRA adapters were used to fine-tune parameters of the Transformer and MLP modules for reducing communication costs.

### 4.3.5. Temporal convolution networks (TCNs)

Compared to RNNs where the prediction for the current step must wait for earlier ones to be completed, TCNs can perform convolutions in parallel as the same filter is used in each convolution layer and extract features from longer sequences using dilated convolutions with exponentially large receptive fields (Lea et al., 2017).

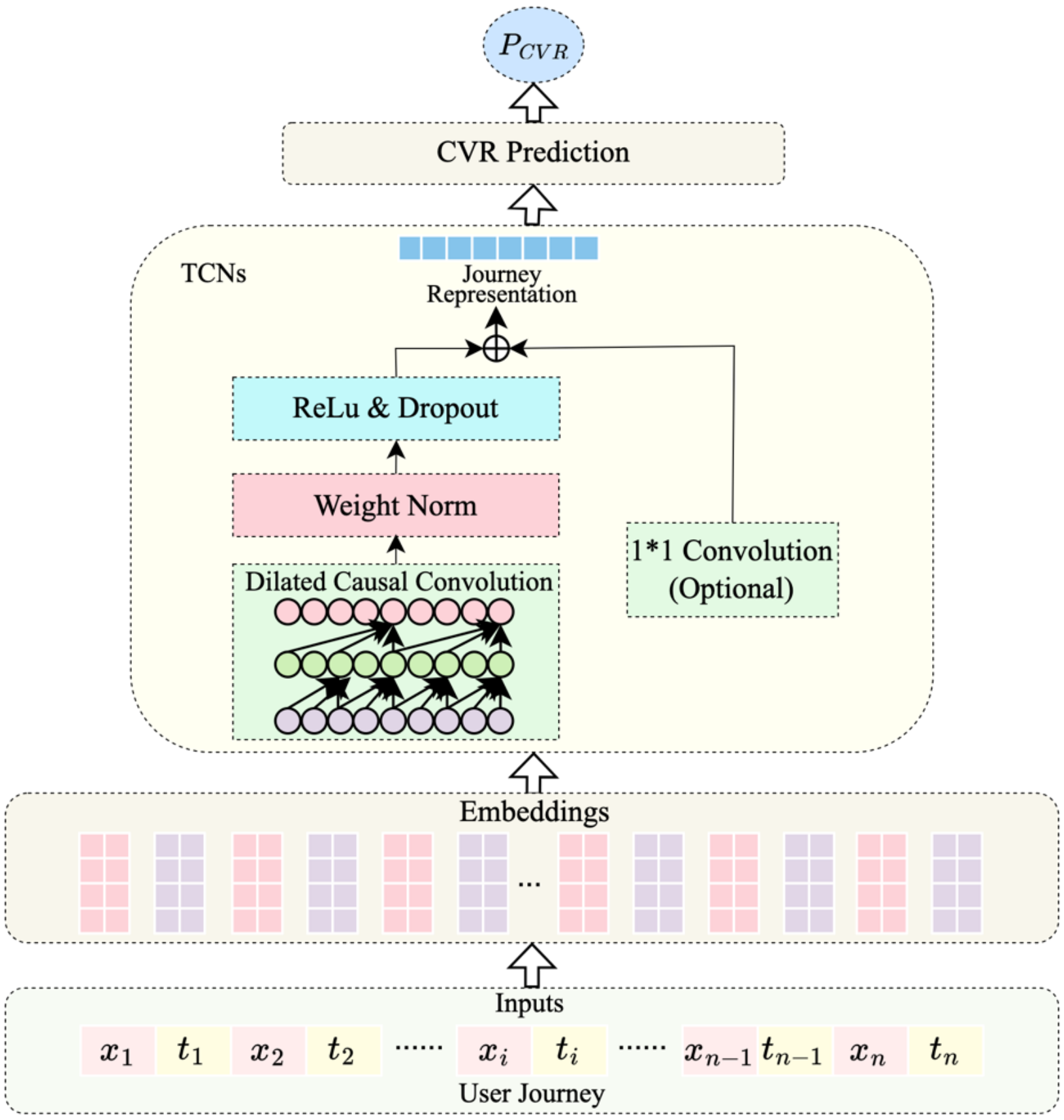

Figure 9. The TCNs modeling framework for CVR prediction.

For a behavioral sequence $X = [\boldsymbol{x}_1, \boldsymbol{x}_2, \dots, \boldsymbol{x}_T]$, TCNs employ causal dilated convolution at $\boldsymbol{x}_t$ to capture the representation of $t$-th behavior, formulated as $F(\boldsymbol{x}_t) = \sum_{i=0}^{k-1} \boldsymbol{f}_i \cdot \boldsymbol{x}_{t-d \cdot i}$, where $\boldsymbol{f}_i \in (\boldsymbol{f}_1, \boldsymbol{f}_2, \dots, \boldsymbol{f}_k)$ is the filter, $k$ is the filter size, and $d$ is the dilated factor which increases exponentially with the network depth. Hence, the residual connection is incorporated to facilitate the training stabilization of deeper TCNs, i.e., $\boldsymbol{h}_t = F(\boldsymbol{x}_t) + \boldsymbol{x}_t$. Finally, the learned representation $\boldsymbol{h}_t$ is fed into the prediction layer to compute conversion probabilities.

Figure 9 presents the TCNs modeling framework for CVR prediction.

TCNs have several advantages: (1) they can capture longer dependencies in behavioral sequences via dilated convolutions and the expanded receptive field, than RNNs; (2) they support parallelization during training and evaluation, thus improving the computational efficiency; (3) they can control the model's memory by adjusting the receptive field size; (4) they provide more stable gradients than RNNs, in that backpropagation is not constrained in the temporal direction while handling long and complex exposure sequences (Agrawal et al., 2022). However, TCNs struggle to extract internal correlation information from behavioral sequences, and the growing receptive field size may affect the model's stability (Hao et al., 2020).

In the B2B advertising, Agrawal et al. (2022) divided the customer journey into multiple stages, employed TCNs to capture long-term sequential dependencies in each stage, and concatenated the output vector of each TCN to predict the CVR. Xie et al. (2022) employed TCNs to extract temporal features from the user journey, employed an attention mechanism to model the importance of different touchpoints for feature enhancement, and fed the augmented representations into a fully connected layer to obtain the conversion probability.

## 4.4. Causal learning models

The performance of CVR prediction models in online advertising is often degraded by three primary types of bias, including the label bias, the sample selection bias and the confounding bias. Among them, the label bias arises from delayed feedback, i.e., conversions usually occur long after clicks, making recently clicked but not-yet-converted samples be incorrectly labeled as negatives (Guo et al., 2022); the sample selection bias stems from the difference in data distributions between the training space and the inference space, i.e., training CVR prediction models on click samples while making inferences on impression samples (Ma et al., 2018); and the confounding bias[4] is referred to the spurious correlation in observed data induced by user preferences, e.g., the observed correlation does not necessarily imply a causal relationship between impressions and conversions (Kumar et al., 2020). In the literature on CVR prediction, two causal learning frameworks have been employed to mitigate these biases, including counterfactual learning and counterfactual recurrent networks (CRNs).

[4] The confounding bias is a systematic error in observed data that arises when there are confounding variables simultaneously influencing both the treatment and the outcome, causing the estimated causal effect to deviate from its true value. In CVR prediction, confounding bias emerges because user preferences may influence both exposures and conversions (Kumar et al., 2020).

#### 4.4.1. Counterfactual learning

Counterfactual learning explicitly models both the factual CVR (i.e., the conversion probability in the observed data) and the counterfactual CVR (i.e., the conversion probability in unobserved data) to alleviate the label bias and the sample selection bias in CVR prediction (Gandhudi et al., 2023; Wang et al., 2023c). Specifically, for addressing the label bias, counterfactual learning methods employ temporal modeling to estimate conversion probabilities in unobserved samples, and incorporate these estimates into the training loss to correct the underestimation of true conversions (Yasui et al., 2020; Gu et al., 2021); for addressing the sample selection bias, counterfactual learning reweights clicked samples to approximate the distribution of the entire impression space, enabling CVR prediction models to generate conversion probabilities that are no longer limited to the click space (Zhang et al., 2020). In the literature, three counterfactual learning methods have been employed to achieve unbiased CVR estimations, including importance weighting, inverse propensity scoring and doubly robust.

**(1) Importance weighting** estimates the true probability distribution from a biased one by adjusting the contribution of individual sample in the loss function (Yang et al., 2021). The loss function for importance weighting models is given as

$$\mathcal{L}(\theta) = -\frac{1}{N}\sum_n w(x_n, y_n) log f_\theta\,(y_n|x_n), \quad (2)$$

where $w(x_n, y_n)$ denotes the weights of positive or negative samples, $f_\theta$ represents the CVR prediction model, $x_n$ is the sample, and $y_n$ denotes whether or not the sample eventually results in a conversion.

**(2) Inverse propensity scoring** provides an unbiased estimation of the ideal CVR loss by assigning a weight to each sample that is inversely proportional to the probability of a sample being clicked (i.e., the propensity score[5]) (Saito et al., 2020). The loss function for inverse propensity scoring models is given as

$$\mathcal{L}_{IPS} = \frac{1}{|\mathcal{D}|}\sum_{(u,a)\in\mathcal{D}}\frac{o_{u,a}e_{u,a}}{\hat{p}_{u,a}}, \quad (3)$$

where $o_{u,a} \in \{0,1\}$ denotes the click indicator, $e_{u,a} = \delta(r_{u,a}, \hat{r}_{u,a})$ is the prediction error, $\delta$ is the cross-entropy loss, $r_{u,a} \in \{0,1\}$ represents the conversion outcome, where $r_{u,a} = 1$ if user $u$ converts after clicking the ad $a$, otherwise 0; $\hat{r}_{u,a} \in [0,1]$ is the predicted CVR, and $\hat{p}_{u,a}$ is the predicted clicked probability.

**(3) Doubly robust** is the augmented inverse propensity scoring designed to address the high variance caused by highly sparse feedback in CVR prediction, by adding an imputation estimator to improve the estimation stability (Saito, 2020). Typically, the loss function for doubly robust models is given as

$$\mathcal{L}_{DR} = \frac{1}{|\mathcal{D}|}\sum_{(u,a)\in\mathcal{D}}(\hat{e}_{u,a} + \frac{o_{u,a}(e_{u,a}-\hat{e}_{u,a})}{\hat{p}_{u,a}}), \quad (4)$$

where $\hat{e}_{u,a}$ is the imputed error predicted by the imputation estimator.

Counterfactual learning methods are effective in addressing the missing-not-at-random problem to obtain the unbiased CVR prediction. However, they heavily rely on the prediction accuracy of auxiliary tasks (e.g., CTR prediction) and imputation error estimations, and the estimation of propensity scores easily suffers from high variance (Fei et al., 2025). Moreover, inverse propensity scoring and doubly robust-based methods for mitigating sample selection bias typically do not incorporate unclicked samples during the model training, as they rely solely on post-click data for estimation (Su et al., 2024).

To mitigate the label bias in CVR prediction, importance weighting has been employed to prevent samples that are unconverted within a specific time window from being incorrectly treated as negative samples (Gu et al., 2021; Yasui & Kato, 2022; Ding et al., 2025). Yasui et al. (2020) proposed an importance weighting-based CVR prediction model to handle delayed feedback by setting a counterfactual deadline, where samples clicked later than the deadline are discarded so as to construct an artificial dataset that has samples converted between click and deadline, which was used to estimate the probabilities that samples are correctly or incorrectly labeled and make CVR prediction. Following this direction, in order to obtain accurate feedback signals while keeping model-freshness, Yang et al. (2021) proposed an elapsed-time distribution to correct labels of samples through importance weighting, which is used to construct a weighted loss function for training the CVR prediction model.

[5] The propensity score measures the probability of observing a sample under specific conditions, e.g., the conditional probability of an ad being clicked in CVR prediction.

Inverse propensity scoring facilitates the CVR prediction by up-weighting samples from users with low clicking propensity and down-weighting those from users with high clicking propensity. Saito et al. (2020) constructed an unbiased estimator learning inverse propensity scores to reweight observed conversion samples, and trained the unbiased estimator with a CVR predictor using only observed positive conversion data to obtain an unbiased CVR estimation.

Doubly robust incorporates an imputation estimator to estimate CVR prediction errors and facilitate the unbiased prediction over the entire impression space. Note that inverse propensity scoring and doubly robust are generally integrated into multi-task learning frameworks for unbiased CVR prediction (Zhang et al., 2020; Dai et al., 2022; Wang et al., 2022; Zhu et al., 2023; Su et al., 2024), which will be discussed in detail in Section 4.6.

### 4.4.2. Counterfactual recurrent networks (CRNs)

In online advertising, user preferences are a common determinant of advertising exposures and conversions, which introduces the confounding bias to CVR prediction (Kumar et al., 2020). Counterfactual learning methods (e.g., inverse probability weighting and propensity scores) focus on debiasing user preferences derived from static attributes (Yao et al., 2022); in contrast, CRNs simultaneously mitigate confounding biases from users' static attributes and dynamic behaviors to capture unbiased causal effects of exposures on conversions.

For a user-ad interaction sequence, CRNs leverage an RNN architecture to learn the latent state representation $\boldsymbol{s}_t^n$ of the $t$ -th touchpoint, which is given as $\boldsymbol{s}_t^n = f(\boldsymbol{x}_t^n, \boldsymbol{x}_{t-1}^n, \dots, \boldsymbol{x}_1^n, \boldsymbol{c}_{t-1}, \dots, \boldsymbol{c}_1, z_t^n, z_{t-1}^n, \dots, z_2^n)$, where $[\boldsymbol{x}_t^n, \boldsymbol{x}_{t-1}^n, \dots, \boldsymbol{x}_t^n]$ are feature vectors, $[\boldsymbol{c}_{t-1}, \dots, \boldsymbol{c}_1]$ represent advertising channels, and $[z_t^n, z_{t-1}^n, \dots, z_2^n]$ are click outcomes. In order to eliminate confounding effects induced by user preferences, CRNs use a gradient reversal layer (GRL)[6] and adversarial learning to obtain unbiased causal representations $\boldsymbol{r}_t^n$ of the $t$-th touchpoint, which is given as $\Phi_t: \boldsymbol{s}_t^n \rightarrow \boldsymbol{r}_t^n$, where $\Phi_t$ is a transformation to minimize confounding bias. For details about CRNs, refer to Appendix B.6. The CRNs modeling framework for CVR prediction is presented in Figure 10.

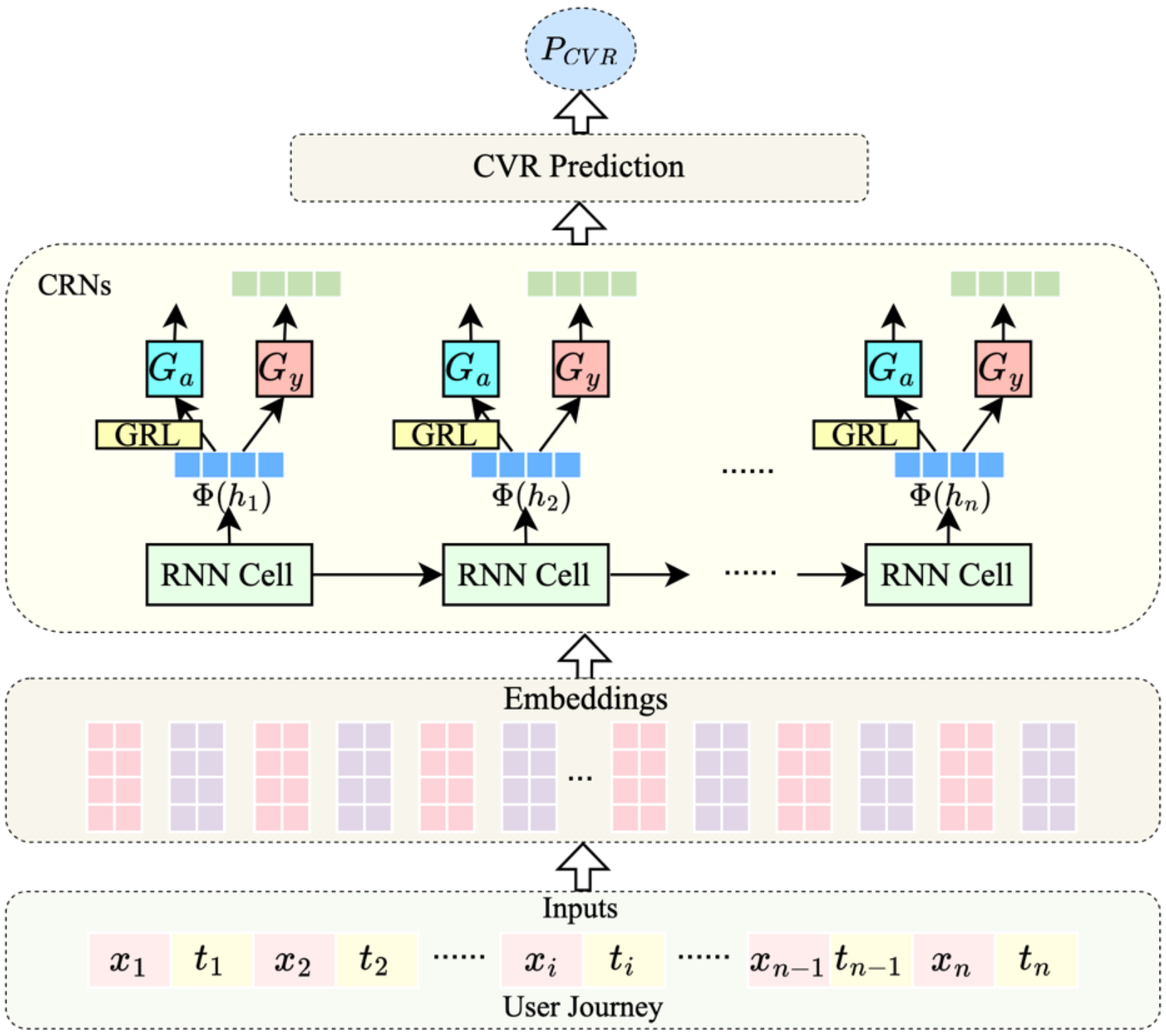

Figure 10. The CRNs modeling framework for CVR prediction.

Although CRNs can reduce confounding bias resulting from user preferences through adversarial learning, they fail to capture long-range dependencies in the conversion journey, due to their adoption of RNN frameworks.

In order to eliminate confounding factors caused by user preferences, Kumar et al. (2020) used CRNs to learn unbiased representations of user-ad interactions, employed an attention layer to compute the attribution credit for each touchpoint and an MLP to obtain conversion probabilities; Yao et al. (2022) decomposed user preferences into static biases stemming from user attributes and dynamic biases arising from the browsing history, employed the variational recurrent auto-encoders to characterize generation probabilities of pure channel sequences and CRNs to learn balancing representations of user journeys for CVR prediction; Tang et al. (2024)

[6] GRL is an adversarial learning technique without learning parameters and serves as an identity function in forward propagation but reverses the sign of the passing gradient in backpropagation. In advertising CVR prediction, GRL is employed to force models to learn representations that are independent of advertising exposures.

employed CRNs and causal attention to learn invariant causal representations of touchpoints and the journey for CVR prediction.

## 4.5. Knowledge distillation models

Although counterfactual learning methods can, to some extent, mitigate the label bias and the sample selection bias, they rely on the prediction accuracy of auxiliary tasks (e.g., CTR prediction), and ignore unclicked samples during the training process (Xu et al., 2022). To this end, knowledge distillation has been proposed to address the label bias and the sample selection bias in CVR prediction.

Knowledge distillation is designed with a teacher-student framework, where a teacher model is first trained on observed conversion data and generates soft pseudo-labels of unobserved samples (i.e., clicked but no-converted samples and unclicked samples), then a lightweight student model (i.e., the CVR estimator) learns from these soft pseudo-labels, effectively distilling knowledge from the teacher model on the full data distribution with fewer parameters. In other words, knowledge distillation mitigates the label bias and the sample selection bias by training and inferring in the entire space with observed and unobserved samples (Guo et al., 2022; Yuan et al., 2025). The distillation loss guides the student model to learn from the teacher model during the training stage, which is given as

$$\begin{cases} \mathcal{L}_{soft} = T^2 \times KL(p^T \parallel q^T) \\ \mathcal{L}_{hard} = -\sum_j^N c_j \, log\left(q_j^T\right) \\ \mathcal{L}_{KD} = \alpha\mathcal{L}_{soft} + \beta\mathcal{L}_{hard} \end{cases}, \quad (5)$$

where $\mathcal{L}_{KD}$ denotes the total loss of the knowledge distillation model, $\mathcal{L}_{hard}$ is the cross-entropy loss of the student model, $\mathcal{L}_{soft}$ is the Kullback-Leibler divergence of the teacher's soft outputs; $T$ is the distillation temperature, $p^T$ and $q^T$ denote softmax outputs of the teacher model and the student model, respectively; $c_j \in \{0,1\}$ denotes the true conversion label, $\alpha$ and $\beta$ are used to control weights of $\mathcal{L}_{hard}$ and $\mathcal{L}_{soft}$, respectively. For details about knowledge distillation, refer to Appendix B.7. The knowledge distillation modeling framework for CVR prediction is illustrated in Figure 11.

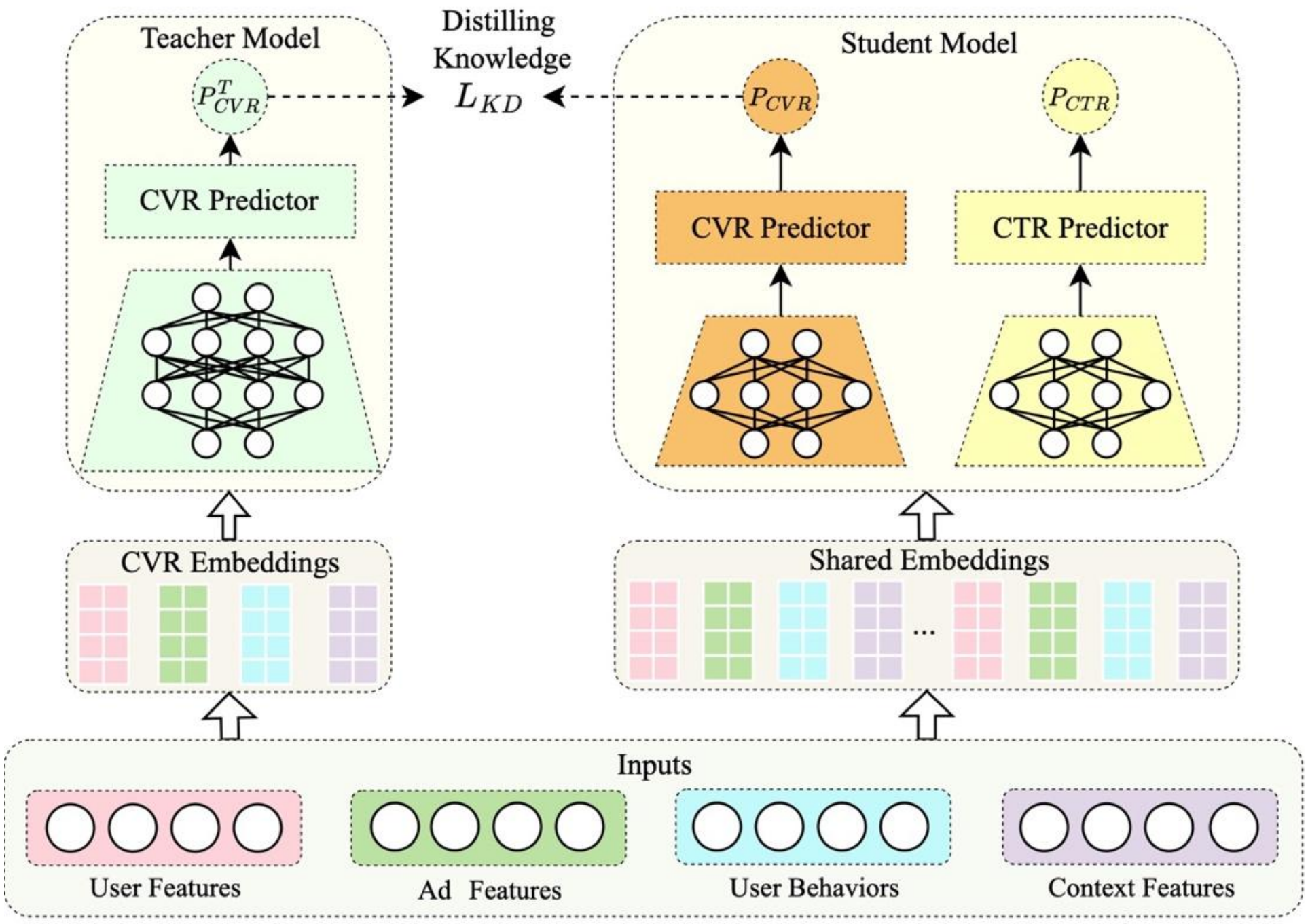

Figure 11. The modeling framework of knowledge distillation for CVR prediction.

Knowledge distillation models can reduce computational costs of CVR prediction by allowing the compact student model to effectively learn valuable knowledge from the larger teacher model via diverse distillation mechanisms (Mansourian et al., 2025). However, the prediction performance of knowledge distillation models critically depends on the quality of the generated pseudo-labels, which are prone to biases. Moreover, when there is a significant capacity gap between the teacher model and the student model in that the latter may fail to fully absorb the knowledge transferred from the former, leading to suboptimal knowledge transfer.

To account for the label bias, Guo et al. (2022) combined a base CVR prediction model and a teacher-student model to estimate CVR, where the base CVR prediction model is trained on freshly observed data, and the teacher model was trained on historical data with complete feedbacks and transferred its knowledge to the student model, allowing to learn from samples with observed labels and pseudo-labels. For addressing the sample selection bias, Xu et al. (2022) proposed a click-adaptive teacher model to generate pseudo-conversion labels for unclicked samples, introduced an uncertainty-aware learning mechanism and a label distillation approach to dynamically distill the knowledge from pseudo-labels and filter out noise, and employed the CVR prediction tower in the student model to predict CVR by leveraging the distilled knowledge and the information from users, ads, and contexts. However, existing pseudo-labeling methods for unclicked samples ignore the bias in label estimation. To address this issue, Fei et al. (2025) trained a teacher model consisting of CTR and CVR estimators to generate unbiased pseudo-labels based on the predicted click propensity and the output of CVR representation learner, and then applied variational information and logit distillation to transfer the information of unclicked samples to the student model for CVR prediction.

In online advertising, some features are collected only for offline training and unavailable in online serving, which heavily degrades the model's generalization. Yuan et al. (2025) trained the teacher model to learn mappings from offline features to conversion labels, and then employed two focal-style distillation losses on online features to train the student model for CVR prediction, avoiding demand for offline features in online training and inference.

## 4.6. Multi-task learning models

In online advertising scenarios, treating CVR prediction as a single task often overlooks the underlying correlations and shared patterns across related behaviors, such as clicks and conversions (Ma et al., 2018; Xi et al., 2021). Hence, researchers have explored multi-task learning models for CVR prediction by incorporating auxiliary tasks (e.g., CTR prediction) to improve representation learning of user behaviors via shared embeddings and joint optimization (Wen et al., 2020; Zhu et al., 2023). In the literature, three multi-task learning models have been employed for CVR prediction, including hard parameter sharing, soft parameter sharing and expert sharing.

### 4.6.1. Hard parameter sharing

Hard parameter sharing is a multi-task learning structure that embeds shared features of CVR prediction and auxiliary tasks (e.g., CTR prediction) in the bottom layers and applies task-specific networks to learn representations for each task in the upper layers. Figure 12 presents the modeling framework of hard parameter sharing.

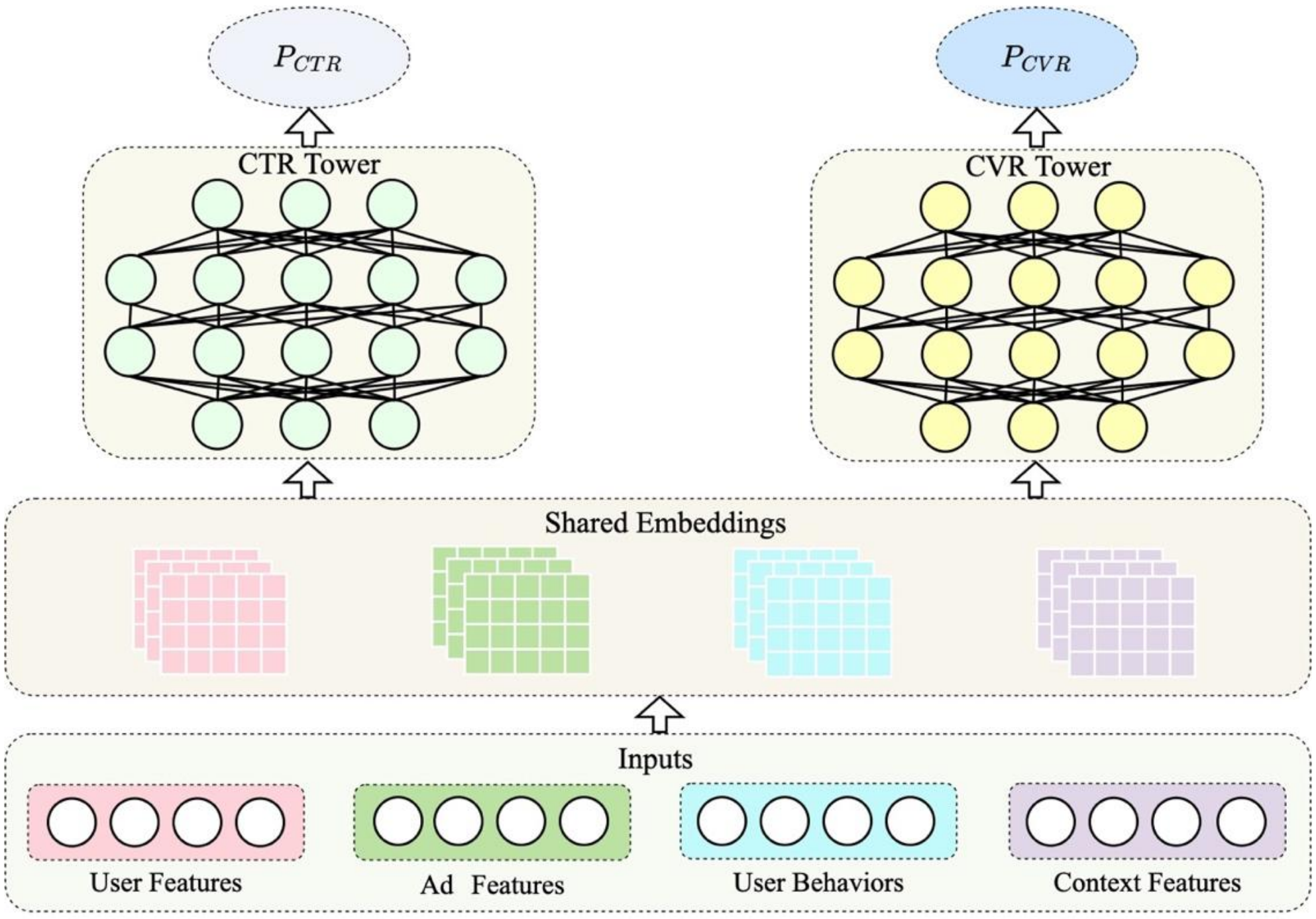

Figure 12. The modeling framework of hard parameter sharing for CVR prediction.

Hard parameter sharing allows CVR prediction tasks to benefit from the data with more frequent signals provided by other tasks such as CTR prediction (Dishi et al., 2025). Moreover, it alleviates the risk of overfitting, while improving the computational efficiency compared to independently learning for each task (Wang et al., 2023b). However, it suffers from the negative transfer and the seesaw phenomenon (i.e., the performance improvement of a task usually leads to the degradation of other tasks) when multiple tasks exhibit weak correlations (Tang et al., 2020).

In order to address sample selection bias, Ma et al. (2018) proposed the entire space multi-task model, where two auxiliary tasks (i.e., CTR prediction and CT-CVR prediction) were introduced to predict CVR through the chain rule. The relationship among CTR, CVR and CT-CVR can be represented as $p_{CT-CVR} = p_{CTR} \times p_{CVR} = p(y = 1|\boldsymbol{x}) \times p(z = 1|y = 1, \boldsymbol{x})$, where $\boldsymbol{x}$ is the shared embeddings. Following this line of research, Wen et al. (2020) decomposed post-click behaviors into deterministic (e.g., adding to cart) and other activities between clicks and purchases (except deterministic ones), inferred probabilities of different behaviors using task-specific networks on the hard parameter shared layer of sparse ID features and dense numerical features, and applied the conditional probability rule to derive the final conversion probabilities; Zhang et al. (2020) constructed two debiased estimators in a multi-task learning framework with a shared embedding lookup table, namely an inverse propensity weighting estimator and a doubly robust estimator, where the former used the predicted CTR as propensity scores to reweight the prediction losses, thereby correcting the selection bias, and the latter incorporated an imputation model to account for prediction errors, enabling the unbiased CVR estimation over the entire impression space, even when propensity estimations are imperfect; Dai et al. (2022) integrated multiple doubly robust estimators into a multi-task learning framework as error imputation networks and optimized with Deep & Cross network-based CTR and CVR prediction networks based on a shared embedding lookup layer for the unbiased CVR estimation via stochastic gradient descents; Ouyang et al. (2023a) enhanced feature embedding vectors by mapping the masked feature embeddings (i.e., element-wise masks on the shared embedding vector) into latent vectors and applying contrastive learning with false negative eliminations and supervised positive inclusion strategies, and employed the entire space multi-task model to predict CVR. Another solution is to generated pseudo conversion labels for unclicked samples that were used to jointly train the CTR and CVR networks on a shared embedding vector (Huang et al., 2024; Su et al., 2024); Cheng et al. (2025) introduced a clicked but un-converted CVR auxiliary task to discriminate factual negative samples and ambiguous negative samples, and soft labels generated from the auxiliary task were used to supervise the alignment of the CVR model on unclicked samples, enabling the model to learn from clicked and unclicked samples. To address the co-variate shift between click and unclick data spaces, Feng et al. (2024) aligned the distribution of covariates on clicked and unclicked samples to learn the shared representations from a hard parameter sharing method, the shared representations are fed into a CTR estimator, CVR estimator and parameter varying doubly robust estimator to obtain the predicted CVR. Liu et al. (2024b) employed a multi-task learning model to achieve unbiased CVR estimation over the entire impression space, where CTR and CVR towers were established on a shared-embedding layer, and used the sample reweighting technique to handle sample selection bias.

For mitigating the label bias resulting from delayed feedback in multi-task CVR prediction, Wang et al. (2020) discretized delay time into day slots, described the probability distribution over these slots using survival analysis, and combined delay modeling with shared embedding parameters to estimate the final CVR in the entire impression space; Li et al. (2021b) and Gao & Yang (2022) quantized conversion delays into multiple time-window prediction tasks, and employed a shared-bottom architecture to generate conversion probability for each task which were aggregated to obtain the final CVR; Dai et al. (2023) developed dual unbiased CVR estimators, one for the observed data and another for the duplicated data, which are integrated into a shared-bottom structure to predict CVR; Zhang et al. (2025a) conducted CVR estimation by interpolating CVR predictions learned on data from short and long optimization windows upon hard parameter sharing.

In online advertising, conversion has multiple types, each of which indicates underlying user intents and decisive factors for CVR prediction. Pan et al. (2019) used the multi-task learning framework to construct prediction models for different types of conversions (i.e., leading, viewing content, signing up and purchasing) through hard parameter sharing and task-specific parameters. Ad creatives are critical for effective communications with users. Kitada et al. (2019) employed a multi-task learning framework to simultaneously predict CTR and CVR by learning shared feature representations of ad creative texts.

### 4.6.2. Soft parameter sharing

Soft parameter sharing allows task-specific networks to implicitly exchange information through mechanisms such as attention mechanisms and regularization, enabling the model to capture differences and underlying correlations between CTR and CVR prediction tasks (Wei et al., 2021). Figure 13 illustrates the modeling framework of soft parameter sharing.

Soft parameter sharing can achieve better optimization effects in the training process than hard parameter sharing, even for tasks with looser connections, as parameters of each task are regularized to encourage parameter similarities among tasks. However, it entails a substantial increase in the number of parameters as the number of tasks grows, resulting in the higher computational complexity and resource demands (Wang et al., 2023b).

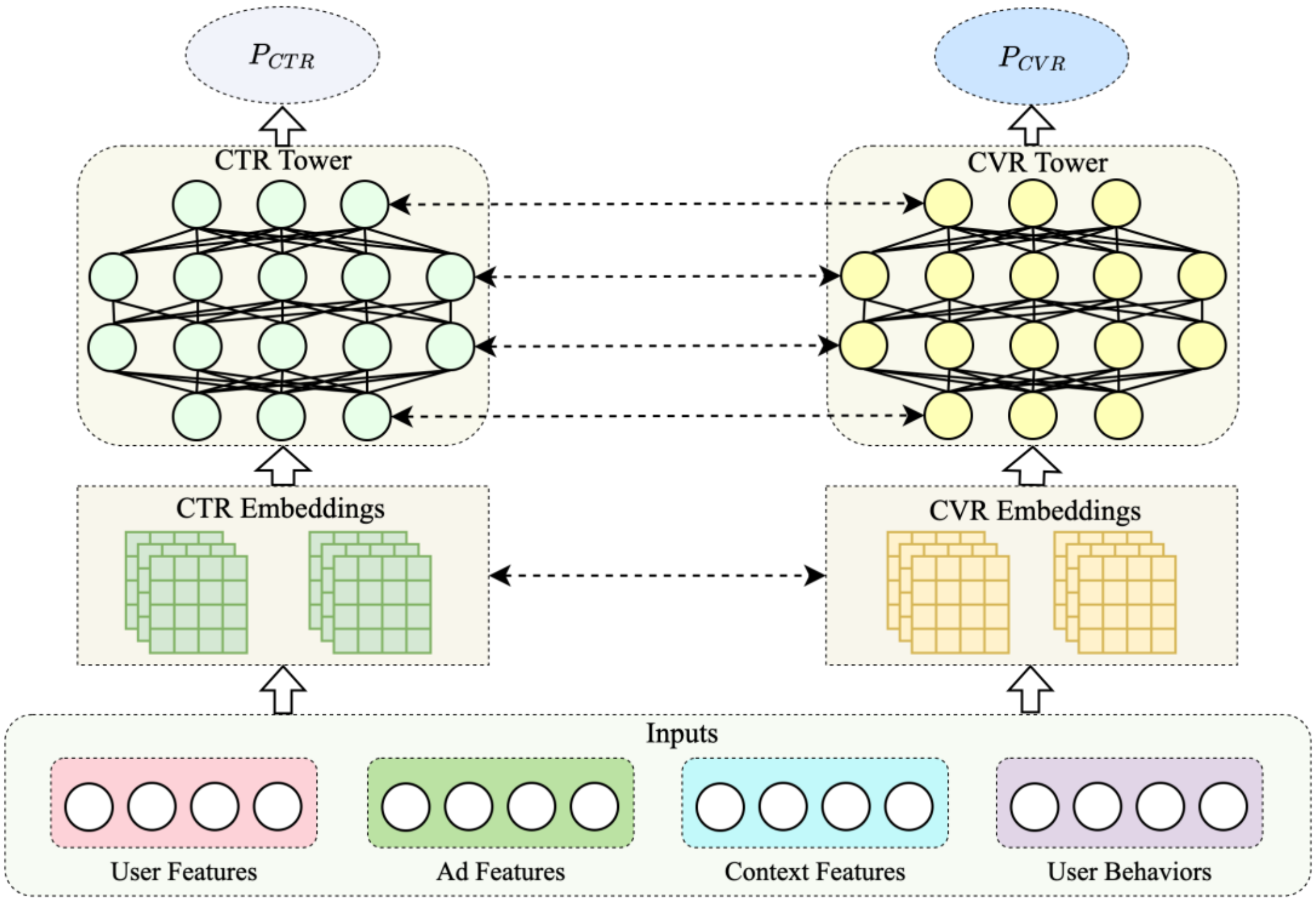

Figure 13. The modeling framework of soft parameter sharing for CVR prediction.

In order to exploit the task hierarchy induced from behavioral funnels, Wei et al. (2021) employed soft parameter sharing to automatically add integration connections between layer-wise representations, used two three-layer MLPs to predict CTR and CVR, and calculated CT-CVR through the chain rule. To mitigate the selection bias inherent in CVR prediction, Zhu et al. (2023) employed the Wide & Deep dual-tower structure to implement soft parameter sharing between CTR and CVR predictions, and designed a counterfactual mechanism to estimate factual and counterfactual CVR, which were incorporated as auxiliary supervisory signals during training to enhance the model's debiasing capability. Yao et al. (2023) decomposed advertising campaigns into three views (i.e., advertising messages, conversion rules and audience targeting), each of which was modeled as a sub-task in a multi-task learning framework, used soft parameter sharing to convey the shared information across tasks, and aggregated predictions from these views for CVR prediction. Zhuang et al. (2025) proposed a unified multi-task learning framework with soft parameter sharing to predict various CVR tasks (i.e., checkout, adding to cart, and signing up), in which a self-supervised auxiliary loss was designed to alleviate the label sparseness issue in CVR prediction.

#### 4.6.3. Expert sharing

Expert sharing is an extension of soft parameter sharing that employs multiple experts to learn patterns of shared feature embeddings, where outputs of experts are selectively fused via task-specific gating mechanisms (Wang et al., 2023b; Hou et al., 2021) and the fused representations are routed to task-specific towers to generate the final probability for each prediction task. Figure 14 depicts the expert sharing framework for CVR prediction. The representation fusion process can be given as

$$\begin{cases} g^k(\boldsymbol{x}) = Softmax(\boldsymbol{W}_k\boldsymbol{x}) \\ f^k(\boldsymbol{x}) = \sum_{i=1}^{n} g_i^k(\boldsymbol{x}) f_i(\boldsymbol{x}) \end{cases}, \quad (6)$$

where $g^k(\boldsymbol{x})$ is the gate of $k-th$ task (i.e., CTR or CVR task) and $g_i^k(\boldsymbol{x})$ is the $i-th$ logit in the output of that gate; $\boldsymbol{x}$ is the input features, $\boldsymbol{W}_k$ is a parameter matrix, $f^k(\boldsymbol{x})$ is the output representation of task $k$, $n$ is the number of experts, and $f_i(\boldsymbol{x})$ represents the learned representation of the $i-th$ expert.

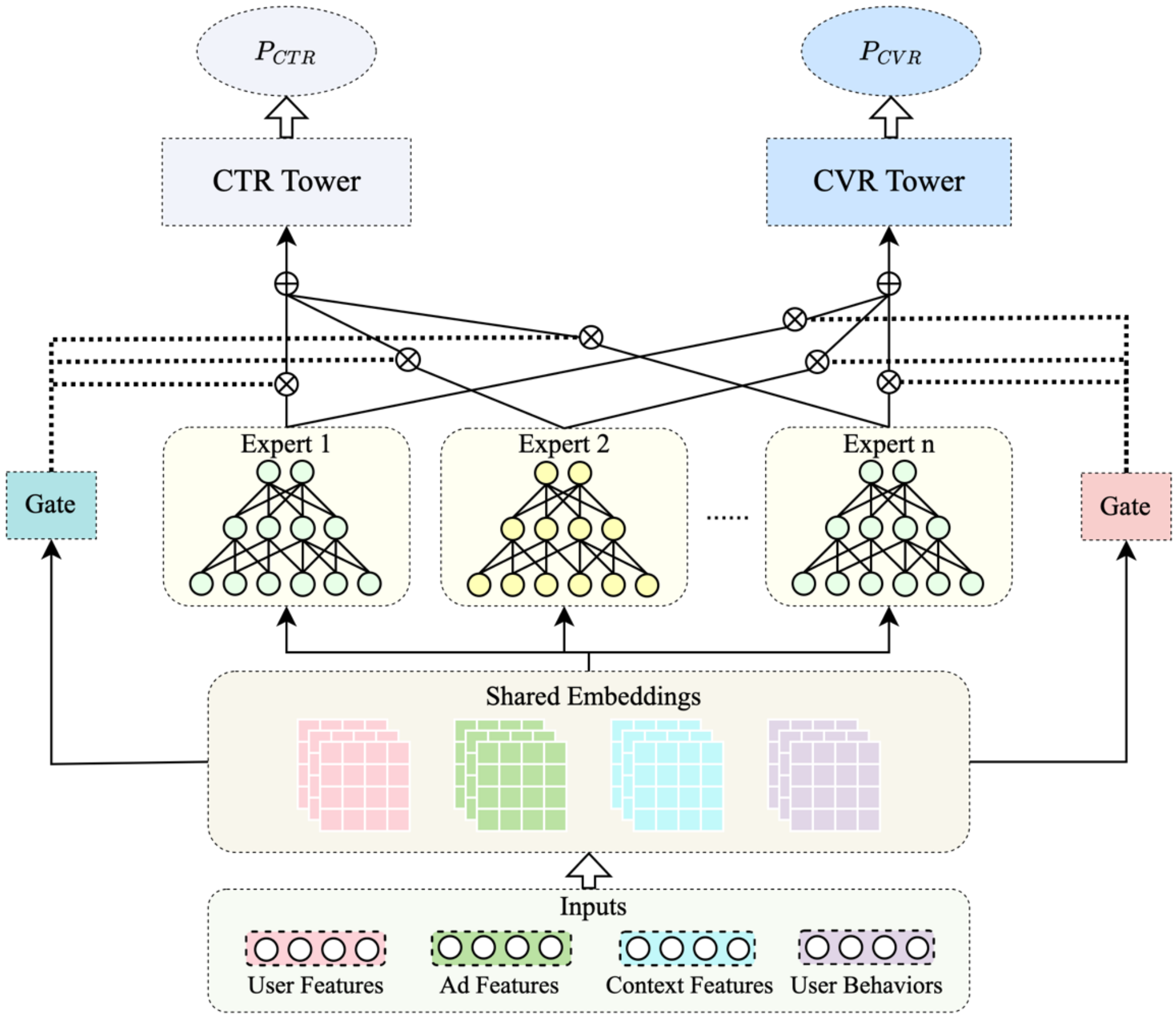

Figure 14. The modeling framework of expert sharing for CVR prediction.

Expert sharing has several advantages: (1) it dynamically selects expert networks for various tasks (Zhang et al., 2024a); (2) it provides an efficient and flexible fusion strategy across multiple tasks via gating mechanisms (Liu et al., 2023b); (3) it partially mitigates the issue of negative transfer via task-specific gates (Tang et al., 2020). However, it incurs significant computational complexity due to the large number of parameters of multiple experts (Oldfield et al., 2024). In addition, due to the random initialization, expert polarization may occur, i.e., some experts initially perform better than others, the gating network may assign them higher weights, in turn they receive more training signals that improve their performance (Li et al., 2025b).

In order to capture features that actually correlated with conversions, Chen et al. (2022b) utilized pseudo-intervention to selectively learn more stable causal features, which are used as shared features feed into hard parameter sharing, soft parameter sharing, and expert sharing structures modules to facilitate CVR prediction. Hou et al. (2021) used mixture-of-experts (MOE) to predict conversion probabilities by leveraging multiple conversion labels under different observation intervals; Cui et al. (2024) proposed long-delayed prediction model for ad conversion volumes, consisting of a bucket classification module with label smoothing and a value regression module with proxy labels, and integrated outputs of the two modules via a MOE framework to obtain the predicted conversion volumes.

Multi-gate mixture-of-experts (MMOE) extends MOE by developing task-specific gating networks that adaptively fuse outputs of various experts to facilitate joint prediction modeling of CTR and CVR (Zhang et al., 2024b). Chen et al. (2022a) used MMOE to learn user behavior patterns and used task-specific gating network to fuse the outputs of multiple experts, which are fed into a fully connected layer to predict CVR; Liu et al. (2023b) employed three groups of experts (i.e., CTR experts, CVR experts, and general experts) to learn multi-grained knowledge from shared feature embeddings. The predictions from multiple experts are aggregated into the final predicted CVR through self-distillation. Zhang et al. (2024a) estimated the probability that observed negatives are fake negatives, reweighted different sample types to eliminate biases arising from delayed feedback, and employed MMOE to compute probabilities of immediate conversion and delay conversion on reweighted samples and obtain the final CVR (i.e., the sum of the two probabilities).

However, MMOE overlooks inter-expert variabilities and task interactions, resulting in the seesaw phenomenon. To mitigate the seesaw phenomenon, Tang et al. (2020) proposed a progressive layered extraction model based on MMOE, which explicitly separates shared and task-specific experts to prevent harmful parameter interference between general and task-specific representations, refines shared knowledge hierarchically while gradually isolating task-specific information through multi-level experts and progressive gating. This model was extended in a later work by Yi et al. (2025) to support multi-scenario, multi-task and personalized CVR prediction.

In online advertising systems, advertisers may achieve higher CTR by over-decorating ads (i.e., eye-catching click-baits), resulting in numerous high CTR and low CVR samples. To address this issue, Zhang et al. (2022) proposed an auxiliary task that predicts the probability of click but no-conversion by penalizing samples with high CTR and low CVR, and incorporated this auxiliary task into three multi-task learning models (i.e., the entire space multi-task model, multi-gate mixture-of-experts, and the customized gate control) to predict the final CVR.

# 5. Datasets and model evaluation

In the extant literature, a wide range of publicly available and proprietary datasets have been used to evaluate advertising CVR prediction models, as presented in Table C1 (see Appendix C). From Table C1, we observe that the Criteo conversion and Ali-CCP datasets are the most commonly adopted public datasets. Notably, proprietary datasets are used in a large portion of CVR prediction research, primarily due to the sensitivity of user conversion information, which is often strictly constrained by privacy issues and business policies. Meanwhile, the limited data sharing results in a scarcity of large-scale public datasets, thereby constraining the reproducibility of CVR research.

Based on the categorization of CVR prediction models in Section 4, we discuss the performance of CVR prediction models on the datasets. It is worth noting that the following performance comparisons rely on experimental results reported in the literature, and exclude those without direct evidence. Table 5 summarizes the performance of CVR prediction models reported in the literature in terms of various evaluation metrics described in Table 3, on public and proprietary datasets presented in Table C1. From Table 5, we can obtain the following observations.

First, we discuss performance comparisons among multivariate statistical models (i.e., LR and SA), tree models (i.e., GBDT and XGBoost) and classical deep learning models (i.e., MLP, DNNs, RNNs, Transformer and TCNs). Deep learning models generally exhibit superior performance compared to multivariate statistical models and tree models. Specifically, (a) among 21 studies conducted on 4 public datasets, 1 synthetic dataset and 11 proprietary datasets, 18 studies reported that deep learning models perform better than multivariate statistical models in AUC, PRAUC, GAUC, Logloss, NLL, accuracy, precision and recall, with the maximum increase of 0.3924 in AUC on the Criteo attribution dataset (Kumar et al., 2020) and the maximum decrease of 0.4729 in Losloss on a synthetic dataset (Yao et al., 2022), while 3 reported inconsistent findings: Ban et al. (2024b) showed that RNNs and GRU outperform LR, and yet LR outperforms LSTM in AUC; Kumar et al. (2020) reported that the performance of LR lies between two LSTM-based models in AUC and Logloss; Li et al. (2018) showed that LSTM outperforms LR by 0.018 in accuracy, and yet the latter outperforms the former by 0.005 in AUC. (b) Among 5 studies conducted on 1 public dataset and 3 proprietary datasets, 3 studies reported that deep learning models outperform tree models in AUC and GAUC, while Xi et al. (2021) showed opposing findings, and Gligorijevic et al. (2020) reported inconsistent results: RNNs outperform random forest by 0.0084 in AUC, 0.0144 in accuracy and 0.0433 in recall, and yet random forest outperforms RNNs by 0.0021 in precision.

Second, we examine performance comparisons between classical deep learning models and their extensions (i.e., causal learning models, knowledge distillation models and multi-task learning models). We can find that extension models outperform deep learning models. Specifically, (a) among 13 studies conducted on 3 public datasets, 1 synthetic dataset and 1 proprietary dataset, most studies reported that CRNs-based models outperform deep learning models in AUC, Logloss, reaching the largest improvements of 0.3483 in AUC and 0.1412 in Logloss on the Criteo attribution dataset (Ren et al., 2018), with a few exceptions (Yang & Zhan, 2022; Chan et al., 2023; Liu et al., 2023a; 2024a; Yu et al., 2024) found that deep learning models (i.e., MLP and DNNs) outperform counterfactual learning models. (b) Among 5 studies conducted on 1 public dataset and 4 proprietary datasets, 4 studies reported that knowledge distillation models outperform deep learning models (i.e., DNNs) in AUC, D-AUC, NLL and D-NLL, with the improvement in AUC ranging from 0.0077 to 0.0136 on the proprietary datasets (Xu et al., 2022), while 1 study by Xu et al. (2022) showed that inconsistent results: knowledge distillation models outperform DNNs by 0.0405 in AUC and 0.0417 in D-AUC, and yet DNNs outperform knowledge distillation models by 0.0013 in NLL and by 0.0012 in D-NLL on the Ali-CCP dataset. (c) Among 39 studies conducted on 4 public datasets and 15 proprietary datasets, 23 studies reported that multi-task learning models outperform deep learning models (i.e., MLP and DNNs) in AUC, GAUC, PRAUC and NLL, with the largest improvement of 0.0719 in AUC on the Ali-CCP dataset (Xi et al., 2021), while 11 studies (Xu et al., 2022; Yang & Zhan, 2022; Chan et al., 2023; Liu et al., 2023a; Jin et al., 2023; Ouyang et al., 2023b; Liu et al., 2024a; Yu et al., 2024; Yi et al., 2025) gave opposing results, and 5 studies reported inconsistent results: 3 studies by Yi et al. (2025) reported that the performance of DNNs ranges between hard parameter sharing models and expert sharing models; and 2 studies by Xu et al. (2022) showed that the entire space multi-task model outperforms DNNs in AUC and D-AUC, and yet DNNs perform better than the entire space multi-task model in NLL and D-DNLL.

Third, performance comparisons among three deep learning extension models illustrate that multi-task learning models outperform causal learning models, and knowledge distillation models perform better than multi-task learning models. Specifically, (a) among 21 studies conducted on 3 public datasets and 5 proprietary datasets, 13 studies reported that multi-task learning models outperform causal learning models in AUC, GAUC, PRAUC, Logloss, NLL, KS, recall, F1, ECE and PCOC, with the improvement in AUC ranging from 0.0007 to 0.0215 on Criteo conversion, Ali-CCP and proprietary datasets, while 4 studies hold inconsistent findings (Yang & Zhan, 2022; Liu et al., 2023a; 2024a; Chan et al., 2023), and 4 studies (Dai et al., 2023; Liu et al., 2023a; 2024a) showed that the performance of causal learning models ranges between multi-task learning models. (b) Among 7 studies conducted on 1 public dataset and 4 proprietary datasets, 2 studies reported that knowledge distillation models outperform multi-task learning models in AUC, D-AUC, NLL and D-NLL, while 5 studies reported inconsistent results: 3 studies by Xu et al. (2022) showed the uncertainty-regularized knowledge distillation model outperforms the entire space multi-task model in AUC and D-AUC, and yet the latter outperforms the former in NLL and D-NLL; Cheng et al. (2025) reported that the uncertainty-regularized knowledge distillation model outperforms the direct entire-space causal multi-task model, and its performance lies between hard parameter sharing models (i.e., the direct dual propensity optimization model, the non-click samples improved semi-supervised model, the entire space counterfactual multi-task model, and the entire space multi-task model) in AUC; and Fei et al. (2025) reported that the entire-space variational information exploitation model outperforms hard parameter sharing models (i.e., the direct dual propensity optimization model, the entire space multi-task model, and the entire space counterfactual multi-task model) and a soft parameter sharing model (i.e., the direct entire-space causal multi-task model), and yet the hard parameter sharing models and the soft parameter sharing model outperform the uncertainty-regularized knowledge distillation model in AUC and NLL.

Fourth, comparisons between multi-task learning models and single-task learning models demonstrate that the former generally exhibits superior performance. Specifically, among 34 studies conducted on 3 public datasets and 15 proprietary datasets, 20 studies reported that multi-task learning models outperform single-task learning models in AUC, GAUC, PRAUC, D-AUC, Logloss, NLL, KS, precision, recall and F1, with the largest improvement of 0.0663 in AUC, while 2 studies by Ouyang et al. (2023b) gave opposing results, and 10 studies gave inconsistent findings: 2 studies by Xu et al. (2022) reported that the entire space multi-task model outperforms DNNs in AUC and D-AUC, and yet the latter outperforms the former in NLL and D-NLL; and 10 studies (Zhang et al., 2020; Wang et al., 2022; Yi et al., 2025) showed that the performance of single-task learning models lies between multi-task learning models.

Fifth, we discuss performance comparisons among deep learning models. We can find that TCNs, Transformer and RNNs' variants (i.e., LSTM and GRU) generally perform better than DNNs and MLP. Specifically, 3 studies conducted on 5 proprietary datasets reported that RNNs' variants (i.e., LSTM and GRU) outperform MLP and DNNs with the largest improvement of 0.1084 in AUC (Su et al., 2021; Ban et al., 2024b); 1 study conducted on 1 proprietary dataset reported that Transformer outperforms MLP with the improvement of 0.0102 in AUC and 0.0005 in Logloss (Li et al., 2021a); and 3 studies conducted on 1 public and 2 proprietary datasets consistently reported that TCNs outperforms RNNs with the largest improvements of 0.12 in AUC, 0.08 in accuracy and 0.1629 in Logloss.

Sixth, we evaluate the performance among multi-task learning models, which reveal that expert sharing and soft parameter sharing models generally perform better than hard parameter sharing models. Specifically, (a) among 15 studies conducted on 1 public dataset and 6 proprietary datasets, 5 studies reported that soft parameter sharing models outperforms hard parameter sharing models in AUC, NLL, precision and recall, with the largest improvement of 0.0572 in AUC, while 3 studies by Lin et al. (2022) and Liu et al. (2023b) gave opposing findings, and 7 studies (Wei et al., 2021; Zhu et al., 2023; Feng et al., 2024; Huang et al., 2024; Su et al., 2024; Cheng et al., 2025) reported inconsistent results, i.e., that the performance of soft parameter sharing models lies between hard parameter sharing models. (b) Among 31 studies conducted on 3 public and 6 proprietary datasets, 12 studies reported that expert sharing models outperforms hard parameter sharing models in AUC, precision and recall, with the improvement in AUC ranging from 0.0001 to 0.074, while 2 studies showed opposing findings (Liu et al., 2023a; 2024a), and 17 studies reported inconsistent results: 7 studies (Ouyang et al., 2023b; Feng et al., 2024; Huang et al., 2024) showed that the performance of expert sharing models lies between hard parameter sharing models, and 10 studies (Lin et al., 2022; Liu et al., 2023b; Ouyang et al., 2023b; Zhu et al., 2023; Yi et al., 2025) showed that the performance of hard parameter sharing models lies between expert sharing models. (c) Among 15 studies conducted on 1 public and 5 proprietary datasets, 9 studies reported that soft parameter sharing models outperforms expert sharing models, with the improvement in AUC ranging from 0.0108 to 0.0742, while 2 studies gave opposing findings (Liu et al., 2023b; Feng et al., 2024), and 4 studies by Lin et al. (2022) and Liu et al. (2023b) reported inconsistent results, i.e., the performance of soft parameter sharing models lies between those of expert sharing models.

In summary, as we can notice, multi-task learning models commonly perform better than single-task learning models, and classical deep learning models generally perform inferior to their extensions, particularly in scenarios involving complex user behaviors and biases (Yang et al., 2021; Wen et al., 2020; Yi et al., 2025). Moreover, inconsistent findings have been found in comparisons between knowledge distillation models and multi-task learning models, between causal learning models and multi-task learning models, and among hard parameter sharing models, soft parameter sharing models and expert sharing models. Even on the same dataset, the performance of a CVR prediction model may not achieve the best performance in all metrics simultaneously (Gligorijevic et al., 2020; Wei et al., 2021; Xu et al., 2022; Yang et al., 2023; Tang et al., 2024). We believe that comparative studies among CVR prediction models are an important avenue for future research.

Table 5. Performance comparison of CVR prediction models on various datasets.

| **Dataset** | **References** | **Model Category** | **Model Performance** |
|---|---|---|---|
| Criteo conversion | Chapelle. (2014) | Multivariate statistical models (DFM) | (1) Entire dataset:<br>NLL: DFM (0.3960) < Short term conversion model (STC) (0.4111);<br>(2) Recent dataset:<br>NLL: DFM (0.4006) < Short term conversion model (STC) (0.4117); |
| | Vasile et al. (2017) | Multivariate statistical models | MSEW: LR + WNLL < LR;<br>Utility ($\beta = 10$): LR + WNLL > LR;<br>Utility ($\beta = 1000$): LR + WNLL > LR; |
| | Yoshikawa & Imai (2018) | Multivariate statistical models (NoDeF, DFM) | (1) Recent campaigns dataset:<br>Accuracy: NoDeF (0.9157) > DFM (0.9151) > LR (0.9124);<br>Logloss: NoDeF (0.2575) < LR (0.2818) < DFM (0.3689);<br>AUC: NoDeF (0.7242) > DFM (0.7213) > LR (0.7187);<br>(2) All campaigns dataset:<br>Accuracy: NoDeF (0.8725) > LR (0.8714) > DFM (0.8702);<br>Logloss: NoDeF (0.3438) < DFM (0.3450) < LR (0.3571);<br>AUC: DFM (0.7423) > NoDeF (0.7387) > LR (0.7349); |
| | Yasui et al. (2020) | Causal learning models (FSIW), Multivariate statistical models (DFM) | Logloss: LR-FSIW (0.3928) < DFM (0.3989) < LR (0.4076);<br>PRAUC: LR-FSIW (0.6482) > DFM (0.6481) > LR (0.6345);<br>Normalized Logloss: LR-FSIW (28.0200) > DFM (27.3300) > LR (25.2100); |
| | Gu et al. (2021) | Causal learning models (DEFER, ES-DFM) | AUC: DEFER (0.8394) > ES-DFM (0.8373) > FNW-RN (0.8372) > FNC-RN (0.8370) > FNW (0.8348) > FNC (0.8347);<br>PRAUC: DEFER (0.6367) > ES-DFM (0.6347) > FNW-RN (0.6326) > FNC-RN (0.6299) > FNW (0.6262) > FNC (0.6189);<br>NLL: DEFER (0.3943) < ES-DFM (0.3956) < FNW-RN (0.3970) < FNW (0.4006) < FNC-RN (0.4103) < FNC (0.4659);<br>RI-AUC: DEFER (0.9011) > ES-DFM (0.8418) > FNW-RN (0.8390) > FNC-RN (0.8333) > FNW (0.7712) > FNC (0.7684);<br>RI-PRAUC: DEFER (0.8825) > ES-DFM (0.8508) > FNW-RN (0.8175) > FNC-RN (0.7746) > FNW (0.7159) > FNC (0.6000);<br>RI-NLL: DEFER (0.9661) > ES-DFM (0.9576) > FNW-RN (0.9485) > FNW (0.9250) > FNC-RN (0.8618) > FNC (0.4993); |
| | Hou et al. (2021) | Multi-task learning models (MM-DFM), Causal learning models (ES-DFM), Multivariate statistical models (DFM) | AUC: MM-DFM (0.7911) > ES-DFM (0.7861) > DFM (0.7832);<br>PRAUC: MM-DFM (0.3134) > ES-DFM (0.3056) > DFM (0.2931);<br>GAUC: MM-DFM (0.7106) > ES-DFM (0.7060) > DFM (0.7046); |
| | Li et al. (2021b) | Multi-task learning models (FTP), Causal learning models (FSIW) | AUC: FTP > FNW > FNC > FSIW;<br>Logloss: FTP < FNW < FSIW < FNC; |
| | Yang et al. (2021) | Causal learning models (ES-DFM), Multivariate statistical models (DFM) | AUC: ES-DFM (0.8402) > FNC (0.8373) = FNW (0.8373) > FSIW (0.8290) > DFM (0.8132);<br>PRAUC: ES-DFM (0.6393) > FNW (0.6313) > FNC (0.6267) > FSIW (0.6189) > DFM (0.5784);<br>NLL: ES-DFM (0.3924) < FNW (0.4033) < FSIW (0.4099) < FNC (0.4382) < DFM (1.2599);<br>RI-AUC: ES-DFM (0.3560) > FNW (-0.0308) > FNC (-0.0393) > FSIW (-1.1432) > DFM (-3.2581);<br>RI-PRAUC: ES-DFM (0.5799) > FNW (0.1400) > FNC (-0.1147) > FSIW (-0.5479) > DFM (-2.7833); |
| | Chen et al. (2022a) | Multi-task learning models (DEFUSE), Causal learning models (DEFER, ES-DFM) | (1) Criteo-30d dataset:<br>AUC: DEFUSE (0.8408) > ES-DFM (0.8396) > DEFER (0.8382) > Bi-DEFUSE (0.8379) > FNW (0.8376) > FNC (0.8373);<br>RI-AUC: DEFUSE (0.5233) > ES-DFM (0.4611) > DEFER (0.3886) > Bi-DEFUSE (0.3731) > FNW (0.3575) > FNC (0.3420);<br>PRAUC: DEFUSE (0.6400) > ES-DFM (0.6384) > DEFER (0.6338) > FNW (0.6310) > Bi-DEFUSE (0.6301) > FNC (0.6222);<br>NLL: DEFUSE (0.3946) < ES-DFM (0.3947) < Bi-DEFUSE (0.3963) < FNW (0.3971) < FNC (0.4688) < DEFER (0.4800);<br>(2) Criteo-1d dataset:<br>AUC: Bi-DEFUSE (0.8467) > DEFUSE (0.8465) > DEFER (0.8463) > ES-DFM (0.8459) > FNW (0.8348) > FNC (0.8343);<br>RI-AUC: Bi-DEFUSE (0.9630) > DEFUSE (0.9524) > DEFER (0.9417) > ES-DFM (0.9206) > FNW (0.3333) > FNC (0.3069);<br>PRAUC: Bi-DEFUSE (0.5499) > ES-DFM (0.5492) > DEFUSE (0.5490) = DEFER (0.5490) > FNW (0.4982) > FNC (0.4806);<br>NLL: ES-DFM (0.2885) < DEFUSE (0.3086) < Bi-DEFUSE (0.3092) < DEFER (0.3098) < FNC (0.3145) < FNW (0.3367); |
| | Gao & Yang (2022) | Multi-task learning models (MHOL), Causal learning models (DEFER, FSIW) | NLL: MHOL (0.2819) < DEFER (0.2833) < FSIW (0.2836);<br>RCE: MHOL (14.7260) > DEFER (14.3070) > FSIW (14.2080); |
| | Guo et al. (2022) | Knowledge distillation models (KD Calibration), Multivariate statistical models | AUC: KD Calibration (0.8463) > LR-Focal Loss (0.8452) > LR (0.8435) = Platt Scaling (0.8435) = Temperature Scaling (0.8435) = Beta Calibration (0.8435) > Isotonic Regression (0.8431) > Histogram Binning (0.8429) > Neural Calibration (0.8272) > FNC (0.7854);<br>Logloss: KD Calibration (0.2667) < Temperature Scaling (0.2684) < LR (0.2685) < Platt Scaling (0.2686) = Beta Calibration (0.2686) < Histogram Binning (0.2699) < Isotonic Regression (0.2719) < Neural Calibration (0.2815) < FNC (0.2949) < LR-Focal Loss (0.3283);<br>BS: KD Calibration (0.0773) < Base (0.0778) = Histogram Binning (0.0778) = Platt Scaling (0.0778) = Temperature Scaling (0.0778) = Beta Calibration (0.0778) < Isotonic Regression (0.0779) < Neural Calibration (0.0821) < FNC (0.0859) < LR-Focal Loss (0.0913);<br>F-ECE: KD Calibration (0.0238) = Histogram Binning (0.0238) < Temperature Scaling (0.0248) < Platt Scaling (0.0252) < Isotonic Regression (0.0254) = Beta Calibration (0.0254) < LR (0.0257) < Neural Calibration (0.0341) < FNC (0.0793) < LR-Focal Loss (0.1181);<br>F-RCE: KD Calibration (0.3371) < Histogram Binning (0.3728) < Isotonic Regression (0.3798) < Beta Calibration (0.3819) < Temperature Scaling (0.3843) < Platt Scaling (0.3849) < LR (0.4124) < Neural Calibration (0.4818) < FNC (2.0771) < LR-Focal Loss (2.6653); |
| | Yang & Zhan (2022) | Deep learning models (GDFM), Causal learning models (ES-DFM), | AUC: GDFM (0.7490) > ES-DFM (0.7140) > MM-DFM (0.6970) > FNW (0.6200);<br>PRAUC: GDFM (0.6810) > ES-DFM (0.6330) > FNW (0.4310) > MM-DFM (0.3920);<br>NLL (absolute performance): GDFM (0.7240) > ES-DFM (0.6620) > MM-DFM (0.5420) > FNW (0.4030); |
| | | Multi-task learning models (MM-DFM) | |
| | Yasui & Kato (2022) | Causal learning models (FSIW, nnDF), Multivariate statistical models (DFM) | (1) Criteo-day54 dataset:<br>AUC: PUTM (0.8890) > TW (0.8830) > FSIW (0.8690) > DFM (0.8670) > Biased LR (0.8640) > nnDF (0.8170);<br>Accuracy: DFM (0.9360) = FSIW (0.9360) = TW (0.9360) = Biased LR (0.9360) > nnDF (0.9350) > PUTW (0.8890);<br>NLL: TW (0.2600) < nnDF (0.2650) < FSIW (0.2740) < DFM (0.2800) < Biased LR (0.2900) < PUTW (0.3200);<br>(2) Criteo-day55 dataset:<br>AUC: TW (0.8780) > PUTW (0.8740) > FSIW (0.8620) > DFM (0.8600) > Biased LR (0.8590) > nnDF (0.8290);<br>Accuracy: DFM (0.9290) = FSIW (0.9290) = TW (0.9290) = Biased LR (0.9290) = nnDF (0.9290) > PUTW (0.9280);<br>NLL: nnDF (0.2690) < TW (0.2840) < FSIW (0.3000) < Biased LR (0.3140) < DFM (0.3200) < PUTW (0.3680);<br>(3) Criteo-day56 dataset:<br>AUC: TW (0.8580) > PUTW (0.8560) > nnDF (0.8420) > FSIW (0.8400) > DFM (0.8390) > Biased LR (0.8380);<br>Accuracy: DFM (0.9170) = FSIW (0.9170) = TW (0.9170) = Biased LR (0.9170) = nnDF (0.9170) > PUTW (0.9170);<br>NLL: nnDF (0.2830) < TW (0.3240) < PUTW (0.3350) < FSIW (0.3400) < Biased LR (0.3550) < DFM (0.3560);<br>(4) Criteo-day57 dataset:<br>AUC: FSIW (0.8270) > TW (0.8220) > nnDF (0.8150) > PUTW (0.8090) > Biased LR (0.7970) > DFM (0.7940);<br>Accuracy: FSIW (0.9080) > TW (0.8880) = nnDF (0.8880) = PUTW (0.8880) = Biased LR (0.8880) = DFM (0.8880);<br>NLL: nnDF (0.3260) < FSIW (0.3740) < TW (0.4160) < Biased LR (0.4400) < PUTW (0.4410) < DFM (0.4440);<br>(5) Criteo-day58 dataset:<br>AUC: FSIW (0.6880) > DFM (0.6840) > Biased LR (0.6810) > TW (0.6680) > PUTW (0.6170) > nnDF (0.4840);<br>Accuracy: DFM (0.7630) = FSIW (0.7630) = TW (0.7630) = Biased LR (0.7630) = nnDF (0.7630) > PUTW (0.7620);<br>NLL: PUTW (0.5700) < FSIW (0.5820) < Biased LR (0.5890) = DFM (0.5890) < TW (0.6030) < nnDF (0.6530);<br>(6) Criteo-day59 dataset:<br>AUC: FSIW (0.9950) > DFM (0.9790) > Biased LR (0.9750) > nnDF (0.9040) > TW (0.8470) > PUTW (0.6070);<br>Accuracy: FSIW (0.9580) > DFM (0.8210) > Biased LR (0.8000) > nnDF (0.7810) > PUTW (0.7630) = TW (0.7630);<br>NLL: FSIW (0.2080) < DFM (0.3150) < Biased LR (0.3400) < nnDF (0.4210) < TW (0.5180) < PUTW (0.6620);<br>(7) Criteo-day60 dataset:<br>AUC: nnDF (0.9940) > FSIW (0.9930) > DFM (0.9920) > Biased LR (0.9900) > TW (0.9230) > PUTW (0.6110);<br>Accuracy: FSIW (0.9950) > nnDF (0.9830) > DFM (0.8400) > Biased LR (0.8260) > PUTW (0.7530) = TW (0.7530);<br>NLL: FSIW (0.1420) < nnDF (0.2330) < DFM (0.2560) < Biased LR (0.2810) < TW (0.4720) < PUTW (0.7350);<br>(8) Criteo-average dataset:<br>AUC: FSIW (0.8680) > Biased LR (0.8590) > DFM (0.8580) > TW (0.8460) > nnDF (0.8100) > PUTW (0.8020);<br>Accuracy: FSIW (0.9160) > nnDF (0.8880) > DFM (0.8720) > Biased LR (0.8670) > TW (0.8520) > PUTW (0.8450);<br>NLL: FSIW (0.3120) < nnDF (0.3470) < DFM (0.3650) < Biased LR (0.3710) < TW (0.4080) < PUTW (0.4870); |
| | Dai et al. (2023) | Multi-task learning models (DDFM, DEFUSE), Causal learning models (DEFER, ES-DFM, FSIW) | AUC: DDFM (0.8413) > DEFUSE (0.8378) > ES-DFM (0.8377) > DEFER (0.8374) > FNW (0.8355) > FNC (0.8351) > FSIW (0.8152);<br>RI-AUC: DDFM (0.9651) > DEFUSE (0.8540) > ES-DFM (0.8508) > DEFER (0.8413) > FNW (0.7810) > FNC (0.7683) > FSIW (0.1365);<br>KS: DDFM (0.5292) > DEFUSE (0.5247) > ES-DFM (0.5244) > DEFER (0.5237) > FNW (0.5210) > FNC (0.5204) > FSIW (0.4855);<br>RI-KS: DDFM (0.9609) > DEFUSE (0.8771) > ES-DFM (0.8715) > DEFER (0.8585) > FNW (0.8082) > FNC (0.7970) > FSIW (0.1471);<br>PRAUC: DDFM (0.6414) > ES-DFM (0.6351) > DEFUSE (0.6349) > DEFER (0.6324) >FNW (0.6303) > FNC (0.6263) > FSIW (0.6022);<br>RI-PRAUC: DDFM (0.9640) > ES-DFM (0.8305) > DEFUSE (0.8263) > DEFER (0.7733) > FNW (0.7288) > FNC (0.6441) > FSIW (0.1335);<br>NLL: DDFM (0.3946) < ES-DFM (0.3963) < DEFUSE (0.3967) < FNW (0.4048) < DEFER (0.4149) < FNC (0.4178) < FSIW (0.5249);<br>RI-NLL: DDFM (0.9670) > ES-DFM (0.9564) > DEFUSE (0.9539) > FNW (0.9035) > DEFER (0.8406) > FNC (0.8225) > FSIW (0.1557); |
| | Liu et al. (2023a) | Deep learning models (DFSN), Causal learning models (ES-DFM), Multi-task learning models (FTP, DEFUSE) | AUC: DFSN-α (0.7300) > DFSN- β (0.7020) > DEFUSE (0.6690) > ES-DFM (0.6460) > FTP (0.5950) > FNW (0.4600) > FNC (0.4490);<br>NLL (absolute performance): DFSN-α (0.7830) > DFSN- β (0.7520) > ES-DFM (0.7400) > FTP (0.5080) > FNW (0.4920) > FNC (0.2280) > DEFUSE (-0.1930);<br>Bias: DFSN- β (0.9070) > FNW (0.8960) > FNC (0.8860) > ES-DFM (0.8830) > DFSN-α (0.8550) > FTP (0.6490) > DEFUSE (- 0.9980); |
| | Ouyang et al. (2023b) | Deep learning models (MMN), Multi-task learning models (ESMM, PLE, MMOE) | AUC: MMN (0.7394) > STAR (0.7359) > DNN (0.7311) > ESMM (0.7298) > PLE (0.7273) > MT-FwFM (0.6963) > MMOE (0.6934); |
| | Wang et al. (2023c) | Causal learning models (ULC, FSIW, nnDF), Multivariate statistical models (DFM) | (1) Backbone-MLP:<br>AUC: ULC (0.8403) > FSIW (0.8335) > DFM (0.8264) > nnDF (0.6859);<br>PRAUC: ULC (0.6543) > FSIW (0.6465) > DFM (0.6398) > nnDF (0.4169);<br>Logloss: ULC (0.4104) < FSIW (0.4178) < DFM (0.4378) < nnDF (0.5969);<br>RI-AUC: ULC (0.8369) > FSIW (0.5450) > DFM (0.2403) > nnDF (-5.7890);<br>RI-PRAUC: ULC (0.7824) > FSIW (0.4560) > DFM (0.1757) > nnDF (-9.1500);<br>RI-Logloss: ULC (0.8354) > FSIW (0.6812) > DFM (0.2645) > nnDF (-3.0500);<br>(2) Backbone-DeepFW:<br>AUC: ULC (0.8393) > FSIW (0.8326) > DFM (0.8266) > nnDF (0.6994);<br>PRAUC: ULC (0.6525) > FSIW (0.6451) > DFM (0.6416) > nnDF (0.4332);<br>Logloss: ULC (0.4104) < FSIW (0.4195) < DFM (0.4319) < nnDF (0.5960);<br>RI-AUC: ULC (0.7792) > FSIW (0.4774) > DFM (0.2072) > nnDF (-5.5220);<br>RI-PRAUC: ULC (0.6606) > FSIW (0.3257) > DFM (0.1674) > nnDF (-9.2620);<br>RI-Logloss: ULC (0.8107) > FSIW (0.5981) > DFM (0.3084) > nnDF (-3.5250);<br>(3) Backbone-AutoInt:<br>AUC: ULC (0.8388) > FSIW (0.8329) > DFM (0.8276) > nnDF (0.6903);<br>PRAUC: ULC (0.6517) > FSIW (0.6460) > DFM (0.6412) > nnDF (0.4214);<br>Logloss: ULC (0.4114) < FSIW (0.4192) < DFM (0.4306) < nnDF (0.6123);<br>RI-AUC: ULC (0.7428) > FSIW (0.4619) > DFM (0.2095) > nnDF (-6.3280);<br>RI-PRAUC: ULC (0.6146) > FSIW (0.3532) > DFM (0.1330) > nnDF (-9.949);<br>RI-Logloss: ULC (0.7990) > FSIW (0.6230) > DFM (0.3656) > nnDF (-3.7350);<br>(4) Backbone-DCNV2:<br>AUC: ULC (0.8391) > FSIW (0.8328) > DFM (0.8272) > nnDF (0.6867);<br>PRAUC: ULC (0.6519) > FSIW (0.6461) > DFM (0.6417) > nnDF (0.4230);<br>Logloss: ULC (0.4105) < FSIW (0.4174) < DFM (0.4325) < nnDF (0.6101);<br>RI-AUC: ULC (0.7431) > FSIW (0.4541) > DFM (0.1972) > nnDF (-6.2470);<br>RI-PRAUC: ULC (0.6045) > FSIW (0.3409) > DFM (0.1409) > nnDF (-9.8000);<br>RI-Logloss: ULC (0.8004) > FSIW (0.6422) > DFM (0.2958) > nnDF (-3.7770); |
| | Liu et al. (2024a) | Deep learning models (MISS), Causal learning models (ES-DFM), Multi-task learning models (FTP, DEFUSE) | AUC: MISS (0.8370) > DEFUSE (0.6690) > MTDFM (0.6580) > ES-DFM (0.6460) > FTP (0.5950) > FNW (0.4600) > FNC (0.4490);<br>PRAUC: MISS (0.7810) > DEFUSE (0.6680) > ES-DFM (0.6640) > MTDFM (0.6540) > FTP (0.5480) > FNW (0.2010) > FNC (0.0740);<br>NLL (absolute performance): MISS (0.8390) > ES-DFM (0.7400) > MTDFM (0.5680) > FTP (0.5080) > FNW (0.4920) > FNC (0.2280) > DEFUSE (-0.1930); |
| | Yu et al. (2024) | Deep learning models (DelayAdapter), Multivariate statistical models (DFM), Causal learning models (ULC, FSIW), Multi-task learning models (MM-DFM) | AUC: DelayAdapter (0.8415) > MM-DFM (0.8352) > DFM (0.8349) > ULC (0.8336) > FSIW (0.8288) > nnDF (0.8279);<br>PRAUC: DelayAdapter (0.6488) > ULC (0.6419) > MM-DFM (0.6415) > DFM (0.6376) > FSIW (0.6339) > nnDF (0.6335);<br>NLL: DelayAdapter (0.4014) < MM-DFM (0.4059) < ULC (0.4119) < nnDF (0.4152) < DFM (0.4157) < FSIW (0.4369); |
| | Ding et al. (2025) | Causal learning models (IF-DFM, ULC, FSIW, nnDF), Multivariate statistical models (DFM) | (1) Backbone-MLP:<br>AUC: IF-DFM (0.8411) > FSIW (0.8369) > DFM (0.8355) > ULC (0.8343) > nnDF (0.6887);<br>PRAUC: IF-DFM (0.6491) > FSIW (0.6433) > ULC (0.6412) > DFM (0.6409) > nnDF (0.4029);<br>Logloss: IF-DFM (0.3958) < ULC (0.3994) < FSIW (0.4020) < DFM (0.4152) < nnDF (0.5639);<br>RI-AUC: IF-DFM (0.8788) > FSIW (0.2424) > DFM (0.0303) > ULC (-0.1515) > nnDF (-22.2100);<br>RI-PRAUC: IF-DFM (0.8087) > FSIW (0.3043) > ULC (0.1217) > DFM (0.0957) > nnDF (-20.6000); |
| | | | RI-Logloss: IF-DFM (0.9662) > ULC (0.7510) > FSIW (0.6498) > DFM (0.1362) > nnDF (-5.6700.);<br>(2) Backbone-DeepFM:<br>AUC: IF-DFM (0.8412) > DFM (0.8381) > FSIW (0.8378) > ULC (0.8376) > nnDF (0.7025);<br>PRAUC: IF-DFM (0.6492) > ULC (0.6446) > FSIW (0.6436) > DFM (0.6434) > nnDF (0.4146);<br>Logloss: IF-DFM (0.3962) < ULC (0.3982) < FSIW (0.4023) < DFM (0.4065) < nnDF (0.5651);<br>RI-AUC: IF-DFM (0.8800) > DFM (0.2600) > FSIW (0.2000) > ULC (0.1600) > nnDF (-26.8600);<br>RI-PRAUC: IF-DFM (0.8265) > ULC (0.3571) > FSIW (0.2551) > DFM (0.2347) > nnDF (-23.1100);<br>RI-Logloss: IF-DFM (0.8512) > ULC (0.7686) > FSIW (0.5992) > DFM (0.4256) > nnDF (-6.1280);<br>(3) Backbone-AutoInt:<br>AUC: IF-DFM (0.8404) > FSIW (0.8376) > DFM (0.8374) = ULC (0.8374) > nnDF (0.6820);<br>PRAUC: IF-DFM (0.6475) > ULC (0.6443) > FSIW (0.6442) > DFM (0.6426) > nnDF (0.3770);<br>Logloss: IF-DFM (0.3965) < ULC (0.3987) < FSIW (0.4026) < DFM (0.4091) < nnDF (0.5789);<br>RI-AUC: IF-DFM (0.7963) > FSIW (0.2778) > DFM (0.2407) = ULC (0.2407) > nnDF (-28.5400);<br>RI-PRAUC: IF-DFM (0.7255) > ULC (0.4118) > FSIW (0.4020) > DFM (0.2451) > nnDF (-25.7900);<br>RI-Logloss: IF-DFM (0.8530) > ULC (0.7633) > FSIW (0.6041) > DFM (0.3388) > nnDF (-6.5920);<br>(4) Backbone-DCNV2:<br>AUC: IF-DFM (0.8410) > FSIW (0.8388) > DFM (0.8372) > ULC (0.8368) > nnDF (0.6850);<br>PRAUC: IF-DFM (0.6501) > FSIW (0.6456) > ULC (0.6441) > DFM (0.6422) > nnDF (0.3878);<br>Logloss: IF-DFM (0.3985) < ULC (0.3989) < FSIW (0.3997) < DFM (0.4087) < nnDF (0.5728);<br>RI-AUC: IF-DFM (0.8511) > FSIW (0.3830) > DFM (0.0426) > ULC (-0.0426) > nnDF (-32.3400);<br>RI-PRAUC: IF-DFM (0.9222) > FSIW (0.4220) > ULC (0.2556) > DFM (0.0440) > nnDF (-28.2200);<br>RI-Logloss: IF-DFM (0.7652) > ULC (0.7490) > FSIW (0.6045) > DFM (0.3522) > nnDF (-6.2910); |
| Ali-CCP | Ma et al. (2018) | Multi-task learning models (ESMM), Deep learning models | CVR-AUC: ESMM (0.6856) > ESMM-NS (0.6825) > ESMM-DIVISION (0.6756) > MLP (0.6600);<br>CTCVR-AUC: ESMM (0.6532) > ESMM-NS (0.6444) > ESMM-DIVISION (0.6362) > MLP (0.6207); |
| | Tang et al. (2020) | Multi-task learning models (MMOE, PLE) | AUC: PLE (0.6097) > MMOE (0.5738); |
| | Wen et al. (2020) | Multi-task learning models ($ESM^2$, ESMM), Deep learning models, Tree models (GBDT) | CVR-AUC: $ESM^2$ (0.8486) > ESMM (0.8398) > DNN-OS (0.8124) > DNN (0.8065) > GBDT (0.7823);<br>CTCVR-AUC: $ESM^2$ (0.8371) > ESMM (0.8270) > DNN-OS (0.8192) > DNN (0.8161) > GBDT (0.8059);<br>CTCVR-GAUC: $ESM^2$ (0.8051) > ESMM (0.7906) > DNN-OS (0.7893) > DNN (0.7864) > GBDT (0.7747); |
| | Zhang et al. (2020) | Multi-task learning models (Multi-IPW/DR, ESMM), Deep learning models (MLP) | CVR-AUC: Multi-DR (0.6929) > Multi-IPW (0.6921) > ESMM (0.6856) > ESMM-NS (0.6825) > MLP (0.6600);<br>CTCVR-AUC: Multi-DR (0.6543) > ESMM (0.6532) > Multi-IPW (0.6530) > ESMM-NS (0.6444) > MLP (0.6207); |
| | Wei et al. (2021) | Multi-task learning models (AutoHERI, AITM, Multi-IPW/DR, ESMM), Deep learning models (DNN) | CVR-AUC: AutoHERI (0.6783) > Multi-DR (0.6774) > AITM (0.6761) > ESMM (0.6725) > DBMTL (0.6691) > DNN (0.6596);<br>CVR-NLL: AITM (0.0330) < AutoHERI (0.0330) < Multi-DR (0.0331) < ESMM (0.0332) < DNN (0.0334) < DBMTL (0.0335);<br>CTCVR-AUC: AutoHERI (0.6379) > DBMTL (0.6336) > AITM (0.6327) > Multi-DR (0.6322) > ESMM (0.6289) > DNN (0.6084);<br>CTCVR-NLL: AutoHERI (0.0020) = AITM (0.0020) < Multi-DR (0.0021) = ESMM (0.0021) < DNN (0.0021) < DBMTL (0.0021); |
| | Xi et al. (2021) | Multi-task learning models (AITM, PLE, | AUC: AITM (0.6525) > MMOE (0.6420) > PLE (0.6417) > OMoE (0.6405) > ESMM (0.6291) > LightGBM (0.5870) > MLP (0.5806); |
| | | MMOE, ESMM), Deep learning models (MLP) | |
| | Lin et al. (2022) | Multi-task learning models (AdaFTR, PLE, MMOE, AutoHERI), Deep learning models (DNN) | AUC: AdaFTR (0.6193) > Cross-Stitch (0.6178) > Shared-Bottom (0.6164) > PLE (0.6050) > AutoHERI (0.6039) > MMOE (0.6011) > DNN (0.5953); |
| | Wang et al. (2022) | Multi-task learning models ($ESCM^2$), Causal learning models (Doubly Robust) | AUC: $ESCM^2$-IPS (0.6163) > $ESCM^2$-DR (0.6142) > MTL-IMP (0.6114) > MTL-IPS (0.6091) > ESMM (0.6071) > MTL-DR (0.6065) > Doubly Robust (0.5987) > MTL-EIB (0.5603);<br>KS: $ESCM^2$-DR (0.1393) > $ESCM^2$-IPS (0.1312) > ESMM (0.1267) > MTL-DR (0.1255) > MTL-IPS (0.1177) > MTL-IMP (0.1163) > Doubly Robust (0.1123) > MTL-EIB (0.0717);<br>F1: $ESCM^2$-DR (0.1315) > $ESCM^2$-IPS (0.1180) > MTL-DR (0.1159) > ESMM (0.1157) > MTL-IMP (0.1135) > Doubly Robust (0.0991) > MTL-IPS (0.0941) > MTL-EIB (0.0825);<br>Recall: $ESCM^2$-DR (0.3095) > $ESCM^2$-IPS (0.3061) > MTL-IPS (0.2975) > ESMM (0.2968) > MTL-IMP (0.2962) > MTL-DR (0.2953) > Doubly Robust (0.2854) > MTL-EIB (0.2372); |
| | Xu et al. (2022) | Knowledge distillation models (UKD), Multi-task learning models (ESMM), Deep learning models (DNN) | CVR-AUC: UKD (0.6919) > CFL (0.6789) > ESMM (0.6711) > Joint Learning (0.6699) > ESMM-Division (0.6647) > DNN (0.6514);<br>CVR-D-AUC: UKD (0.6864) > CFL (0.6738) > Joint Learning (0.6644) > ESMM (0.6627) > ESMM-Division (0.6598) > DNN (0.6447);<br>CVR-NLL: DNN (0.0442) < ESMM (0.0449) < CFL (0.0451) < UKD (0.0455) = Joint Learning (0.0455) < ESMM-Division (0.0461);<br>CVR-D-NLL: DNN (0.0551) < ESMM (0.0554) < CFL (0.0558) < Joint Learning (0.0560) < UKD (0.0563) < ESMM-Division (0.0567);<br>CTCVR-AUC: UKD (0.6493) > CFL (0.6371) > ESMM (0.6290) > Joint Learning (0.6206) > ESMM-Division (0.6172) > DNN (0.6156);<br>CTCVR-NLL: UKD (0.0020) < CFL (0.0021) = ESMM-Division (0.0021) = Joint Learning (0.0021) < ESMM (0.0021) < DNN (0.0021); |
| | Zhang et al. (2022) | Multi-task learning models (CTnoCVR) | AUC: ESMM + CTnoCVR (0.6802) > ESMM (0.6711);<br>AUC: MMOE+ CTnoCVR (0.6961) > MMOE (0.6913);<br>AUC: CGC + CTnoCVR (0.6980) > CGC (0.6903); |
| | Liu et al. (2023b) | Multi-task learning models (TAML, PLE, MMOE, ESMM, AITM), Deep learning models (MLP) | CVR-AUC: TAML (0.6813) > PLE (0.6712) > ESMM (0.6706) > MMOE (0.6703) > AITM (0.6693) > MLP (0.6519);<br>CTCVR-AUC: TAML (0.6544) >AITM (0.6518) > ESMM (0.6485) > PLE (0.6477) > MMOE (0.6400) > MLP (0.6207); |
| | Ouyang et al. (2023a) | Multi-task leaning models (CL4CVR, ESMM) | AUC: CL4CVR (0.6590) > ESMM (0.6524); |
| | Zhu et al. (2023) | Multi-task learning models (DCMT, AITM, ESMM, $ESCM^2$, PLE, MMOE) | CVR-AUC: DCMT (0.6486) > AITM (0.6324) > ESMM (0.6291) > $ESCM^2$-IPW (0.6156) > MMOE (0.6041) > $ESCM^2$-DR (0.5914) > PLE (0.5871) > Cross Stitch (0.5637);<br>CTCVR-AUC: DCMT (0.6341) > ESMM (0.6243) > $ESCM^2$-DR (0.6195) > AITM (0.6121) > MMOE (0.5947) > $ESCM^2$-IPW (0.5932) > PLE (0.5599) > Cross Stitch (0.5584); |
| | Feng et al. (2024) | Multi-task learning models (EGEAN, DDPO, ESMM, DCMT, AECM) | CVR-AUC: EGEAN + PVDR (0.6604) > EGEAN + DR (0.6527) > DDPO (0.6401) > DCMT (0.6398) > AECM (0.6192) > MTL-DR (0.6127) > ESMM (0.5813);<br>CTCVR-AUC: EGEAN + PVDR (0.6550) > EGEAN + DR (0.6455) > DCMT (0.6384) > DDPO (0.6336) > AECM (0.6133) > MTL-DR (0.6117) > ESMM (0.5799); |
| | Guo et al. (2024) | Multi-task learning models (MCAC, AITM, ESMM, PLE, MMOE), Tree models (XGBoost) | AUC: MCAC (0.6561) > AITM (0.6468) > MMOE (0.6459) > PLE (0.6425) > ESMM (0.6221) > XGBoost (0.6122) > LightGBM (0.6070); |
| | Huang et al. (2024) | Multi-task learning models (NISE, ESMM, $ESCM^2$, MMOE, DCMT) | (1) Backbone-MLP:<br>AUC: NISE (0.6498) > $ESCM^2$-IPS (0.6411) > DCMT (0.6407) > ESMM (0.6287) > MMOE (0.6216) > $ESCM^2$-DR (0.6182);<br>Logloss: NISE (0.0024) < $ESCM^2$-IPS (0.0092) < DCMT (0.0101) < MMOE (0.0108) < ESMM (0.0113) < $ESCM^2$-DR (0.0119);<br>KS: NISE (0.2137) > $ESCM^2$-IPS (0.2034) > DCMT (0.2000) > ESMM (0.1874) > MMOE (0.1758) > $ESCM^2$-DR (0.1700); |
| | | | (2) Backbone-DeepFM:<br>AUC: NISE (0.6487) > DCMT (0.6420) > ESCM$^2$-IPS (0.6394) > ESMM (0.6306) > MMOE (0.6192) > ESCM$^2$-DR (0.6003);<br>Logloss: NISE (0.0023) < DCMT (0.0090) < ESMM (0.0094) < ESCM$^2$-IPS (0.0104) = MMOE (0.0104) < ESCM$^2$-DR (0.0131);<br>KS: NISE (0.2153) > DCMT (0.2094) > ESCM$^2$-IPS (0.1986) > ESMM (0.1874) > MMOE (0.1728) > ESCM$^2$-DR (0.1452);<br>(3) Backbone-DCNV2:<br>AUC: NISE (0.6520) > DCMT (0.6431) > ESCM$^2$-IPS (0.6425) > ESMM (0.6280) > MMOE (0.6250) > ESCM$^2$-DR (0.6208);<br>Logloss: NISE (0.0024) < ESMM (0.0096) < ESCM$^2$-IPS (0.0106) < DCMT (0.0110) < MMOE (0.0113) < ESCM$^2$-DR (0.0120);<br>KS: NISE (0.2188) > ESCM$^2$-IPS (0.2089) > DCMT (0.2071) > ESMM (0.1855) > MMOE (0.1787) > ESCM$^2$-DR (0.1758); |
| | Liu et al. (2024b) | Multi-task learning models (EWAUC, Multi-IPW/DR, ESCM$^2$, CTnoCVR) | CVR-AUC: EWAUC (0.6550) > ESCM$^2$-DR (0.6510) > CTnoCVR (0.6500) > Multi-IPW (0.6490) > Multi-DR (0.6480) = ESCM$^2$-IPW (0.6480) > ESMM (0.6440);<br>CTCVR-AUC: EWAUC (0.6360) > Multi-IPW (0.6330) > ESCM$^2$-DR (0.6310) = CTnoCVR (0.6310) > ESCM$^2$-IPW (0.6300) = ESMM (0.6300) > Multi-DR (0.6290); |
| | Su et al. (2024) | Multi-task learning models (DDPO, Multi-IPW/DR, ESCM$^2$, ESMM, DCMT) | CVR-AUC: DDPO (0.6761) > MMOE + DDPO (0.6731) > DCMT (0.6318) > ESCM$^2$-IPW (0.6287) > ESCM$^2$-DR (0.6268) > Multi-DR (0.6191) > DR-JL (0.6153) > MRDR (0.6152) > ESMM (0.5879);<br>CTCVR-AUC: MMOE + DDPO (0.6541) > DDPO (0.6487) > ESCM$^2$-DR (0.6116) > MRDR (0.6047) > ESCM$^2$-IPW (0.6001) > Multi-DR (0.5992) > DCMT (0.5971) > DR-JL (0.5962) > ESMM (0.5721); |
| | Zhang et al. (2024b) | Multi-task learning models (AECM, ESCM-DR/IPW, ESMM) | (1) Dataset in the click space:<br>CVR-AUC: AECM (0.6100) > ESCM-DR (0.5922) > DR-JL (0.5855) > ESCM-IPW (0.5852) > MRDR (0.5836) > ESMM (0.5808) > MTL-DR (0.5789) > MTL-IPW (0.5738);<br>CTCVR-AUC: AECM (0.5938) > MRDR (0.5872) > ESCM-DR (0.5855) > DR-JL (0.5842) > ESCM-IPW (0.5791) > MTL-DR (0.5726) > ESMM (0.5712) > MTL-IPW (0.5685);<br>(2) Dataset in the impression space:<br>CVR-AUC: AECM (0.6157) > ESCM-DR (0.5941) > DR-JL (0.5919) > MRDR (0.5915) > MTL-DR (0.5898) > ESMM (0.5860) > ESCM-IPW (0.5858) > MTL-IPW (0.5746);<br>CTCVR-AUC: AECM (0.5960) > ESCM-DR (0.5947) > DR-JL (0.5945) > MRDR (0.5939) > ESCM-IPW (0.5791) > ESMM (0.5729) > MTL-DR (0.5715) > MTL-IPW (0.5677); |
| | Cheng et al. (2025) | Multi-task learning models (ChorusCVR, DDPO, NISE, ESMM, ESCM$^2$, DCMT), Knowledge distillation models (UKD) | CVR-AUC: ChorusCVR (0.6589) > DDPO (0.6496) > UKD (0.6451) > DCMT (0.6447) > NISE (0.6418) > ESCM$^2$-IPW (0.6385) > ESCM$^2$-DR (0.6354) > ESMM (0.5963);<br>CTCVR-AUC: ChorusCVR (0.6401) > DCMT (0.6375) > DDPO (0.6326) > NISE (0.6291) > UKD (0.6282) > ESCM$^2$-DR (0.6203) > ESCM$^2$-IPW (0.6126) > ESMM (0.5802); |
| | Fei et al. (2025) | Knowledge distillation models (EVI, UKD), Multi-task learning models (DDPO, ESMM, ESCM$^2$, DCMT) | AUC: EVI (0.6802) > DDPO (0.6711) > DCMT (0.6705) > ESMM (0.6703) > ESCM$^2$-IPW (0.6622) > UKD (0.6594) > ESCM$^2$-DR (0.6494) > MRDR-GPL (0.6195) > DR-V2 (0.6093);<br>NLL: EVI (0.0321) < DDPO (0.0330) < ESMM (0.0331) < DCMT (0.0333) < ESCM$^2$-IPW (0.0335) < UKD (0.0375) < DR-V2 (0.0580) < ESCM$^2$-DR (0.0610) < MRDR-GPL (0.1076); |
| | Yi et al. (2025) | Multi-task learning models (AEM$^2$TL, PLE, ESMM), Deep learning models (DNN) | (1) Scenario-B1 dataset:<br>CTCVR-AUC: AEM$^2$TL (0.6245) > LLM4MSR (0.6225) > M$^3$oE (0.6211) > STAR (0.6207) > Shared Bottom (0.6203) > PLE (0.6199) > M-PLE (0.6196) > AdvLoss (0.6188) > MIND (0.6187) > ESMM (0.6186) > DCN (0.6185) > xDeepFM (0.6184) > YoutubeDNN = M2M (0.6183) > DeepFM (0.6177) > DSSM (0.6176) > Wide & Deep (0.6175) > Cross-stitch (0.6171) > MMoE (0.6169) > HMoE (0.6164);<br>RelaImpr: AEM$^2$TL (1.63%) > LLM4MSR (0.00%) > M$^3$oE (-1.14%) > STAR (-1.47%) > Shared Bottom (-1.80%) > PLE (-2.12%) > M-PLE (-2.37%) > AdvLoss (-3.02%) > MIND (-3.10%) > ESMM (-3.18%) > DCN (-3.27%) > xDeepFM (-3.35%) > YoutubeDNN = M2M (-3.43%) > DeepFM (-3.92%) > DSSM (-4.00%) > Wide & Deep (-4.08%) > Cross-stitch (-4.41%) > MMoE (-4.57%) > HMoE (-4.98%);<br>(2) Scenario-B2 dataset: |
| | | | CTCVR-AUC: AEM$^2$TL (0.6176) > M$^3$oE (0.6157) > LLM4MSR (0.6155) > STAR (0.6151) > MIND (0.6147) > HMoE (0.6145) > M-PLE (0.6142) > PLE (0.6137) > Cross-stitch (0.6135) > Shared Bottom (0.6134) > MMoE (0.6133) > M2M = ESMM (0.6132) > xDeepFM (0.6123) > DeepFM (0.6121) > DCN (0.6119) > DSSM (0.6112) > Wide & Deep (0.6103) > AdvLoss (0.6099) > YoutubeDNN (0.6093);<br>RelaImpr: AEM$^2$TL (1.64%) > M$^3$oE (-0.00%) > LLM4MSR (-0.17%) > STAR (-0.52%) > MIND (-0.86%) > HMoE (-1.04%) > M-PLE (-1.64%) > PLE (-1.73%) > Cross-stitch (-1.90%) > Shared Bottom (-1.99%) > MMoE (-2.07%) > M2M = ESMM (-2.16%) > xDeepFM (-2.94%) > DeepFM (-3.11%) > DCN (-3.28%) > DSSM (-3.89%) > Wide & Deep (-4.67%) > AdvLoss (-5.01%) > YoutubeDNN (-5.53%);<br>(3) Scenario-B3 dataset:<br>CTCVR-AUC: AEM$^2$TL (0.6129) > M2M (0.6116) > PLE (0.6102) > LLM4MSR (0.6095) > STAR = Cross-stitch (0.6092) > M$^3$oE (0.6085) > M-PLE (0.6083) > HMoE (0.6075) > DSSM (0.6068) > MIND (0.6067) > Shared Bottom (0.6059) > YoutubeDNN (0.6058) > ESMM (0.6051) > AdvLoss (0.6048) > Wide & Deep (0.6047) > DeepFM (0.6042) > MMoE (0.6039) > xDeepFM (0.6023) > DCN (0.6017);<br>RelaImpr: AEM$^2$TL (0.0199) > M2M (0.0151) > PLE (-0.0045) > LLM4MSR (-0.0108) > STAR = Cross-stitch (-0.0136) > M$^3$oE (-0.0199) > M-PLE (-0.0217) > HMoE (-0.0289) > DSSM (-0.0352) > MIND (-0.0361) > Shared Bottom (-0.0434) > YoutubeDNN (-0.0443) > ESMM (-0.0506) > AdvLoss (-0.0533) > Wide & Deep (-0.0542) > DeepFM (-0.0587) > MMoE (-0.0614) > xDeepFM (-0.0759) > DCN (-0.0813);<br>(4) Entire dataset:<br>CTCVR-AUC: AEM$^2$TL (0.6478) > LLM4MSR (0.6456) > STAR (0.6441) > MIND (0.6431) > PLE (0.6430) > M$^3$oE (0.6422) > DSSM (0.6412) > HMoE (0.6409) > YoutubeDNN (0.6398) = AdvLoss (0.6398) > M2M (0.6397) > xDeepFM (0.6394) > Cross-stitch (0.6393) > ESMM (0.6352) > Shared Bottom (0.6339) > MMoE (0.6310) > DCN (0.6304) > M-PLE (0.6283) > DeepFM (0.6278) > Wide & Deep (0.6272); |
| Criteo attribution | Ren et al. (2018) | Deep learning models (DARNN), Multivariate statistical models (LR, AMTA, AH) | AUC: DARNN (0.9799) > ARNN (0.9793) > LR (0.9286) > AMTA (0.8465) > AH (0.6791) > SP (0.6718);<br>Logloss: DARNN (0.1591) < ARNN (0.1850) < AMTA (0.3897) < LR (0.3981) < AH (0.5067) < SP (0.5535); |
| | Kumar et al. (2020) | Causal learning models (CAMTA), Deep learning models (DNAMTA, DRANN), Multivariate statistical models (LR, AMTA, AH,) | AUC: CAMTA (0.9591) > CAMTA( $\lambda$ =0) (0.9469) > DNAMTA (0.9119) > LR (0.8390) > DARNN (0.6108) > AMTA (0.5309) > AH (0.5195);<br>Logloss (conversion): CAMTA (0.0977) < CAMTA( $\lambda$ =0) (0.1100) < DNAMTA (0.1365) < LR (0.1950) < DARNN (0.2389) < AMTA (0.2418) < AH (0.2620);<br>Logloss (click): CAMTA (0.0639) < CAMTA( $\lambda$ =0) (0.0785) < DARNN (0.3195); |
| | Xie et al. (2022) | Deep learning models (Att-TCN, DNAMTA), Multivariate statistical models (LR, AMTA, AH) | AUC: Att-TCN (0.9525) > TCN (0.9515) > DNAMTA (0.9012) > RNN (0.8617) > AMTA (0.8412) > LR (0.8300) > AH (0.6826) > SP (0.6718);<br>Logloss: Att-TCN (0.1643) < TCN (0.1910) < DNAMTA (0.2043) < RNN (0.3272) < AMTA (0.3615) < LR (0.3805) < AH (0.5029) < SP (0.5535); |
| | Yao et al. (2022) | Causal learning models (CausalMTA, CAMTA), Deep learning models (DNAMTA, DeepMTA, JDMTA, DARNN), Multivariate statistical models (LR, AH) | AUC: CausalMTA (0.9659) > CAMTA (0.9347) > JDMTA (0.9127) = DNAMTA (0.9127) > DeepMTA (0.9104) > DARNN (0.8726) > LR (0.8370) > SP (0.7616) > AH (0.7264);<br>Logloss: CausalMTA (0.0517) < CAMTA (0.0715) < JDMTA (0.0838) < DeepMTA (0.1120) < DNAMTA (0.1360) < DARNN (0.1650) < LR (0.1910) < AH (0.2860) < SP (0.3170); |
| | Tang et al. (2024) | Causal learning models | AUC: DCRMTA (0.7991) > CausalMTA (0.7974) > DNAMTA (0.7951) > Nlinear (0.7733) > LR (0.7021); |
| | | (DCRMTA, CausalMTA), Deep learning models (DNAMTA), Multivariate statistical models (LR) | Logloss: DCRMTA (0.1489) < DNAMTA (0.1528) < CausalMTA (0.1534) < Nlinear (0.1949) < LR (0.2866); |
| Tencent advertising algorithm competition 2017 | Li et al. (2021b) | Multi-task learning models (FTP) | AUC: FTP > FNW > FNC;<br>Logloss: FTP < FNW < FNC; |
| | Guo et al. (2022) | knowledge distillation models (KD Calibration), Multivariate statistical models (LR) | AUC: KD Calibration (0.8459) > LR-Focal Loss (0.8439) > Neural Calibration (0.8438) > LR = Isotonic Regression = Platt Scaling = Temperature Scaling = Beta Calibration (0.8389) > FNC (0.8206) > Histogram Binning (0.8201);<br>Logloss: KD Calibration (0.0733) < Neural Calibration (0.0737) < Beta Calibration (0.0742) < Isotonic Regression = Platt Scaling (0.0745) < Temperature Scaling (0.0753) < LR (0.0760) < Histogram Binning (0.0777) < FNC (0.0780) < LR-Focal Loss (0.1499);<br>BS: KD Calibration (0.0165) < LR = Isotonic Regression = Platt Scaling = Temperature Scaling = Beta Calibration = Neural Calibration (0.0166) < Histogram Binning (0.0167) < FNC (0.0169) < LR-Focal Loss (0.0270);<br>F-ECE: FNC (0.0151) < KD Calibration (0.0161) < Neural Calibration (0.0166) < Histogram Binning (0.0167) < Beta Calibration (0.0168) < Isotonic Regression = Platt Scaling (0.0170) < Temperature Scaling (0.0176) < LR (0.0187) < LR-Focal Loss (0.0939);<br>F-RCE: KD Calibration (0.8135) < Neural Calibration (0.8466) < Histogram Binning (0.9111) < Beta Calibration (0.9186) < Isotonic Regression = Platt Scaling (0.9249) < Temperature Scaling (0.9738) < FNC (0.9782) < LR (1.1231) < LR-Focal Loss (7.1861); |
| | Liu et al. (2023a) | Deep learning models (DFSN), Multi-task learning models (FTP, DEFUSE), Causal learning models (ES-DFM) | AUC: DFSN-α (0.8530) > DFSN-β (0.8477) > FTP (0.7990) > FNW (0.7430) > FNC (0.7300) > DEFUSE (0.7200) > ES-DFM (0.7000);<br>NLL: DFSN-α (0.8280) < DFSN-β (0.7930) < FTP (0.7240) < FNW (0.7070) < FNC (0.6720) < ES-DFM (0.6550) < DEFUSE (-0.7070);<br>Bias: DFSN-β (0.8990) > DFSN-α (0.8840) > FNW (0.8640) > FNC (0.8560) > FTP (0.8500) > ES-DFM (0.8080) > DEFUSE (-0.6570); |
| | Liu et al. (2024a) | Deep learning models (MISS), Multi-task learning models (FTP, DEFUSE), Causal learning models (ES-DFM) | AUC: MISS (0.8600) > FTP (0.7990) > MTDFM (0.7560) > FNW (0.7430) > FNC (0.7300) > ES-DFM (0.7200) > DEFUSE (0.7000);<br>NLL: MISS (0.8280) < FTP (0.7240) < FNW (0.7070) < MTDFM (0.6840) < FNC (0.6720) < ES-DFM (0.6550) < DEFUSE (-0.7070);<br>PRAUC: MISS (0.8820) > FTP (0.7280) > MTDFM (0.7080) > FNW (0.6110) > DEFUSE (0.5840) > ES-DFM (0.5660) > FNC (0.5570); |
| | Yu et al. (2024) | Deep learning models (DelayAdapter), Multivariate statistical models (DFM); Causal learning models (ULC, FSIW, nnDF), Multi-task learning models (MM-DFM) | AUC: DelayAdapter (0.7706) > MM-DFM (0.7704) > ULC (0.7688) > DFM (0.7687) > nnDF (0.7643) > FSIW (0.7629);<br>PRAUC: ULC (0.0950) > FSIW (0.0947) > DFM (0.0942) = nnDF (0.0942) > DelayAdapter (0.0923) > MM-DFM (0.0915);<br>NLL: DelayAdapter (0.1120) < ULC (0.1123) < MM-DFM (0.1124) = DFM (0.1124) < FSIW (0.1128) < nnDF (0.1131); |
| Miaozhen | Zhang et al. (2014) | Multivariate statistical models (AH, LR) | F1: AdditiveHazard > Simple probabilistic & Causal > LR > Last touch; |
| | Ji et al. (2016) | Multivariate statistical models (PMTA, AH, DFM) | (1) Excluding search ads:<br>AUC: PMTA > AH > SP > DFM > LR;<br>(2) Including search ads:<br>AUC: AH > SP > PMTA > DFM > LR; |
| | Ji & Wang (2017) | Multivariate statistical models (AMTA, PMTA, AH, SP, DFM) | AUC: AMTA > PMTA > AH > SP > DFM > LR; |
| | Ren et al. (2018) | Deep learning models (DARNN), Multivariate statistical models (AMTA, AH, LR) | AUC: DARNN (0.9123) > ARNN (0.8914) > AH (0.8693) > LR (0.8418) > AMTA (0.8357) > SP (0.7739);<br>Logloss: DARNN (0.1095) < ARNN (0.1610) < AMTA (0.1636) < LR (0.3496) < SP (0.5617) < AH (0.6791); |
| Alimama | Yi et al. (2025) | Multi-task learning models (AEM$^2$TL, PLE, ESMM), Deep learning models (DNN) | (1) Scenario-A1 dataset:<br>CTCVR-AUC: AEM$^2$TL (0.8608) > M$^3$oE (0.8502) > LLM4MSR (0.8498) > STAR (0.8479) > M2M (0.8475) > HMoE (0.8456) > MMOE (0.8455) > PLE (0.8405) > Cross-stitch (0.8369) > Shared Bottom (0.8358) > M-PLE (0.8357) > MIND (0.8346) > AdvLoss (0.8327) > ESMM (0.8319) > DSSM (0.8313) > Wide & Deep (0.8311) > xDeepFM (0.8267) > YoutubeDNN (0.8261) > DeepFM (0.8255) > DCN (0.8227);<br>RelaImpr: AEM$^2$TL (3.03%) > M$^3$oE (0.00%) > LLM4MSR (-0.11%) > STAR (-0.66%) > M2M (-0.77%) > HMoE (-1.31%) > MMOE (-1.34%) > PLE (-2.77%) > Cross-stitch (-3.80%) > Shared Bottom (-4.11%) > M-PLE (-4.14%) > MIND (-4.45%) > AdvLoss (-5.00%) > ESMM (-5.23%) > DSSM (-5.40%) > Wide & Deep (-5.45%) > xDeepFM (-6.71%) > YoutubeDNN (-6.88%) > DeepFM (-7.05%) > DCN (-7.85%);<br>(2) Scenario-A2 dataset:<br>CTCVR-AUC: AEM$^2$TL (0.8626) > M$^3$oE (0.8575) > LLM4MSR (0.8568) > M2M (0.8551) > STAR (0.8544) > HMoE (0.8537) > PLE (0.8536) > ESMM (0.8498) > M-PLE (0.8492) > MMOE (0.8485> YoutubeDNN (0.8378) > DSSM (0.8353) > AdvLoss (0.8339) > MIND (0.8329) > Shared Bottom (0.8326) > Wide & Deep (0.8323) > DeepFM (0.8317) > xDeepFM (0.8309) > DCN (0.8305) > Cross-stitch (0.8304);<br>RelaImpr: AEM$^2$TL (1.43%) > M$^3$oE (0.00%) > LLM4MSR (-0.20%) > M2M (-0.67%) > STAR (-0.87%) > HMoE (-1.06%) > PLE (-1.09%) > ESMM (-2.15%) > M-PLE (-2.32%) > MMOE (-2.52%) > YoutubeDNN (-5.51%) > DSSM (-6.21%) > AdvLoss (-6.60%) > MIND (-6.88%) > Shared Bottom (-6.97%) > Wide & Deep (-7.05%) > DeepFM (-7.22%) > xDeepFM (-7.44%) > DCN (-7.55%) > Cross-stitch (-7.58%);<br>(3) Scenario-A3 dataset:<br>CTCVR-AUC: AEM$^2$TL (0.8795) > LLM4MSR (0.8735) > M2M (0.8723) > STAR (0.8719) > M-PLE (0.8702) > HMoE (0.8695) > PLE (0.8693) > M$^3$oE (0.8687) > MMOE (0.8618) > Wide & Deep (0.8572) > AdvLoss (0.8562) > ESMM (0.8552) > DCN (0.8524) > Cross-stitch (0.8516) > xDeepFM (0.8465) > YoutubeDNN (0.8461) > Shared Bottom (0.8458) > DeepFM (0.8454) > DSSM (0.8427);<br>RelaImpr: AEM$^2$TL (1.61%) > LLM4MSR (0.00%) > M2M (-0.32%) > STAR (-0.43%) > M-PLE (-0.88%) > HMoE (-1.07%) > PLE (-1.12%) > M$^3$oE (-1.29%) > MMOE (-3.13%) > Wide & Deep (-4.36%) > AdvLoss (-4.63%) > ESMM (-4.90%) > DCN (-5.65%) > Cross-stitch (-5.86%) > xDeepFM (-7.23%) > YoutubeDNN (-7.34%) > Shared Bottom (-7.42%) > DeepFM (-7.52%) > DSSM (-8.25%);<br>(4) Scenario-A4 dataset:<br>CTCVR-AUC: AEM$^2$TL (0.8621) > M2M (0.8517) > LLM4MSR (0.8515) > M$^3$oE (0.8513) > STAR (0.8509) > M-PLE (0.8462) > HMoE (0.8445) > ESMM (0.8432) > PLE (0.8414) > DSSM (0.8413) > xDeepFM (0.8412) > MIND (0.8361) > MMOE (0.8357) > YoutubeDNN (0.8298) > Wide & Deep (0.8289) > AdvLoss (0.8271) > DCN (0.8268) > Cross-stitch (0.8252) > DeepFM (0.8245) > Shared Bottom (0.8229);<br>RelaImpr: AEM$^2$TL (2.96%) > M2M (0.00%) > LLM4MSR (-0.06%) > M$^3$oE (-0.11%) > STAR (-0.23%) > M-PLE (-1.56%) > HMoE (-2.05%) > ESMM (-2.42%) > PLE (-2.93%) > DSSM (-2.96%) > xDeepFM (-2.99%) > MIND (-4.44%) > MMOE (-4.55%) > YoutubeDNN (-6.23%) > Wide & Deep (-6.48%) > AdvLoss (-6.99%) > DCN (-7.08%) > Cross-stitch (-7.53%) > DeepFM (-7.73%) > Shared Bottom (-8.19%);<br>(5) Scenario-A5 dataset:<br>CTCVR-AUC: AEM$^2$TL (0.8893) > M$^3$oE (0.8782) > M2M (0.8755) > LLM4MSR (0.8752) > STAR (0.8734) > HMoE (0.8729) > M-PLE (0.8711) > MMOE (0.8698) > PLE (0.8669) > AdvLoss (0.8628) > DSSM (0.8594) > Shared Bottom (0.8584) > YoutubeDNN (0.8573) > xDeepFM (0.8552) > MIND (0.8537) > Cross-stitch (0.8535) > ESMM (0.8493) > Wide & Deep (0.8491) > DCN (0.8483) > DeepFM (0.8482);<br>RelaImpr: AEM$^2$TL (2.93%) > M$^3$oE (0.00%) > M2M (-0.71%) > LLM4MSR (-0.79%) > STAR (-1.27%) > HMoE (-1.40%) > M-PLE (-1.88%) > MMOE (- |
| | | | 2.22%) > PLE (-2.99%) > AdvLoss (-4.07%) > DSSM (-4.97%) > Shared Bottom (-5.24%) > YoutubeDNN (-5.53%) > xDeepFM (-6.08%) > MIND (-6.48%) > Cross-stitch (-6.53%) > ESMM (-7.64%) > Wide & Deep (-7.69%) > DCN (-7.91%) > DeepFM (-7.93%);<br>(6) Entire dataset:<br>AEM$^2$TL (0.8713) > M$^3$oE (0.8702) > LLM4MSR (0.8701) > PLE (0.8692) > M2M (0.8689) > STAR (0.8685) > M-PLE (0.8641) > HMoE (0.8618) > MMOE (0.8596) > ESMM (0.8434) > Cross-stitch (0.8421) > AdvLoss (0.8402) > MIND (0.8357) > DSSM (0.8246) > Shared Bottom (0.8157) > YoutubeDNN (0.8147) > Wide & Deep (0.8034) > DCN (0.8013) > xDeepFM (0.8005) > DeepFM (0.8002); |
| iPinYou | Shan et al. (2018) | Multivariate statistical models (LR) | (1) Baseline-logistic regression (the first season data):<br>RMSE: Regression-only (0.0112) < CRR (0.012) < CRT (0.0286) < Tripletwise Ranking-only (0.4522) < Pairwise Ranking-only (0.8965);<br>AUC: Pairwise Ranking-only (0.7625) > CRT (0.7622) > Tripletwise Ranking-only (0.7612) > Regression-only (0.7541) > CRR (0.7453);<br>Multi-AUC: Tripletwise Ranking-only (0.6653) > CRT (0.6652) > CRR (0.6552) > Regression-only (0.6436) > Pairwise Ranking-only (0.6349);<br>NDCG@10: CRT (0.1512) > Tripletwise Ranking-only (0.1511) > CRR (0.1457) > Regression-only (0.1398) > Pairwise Ranking-only (0.1372);<br>NDCG@20: CRT (0.1358) > CRR (0.1356) > Tripletwise Ranking-only (0.1352) > Pairwise Ranking-only (0.1270) > Regression-only (0.1248);<br>(2) Baseline-logistic regression (the second season data):<br>RMSE: CRR (0.0123) < Regression-only (0.0117) < CRT (0.0253) < Tripletwise Ranking-only (0.4565) < Pairwise Ranking-only (0.8987);<br>AUC: Tripletwise Ranking-only (0.7500) > CRT (0.7489) > CRR (0.7453) > Pairwise Ranking-only (0.7422) > Regression-only (0.7095);<br>Multi-AUC: CRT (0.6635) > Tripletwise Ranking-only (0.6621) > Pairwise Ranking-only (0.6581) > CRR (0.6552) > Regression-only (0.6349);<br>NDCG@10: Tripletwise Ranking-only (0.1503) > CRT (0.1502) > CRR (0.1457) > Pairwise Ranking-only (0.1434) > Regression-only (0.1420);<br>NDCG@20: CRT (0.1365) > Tripletwise Ranking-only (0.1361) > CRR (0.1356) > Pairwise Ranking-only (0.1303) > Regression-only (0.1287);<br>(3) Baseline-logistic regression (the third season data):<br>RMSE: CRR (0.0168) < Regression-only (0.0175) < CRT (0.0460) < Tripletwise Ranking-only (0.5619) < Pairwise Ranking-only (1.0001);<br>AUC: CRR (0.9598) > Tripletwise Ranking-only (0.9516) = CRT (0.9516) > Regression-only (0.9104) > Pairwise Ranking-only (0.8722);<br>Multi-AUC: Tripletwise Ranking-only (0.8364) > CRT (0.8262) > CRR (0.8156) > Regression-only (0.7493) > Pairwise Ranking-only (0.7165);<br>NDCG@10: CRT (0.4512) > Tripletwise Ranking-only (0.4361) > CRR (0.3587) > Regression-only (0.3217) > Pairwise Ranking-only (0.3026);<br>NDCG@20: CRT (0.3756) > CRR (0.3597) > Tripletwise Ranking-only (0.3684) > Regression-only (0.3169) > Pairwise Ranking-only (0.3124);<br>(4) Baseline-matrix factorization (the first season data):<br>RMSE: Regression-only (0.0193) < CRR (0.0198) < CRT (0.0256) < Tripletwise Ranking-only (0.4591) < Pairwise Ranking-only (0.9709);<br>AUC: CRT (0.9616) > Pairwise Ranking-only (0.9601) > Tripletwise Ranking-only (0.9600) > CRR (0.9598) > Regression-only (0.9125);<br>Multi-AUC: CRT (0.8789) > Tripletwise Ranking-only (0.8735) > CRR (0.8156) > Pairwise Ranking-only (0.8045) > Regression-only (0.7795);<br>NDCG@10: CRT (0.5789) > Tripletwise Ranking-only (0.5749) > CRR (0.5587) > Pairwise Ranking-only (0.5467) > Regression-only (0.5371);<br>NDCG@20: CRT (0.5759) > Tripletwise Ranking-only (0.5750) > CRR (0.5597) > Pairwise Ranking-only (0.5467) > Regression-only (0.5276);<br>(5) Baseline-matrix factorization (the second season data):<br>RMSE: CRR (0.0232) < Regression-only (0.025) < CRT (0.0503) < Tripletwise Ranking-only (0.4306) < Pairwise Ranking-only (0.7361);<br>AUC: CRR (0.7696) > Tripletwise Ranking-only (0.7688) > CRT (0.7687) > Regression-only (0.7653) > Pairwise Ranking-only (0.7694);<br>Multi-AUC: Tripletwise Ranking-only (0.6956) > CRT (0.6823) > CRR (0.6622) > Pairwise Ranking-only (0.6505) > Regression-only (0.6479);<br>NDCG@10: Tripletwise Ranking-only (0.3532) > CRT (0.3524) > CRR (0.3508) > Regression-only (0.3500) > Pairwise Ranking-only (0.3411);<br>NDCG@20: CRT (0.3521) > CRR (0.3505) > Tripletwise Ranking-only (0.3501) > Regression-only (0.3497) > Pairwise Ranking-only (0.3356);<br>(6) Baseline-matrix factorization (the third season data):<br>RMSE: CRR (0.0232) < Regression-only (0.0281) < CRT (0.03563) < Tripletwise Ranking-only (0.4934) < Pairwise Ranking-only (0.9891);<br>AUC: CRR (0.7696) > CRT (0.7683) > Pairwise Ranking-only (0.7673) > Tripletwise Ranking-only (0.7671) > Regression-only (0.6597);<br>Multi-AUC: Tripletwise Ranking-only (0.6961) > CRT (0.6959) > CRR (0.6622) > Pairwise Ranking-only (0.6592) > Regression-only (0.5951);<br>NDCG@10: CRT (0.3528) > Tripletwise Ranking-only (0.3515) > CRR (0.3508) > Pairwise Ranking-only (0.3498) > Regression-only (0.3373); |
| | | | NDCG@20: CRT (0.3526) > Tripletwise Ranking-only (0.3512) > CRR (0.3505) > Pairwise Ranking-only (0.3495) > Regression-only (0.337); |
| Tenc_UnionAds | Zhang et al. (2024b) | Multi-task learning models (AECM, ESMM, $ESCM^2$) | CVR-AUC: AECM (0.8240) > $ESCM^2$-DR (0.8183) > $ESCM^2$-IPW (0.8162)> MRDR (0.8136) > DR-JL (0.8129) > MTL-DR (0.8083) > ESMM (0.8081) > MTL-IPW (0.8058);<br>CVR-Precision: AECM (0.5026) > $ESCM^2$-DR (0.4852) > MTL-DR (0.4800) > $ESCM^2$-IPW (0.4772) > MTL-IPW (0.4762) > MRDR (0.4735) > DR-JL (0.4560) > ESMM (0.4475);<br>CTCVR-AUC: AECM (0.8240) > $ESCM^2$-DR (0.8181) > $ESCM^2$-IPW (0.8161) > MRDR (0.8126) > DR-JL (0.8089) > ESMM (0.8079) > MTL-DR (0.8056) > MTL-IPW (0.8030);<br>CTCVR-Precision: AECM (0.5031) > $ESCM^2$-DR (0.4886) > MTL-DR (0.4878) > MTL-IPW (0.4840) > $ESCM^2$-IPW (0.4791) > MRDR (0.4716) > DR-JL (0.4589) >ESMM (0.4528); |
| Synthetic dataset | Saito et al. (2020) | Causal learning models (nnDLA), Multivariate statistical models (DFM) | RI-Logloss: nnDLA-DF < DFM < LR; |
| | Yao et al. (2022) | Causal learning models (CausalMTA, CAMTA), Deep learning models (DNAMTA, DeepMTA), Multivariate statistical models (LR, AH) | AUC: CausalMTA (0.6814) > CAMTA (0.6485) > DNAMTA (0.6147) > DeepMTA (0.6073) > LR (0.5883) > AH (0.5832) > SP (0.5210);<br>Logloss: CausalMTA (0.6424) < DNAMTA (0.6497) < DeepMTA (0.6519) < CAMTA (0.6872) < AH (1.1136) < LR (1.1226) < SP (1.8853); |
| | Tang et al. (2024) | Causal learning models (DCRMTA, CausalMTA), Deep learning models (DNAMTA), Multivariate statistical models (LR) | AUC: DCRMTA (0.8009) > DNAMTA (0.7854) > CausalMTA (0.7749) > Nlinear (0.5609) > LR (0.5671) > SP (0.5001);<br>Logloss: DNAMTA (0.1034) < CausalMTA (0.1125) < DCRMTA (0.1154) < Nlinear (0.2465) < LR (0.3052) < SP (0.3076); |
| Proprietary datasets | Lee et al. (2012) | Multivariate statistical models (LR) | AUC: LR (0.7410) > User Cluster (0.7100) > Campaign Cluster (0.5830); |
| | Yang et al. (2016) | Multivariate statistical models (LR) | AUC: Trans-RWM > Transfer_Learning > Trans-RWM (without semantic information) > LR; |
| | Li et al. (2018) | Deep learning modes (DNAMTA), Multivariate statistical models (LR) | AUC: DNAMTA (0.8550) > LR (0.8460) > LSTM (0.8410) > HMM (0.8010) > LTA (0.8000);<br>Accuracy: DNAMTA (0.8070) = LSTM (0.8070) > LR (0.7890) > HMM (0.7660) > LTA (0.7650); |
| | Cui et al. (2018) | Deep learning modes (LSTM) | Accuracy: Ensemble CNN + LSTM (0.6580)> CNN + LSTM (0.6530) > LSTM (0.6350);<br>Logloss: Ensemble CNN + LSTM < CNN + LSTM < LSTM; |
| | Ma et al. (2018) | Hard parameter sharing (ESMM), Deep learning models (MLP) | CVR-AUC: ESMM (0.6856) > ESMM-NS (0.6825) > ESMM-DIVISION (0.6756) > MLP (0.6600);<br>CTCVR-AUC: ESMM (0.6532) > ESMM-NS (0.6444) > ESMM-DIVISION (0.6362) > MLP (0.6207); |
| | Du et al. (2019a) | Tree models (GBDT) | Logloss: BTMFE-FD + FF + FC (0.0833) < BTMFE-FD (0.0835) < BTMFE-FF (0.0836) < BTMFE-FC (0.0841) < BTMFE (0.0842) < FFM (0.0909); |
| | Du et al. (2019b) | Deep learning modes (LSTM), Multivariate statistical models (LR) | AUC: BiLSTM > LSTM > LR;<br>Accuracy: BiLSTM > LSTM > LR;<br>Precision: BiLSTM > LSTM > LR;<br>Recall: BiLSTM > LSTM > LR; |
| | Kitada et al. (2019) | Multi-task learning models, Deep learning models (GRU, MLP) | (1) Multi-task:<br>MSE: GRU + Conditional attention (0.0168) < GRU + Attention (0.0169) < GRU (0.0170) < MLP (0.0170);<br>NDCG: GRU + Conditional attention (0.9720) > MLP (0.9718) > GRU + Attention (0.9711) > GRU (0.9700) > SVM (0.9672);<br>(2) Single-task: |
| | | | MSE: GRU + Conditional attention (0.0168) < GRU + Attention (0.0169) < GRU (0.0170) < MLP (0.0171);<br>NDCG: GRU + Conditional attention (0.9677) > GRU + Attention (0.9676) > SVM (0.9672) > MLP (0.9668) > GRU (0.9654); |
| | Pan et al. (2019) | Multi-task learning models (MT-FwFM) | AUC: MT-FwFM (0.9046) > FwFM (0.9027) > FM (0.9012);<br>Weighted AUC: MT-FwFM (0.8450) > FwFM (0.8400) > FM (0.8383); |
| | Zhou et al. (2019) | Deep learning models (RNN), Multivariate statistical models (LR) | (1) ADV1 dataset:<br>AUC (Relative improvement): Global and Local Attention-based RNN (0.2095) > LR Attention Mechanism + Local attention in GRU based RNN (0.2082) = Global Attention-based RNN (0.2082) > LR Attention Mechanism (0.2068) > RNN (0.1918) > RNN + Local Attention Mechanism (0.1878) > Multi-head Self-attention (0.1741) > LR;<br>(2) ADV2 dataset:<br>AUC (Relative improvement): LR Attention Mechanism (0.0774) > LR Attention Mechanism + Local attention in GRU based RNN (0.0761) > Global and Local Attention-based RNN (0.0748) = Global Attention-based RNN (0.0748) > RNN + Local Attention Mechanism (0.0555) = RNN (0.0555) > Multi-head Self-attention (0.0542) > LR;<br>(3) ADV3 dataset:<br>AUC (Relative improvement): LR Attention Mechanism + Local attention in GRU based RNN (0.1508) > Global Attention-based RNN (0.1480) > LR Attention Mechanism (0.1466) > Global and Local Attention-based RNN (0.1411) > RNN + Local Attention Mechanism (0.1148) > RNN (0.1120) > Multi-head Self-attention (0.0982) > LR; |
| | Gligorijevic et al. (2020) | Deep learning models (DTAIN) | (1) Dataset-1:<br>AUC: DTAIN (0.9519) > RNN + Self-Attention (0.9440) > RNN (0.9436) > RNN + Attention (0.9424) > CNN (0.9408) > RF (0.9323);<br>Accuracy: DTAIN (0.9031) > RNN + Attention (0.8937) > RNN (0.8855) > RNN + Self-Attention (0.8846) > CNN (0.8757) > RF (0.8494);<br>Precision: DTAIN (0.6478) > RNN + Attention (0.6231) > RNN (0.6010) > RNN + Self-Attention (0.6002) > CNN (0.5773) > RF (0.5223);<br>Recall: DTAIN (0.8854) > CNN (0.8808) > RNN (0.8804) > RNN + Attention (0.8764) > RNN + Self-Attention (0.8691) > RF (0.8643);<br>(2) Dataset-2:<br>AUC: CNN (0.9034) > DTAIN (0.9031) > RNN + Self-Attention (0.8943) > RNN (0.8929) > RNN + Attention (0.8865) > RF (0.8845);<br>Accuracy: DTAIN (0.8857) > RNN + Attention (0.8525) > RNN (0.8474) > RNN + Self-Attention (0.8461) > RF (0.8330) > CNN (0.8225);<br>Precision: DTAIN (0.2434) > RNN + Attention (0.1974) > RF (0.1963) > RNN (0.1942) > RNN + Self-Attention (0.1921) > CNN (0.1771);<br>Recall: CNN (0.8378) > RNN (0.7902) > RNN + Self-Attention (0.7851) > RNN + Attention (0.7735) > DTAIN (0.7651) > RF (0.7469); |
| | Yasui et al. (2020) | Causal learning models (FFMIW) | Logloss: FFMIW (0.3500) < FFM (0.3523);<br>PRAUC: FFMIW (0.1660) > FFM (0.1612);<br>Normalized Logloss: FFMIW (2.304) > FFM (1.7197); |
| | Zhang et al. (2020) | Multi-task learning models (Multi-IPW/DR), Deep learning models (MLP) | (1) Dataset-A:<br>CTCVR-AUC: Multi-DR (0.7445) > Multi-IPW (0.7399) > ESMM (0.7386) > Naive IPW (0.7382) > Joint Learning DR (0.7367) > Heuristic DR (0.7345) > MLP (0.7312);<br>CTCVR-GAUC: Multi-DR (0.6090) > Multi-IPW (0.6070) > ESMM (0.6053) > Naive IPW (0.6051) > Joint Learning DR (0.6043) > Heuristic DR (0.6001) > MLP (0.5969);<br>(2) Dataset-B:<br>CTCVR-AUC: Multi-DR (0.7491) > Multi-IPW (0.7481) > Joint Learning DR (0.7451) > Naive IPW (0.7434) > ESMM (0.7433) > Heuristic DR (0.7399) > MLP (0.7386);<br>CTCVR-GAUC: Multi-IPW (0.6109) > Multi-DR (0.6099) > Naive IPW (06095) > ESMM (0.6090) > Joint Learning DR (0.6083) > Heuristic DR (0.6030) > MLP (0.6016)<br>(3) Dataset-C:<br>CTCVR-AUC: Multi-DR (0.7539) > Multi-IPW (0.7501) > ESMM (0.7497) > Naive IPW (0.7492) > Joint Learning DR (0.7490) > MLP (0.7470) > Heuristic DR (0.7418);<br>CTCVR-GAUC: Multi-DR (0.6152) > Multi-IPW (0.6125) > ESMM (0.6113) > Naive IPW (0.6109) > Joint Learning DR (0.6097) > Heuristic DR (0.6065) > MLP (0.6058);<br>(4) Dataset-D:<br>CTCVR-AUC: Multi-DR (0.7723) > Multi-IPW (0.7689) > Joint Learning DR (0.7661) > ESMM (0.7655) > Naive IPW (0.7645) > Heuristic DR (0.7640) > MLP (0.7628);<br>CTCVR-GAUC: Multi-DR (0.6228) > Multi-IPW (0.6198) > Naive IPW (0.6177) > ESMM (0.6176) > Joint Learning DR (0.6167) > Heuristic DR (0.6135) > MLP (0.6127); |
| | Li et al. (2021b) | Multi-task learning models (FTP), Causal learning models (FSIW) | AUC: FTP > FNW > FNC > FSIW;<br>Logloss: FTP < FNW < FNC < FSIW; |
| | Gu et al. (2021) | Causal learning models (DEFER, ES-DFM) | AUC: DEFER (0.6483) > FNW-RN (0.6458) > ES-DFM (0.6453) > FNW (0.6440) > FNC-RN (0.6411) > FNC (0.6368);<br>PRAUC: FNW-RN (0.6499) > DEFER (0.6497) > FNW (0.6477) > ES-DFM (0.6476) > FNC-RN (0.6447) > FNC (0.6395);<br>NLL: DEFER (0.6550) < ES-DFM (0.6560) < FNW (0.6589) < FNW-RN (0.6592) < FNC-RN (0.6710) < FNC (0.7196);<br>RI-AUC: DEFER (0.8800) > FNW-RN (0.8244) > ES-DFM (0.8133) > FNW (0.7844) > FNC-RN (0.7200) > FNC (0.6244);<br>RI-PRAUC: FNW-RN (0.8970) > DEFER (0.8920) > FNW (0.8417) > ES-DFM (0.8392) > FNC-RN (0.7663) > FNC (0.6357);<br>RI-NLL: DEFER (0.9093) > ES-DFM (0.9014) > FNW (0.8783) > FNW-RN (0.8759) > FNC-RN (0.7820) > FNC (0.3954); |
| | Hou et al. (2021) | Multi-task learning models (MM-DFM), Causal learning models (ES-DFM), Multivariate statistical models (DFM) | AUC: MM-DFM (0.7966) > ES-DFM (0.7819) > DFM (0.7763);<br>GAUC: MM-DFM (0.6547) > ES-DFM (0.6508) > DFM (0.6461);<br>PRAUC: MM-DFM (0.2779) > ES-DFM (0.2531) > DFM (0.2504); |
| | Li et al. (2021a) | Deep learning models (ACN, MLP) | AUC: ACN (0.7911) > DIEN (0.7877) > DIN (0.7861) > MLP (0.7809) > DCN (0.7800) > xDeepFM (0.7784) > DeepFM (0.7759);<br>Logloss: ACN (0.1104) < DIEN (0.1109) < DIN (0.1113) < MLP (0.1119) < DCN (0.1120) < xDeepFM (0.1123) < DeepFM (0.1126); |
| | Liu & Liu (2021) | Tree models (XGBoost), Multivariate statistical models (LR) | Logloss: XGBoost < FFM < FM < GBDT + LR < LR < GBDT; |
| | Su et al. (2021) | Deep learning models (TA-DL), Multivariate statistical models (DFM) | (1) WP1 dataset:<br>AUC: TS-DL (0.7090) > DIN (0.6986) > CRU + Attention (0.6975) > Wide & Deep (0.6573) > DFM (0.6560);<br>RelaImp: TS-DL (0.0524) > DIN (0.000) >CRU + Attention (-0.0055) > Wide & Deep (-0.2080) > DFM (-0.2145);<br>(2) WP2 dataset:<br>AUC: TS-DL (0.6478) > CRU + Attention (0.6205) > DIN (0.6021) > Wide & Deep (0.6020) > DFM (0.5795);<br>RelaImp: TS-DL (0.4476) > CRU + Attention (0.1802) > DIN (0.0000) > Wide & Deep (-0.0010) > DFM (-0.2214);<br>(3) JD-MP dataset:<br>AUC: TS-DL (0.6589) > DIN (0.6471) > GRU + Attention (0.6210) > DFM (0.6181) > Wide & Deep (0.5621);<br>RelaImp: TS-DL (0.0802) > DIN (0.0000) > GRU + Attention (-0.1774) > DFM (-0.1971) > Wide & Deep (-0.5778); |
| | Wei et al. (2021) | Multi-task learning models (AutoHERI, AITM, ESMM, $ESM^2$), Deep learning models (DNN) | CVR-AUC: AutoHERI (0.6874) > AITM (0.6822) > DBMTL (0.6799) > Multi-DR (0.6791) > $ESM^2$ (0.6781) > ESMM (0.6767) > DNN (0.6678);<br>CVR-NLL: AutoHERI (0.0131) < AITM (0.0131) < DBMTL (0.0131) < ESMM (0.0131) < $ESM^2$ (0.0132) < Multi-DR (0.0132) < DNN (0.0132);<br>CTCVR-AUC: AutoHERI (0.7447) > AITM (0.7429) > DBMTL (0.7401) > $ESM^2$ (0.7400) > ESMM (0.7372) > Multi-DR (0.7370) > DNN (0.7326);<br>CTCVR-NLL: AutoHERI (0.0028) = AITM (0.0028) < DBMTL (0.0028) = $ESM^2$ (0.0028) = Multi-DR (0.0028) < ESMM (0.0028) = DNN (0.0028); |
| | Agrawal et al. (2022) | Deep learning models (Stage-TCN, RNN), Multivariate statistical models (LR, AMTA) | (1) Dataset-1:<br>AUC: Stage-TCN (0.7900) > ARNN (0.7400) > LR (0.6800) > AMTA (0.6700) > LRATT (0.6600);<br>Accuracy: Stage-TCN (0.7200) > ARNN (0.6700) = LR (0.6700) > AMTA (0.6300) > LRATT (0.6000);<br>(2) Dataset-2:<br>AUC: Stage-TCN (0.8900) > LRATT (0.7800) > ARNN (0.7700) > AMTA (0.6800) > LR (0.6700);<br>Accuracy: Stage-TCN (0.8500) > ARNN (0.7700) = AMTA (0.7700) = LR (0.7700) > LRATT (0.5900); |
| | Chen et al. (2022b) | Multi-task learning models (CFS-MTL) | (1) Backbone-Shared-Bottom:<br>AUC1: CFS-MTL (0.7885) > MSSM (0.7799) > MTAN (0.7794) > SENet (0.7784);<br>AUC2: CFS-MTL (0.7854) > SENet (0.7729) > MSSM (0.7709) > MTAN (0.7707);<br>AUC3: CFS-MTL (0.8558) > SENet (0.8525) > MSSM (0.8508) > MTAN (0.8482); |
| | | | (2) Backbone-PLE:<br>AUC1: CFS-MTL (0.7876) > MSSM (0.7806) > MTAN (0.7803) > SENet (0.7799);<br>AUC2: CFS-MTL (0.7864) > SENet (0.7785) > MSSM (0.7746) > MTAN (0.7735);<br>AUC3: CFS-MTL (0.8575) > SENet (0.7546) > MTAN (0.8519) > MSSM (0.8508);<br>(3) Backbone-Cross-stitch:<br>AUC1: CFS-MTL (0.7869) > SENet (0.7804) > MSSM (0.7708) > MTAN (0.7678);<br>AUC2: CFS-MTL (0.7804) > SENet (0.7736) > MSSM (0.7544) > MTAN (0.7519);<br>AUC3: CFS-MTL (0.8511) > SENet (0.8491) > MTAN (0.8421) > MSSM (0.8408); |
| | Dai et al. (2022) | Multi-task learning models (DR-MSE/BIAS, Multi-IPW/DR, ESMM) | CVR-AUC： DR-MSE (0.8207) > DR-BIAS (0.8197) > Multi-IPW (0.8191) > Multi-DR (0.8186) > MRDR (0.8181) > DR-JL (0.8177) > ESMM (0.8165) > DCN (0.7571);<br>CTCVR-AUC： DR-MSE (0.9565) > DR-BIAS (0.9563) > Multi-IPW (0.9557) > Multi-DR (0.9557) > DR-JL (0.9555) > MRDR (0.9554) > DCN (0.9525) > ESMM (0.9551); |
| | Ding et al. (2022) | Deep learning models (DNN) | AUC: Gflow-FT (0.7769) > Child-TuningF (0.771) > Child-TuningD (0.7701) > IMP (0.7687) > Mixout (0.7676) > Layer-wise routing (0.7665) > Fine-Tuning (0.7672) > Regularization-based fine-tuning (0.7656) > Target-only (0.7618);<br>Logloss: Gflow-FT (0.3346) < Child-TuningF (0.3399) < Child-TuningD (0.3412) < IMP (0.3431) < Mixout (0.3465) < Fine-Tuning (0.3486) < Layer-wise routing (0.3488) < Regularization-based fine-tuning (0.3529) < Target-only (0.3548); |
| | Lin et al. (2022) | Multi-task learning models (AdaFTR, MMOE, PLE, AutoHERI), Deep learning models (DNN) | AUC: AdaFTR (0.8757) > MMOE (0.8739) > Cross-Stitch (0.8737) > Shared-Bottom (0.8735) > AutoHERI (0.8734) > PLE (0.8719) > DNN (0.8693);<br>GAUC: AdaFTR (0.8329) > Shared-Bottom (0.8303) > Cross-Stitch (0.8301) > MMOE (0.8301) > AutoHERI (0.8297) > PLE (0.8280) > DNN (0.8234); |
| | Xu et al. (2022) | Knowledge distillation models (UKD), Multi-task learning models (ESMM), Deep learning models (DNN) | (1) EC-Small dataset:<br>CTCVR-AUC: UKD (0.7513) > CFL (0.7453) > Joint Learning (0.7445) > ESMM (0.7441) > ESMM-Division (0.7434) > DNN (0.7401);<br>CTCVR-NLL: UKD (0.0039) < CFL (0.0039) = ESMM (0.0039) = Joint Learning (0.0039) < ESMM-Division (0.0039) < DNN (0.0039);<br>CVR-AUC: UKD (0.6699) > CFL (0.6600) > Joint Learning (0.6584) = ESMM (0.6584) > ESMM-Division (0.6559) > DNN (0.6531);<br>CVR-D-AUC: UKD (0.6732) > CFL (0.6587) > ESMM (0.6585) > Joint Learning (0.6582) > ESMM-Division (0.6572) > DNN (0.6558);<br>CVR-NLL: UKD (0.0208) < ESMM (0.0200) < Joint Learning (0.0209) < DNN (0.0210) < ESMM-Division (0.0210) < CFL (0.0211);<br>CVR-D-NLL: UKD (0.0235) < ESMM (0.0235) < Joint Learning (0.0236) < ESMM-Division (0.0237) < DNN (0.0237) < CFL (0.0238);<br>(2) EC-Large dataset:<br>CTCVR-AUC: UKD (0.7531) > CFL (0.7486) > ESMM (0.7480) > ESMM-Division (0.7471) > Joint Learning (0.7470) > DNN (0.7454);<br>CTCVR-NLL: UKD (0.0039) < CFL (0.0039) < ESMM (0.0039) < ESMM-Division (0.0039) < Joint Learning (0.0039) = DNN (0.0039);<br>CVR-AUC: UKD (0.6741) > ESMM (0.6686) > CFL (0.6685) = Joint Learning (0.6685) > ESMM-Division (0.6635) > DNN (0.6634);<br>CVR-D-AUC: UKD (0.6752) > CFL (0.6722) > Joint Learning (0.6705) > ESMM (0.6697) > ESMM-Division (0.6625) > DNN (0.6623);<br>CVR-NLL: UKD (0.0207) < CFL (0.0207) < Joint Learning (0.0209) < DNN (0.0209) < ESMM (0.0209) < ESMM-Division (0.0210);<br>CVR-D-NLL: UKD (0.0232) < CFL (0.0232) < Joint Learning (0.0232) < ESMM-Division (0.0233) < DNN (0.0234) < ESMM (0.0236);<br>(3) LS-Small dataset:<br>CTCVR-AUC: UKD (0.7937) > ESMM (0.7864) > Joint Learning (0.7856) > CFL (0.7839) > ESMM-Division (0.7822) > DNN (0.7801);<br>CTCVR-NLL: ESMM-Division (0.0041) < UKD (0.0041) = ESMM (0.0041) < CFL (0.0041) < Joint Learning (0.0041) = DNN (0.0041);<br>CVR-AUC: UKD (0.6958) > ESMM (0.6936) > Joint Learning (0.6927) > ESMM-Division (0.6851) > CFL (0.6823) > DNN (0.6773);<br>CVR-D-AUC: UKD (0.6831) > ESMM (0.6801) > Joint Learning (0.6792) > CFL (0.6783) > ESMM-Division (0.6725) > DNN (0.6711);<br>CVR-NLL: ESMM-Division (0.0842) < CFL (0.0844) < ESMM (0.0849) < UKD (0.0850) < DNN (0.0851) < Joint Learning (0.0856);<br>CVR-D-NLL: ESMM-Division (0.1000) < ESMM (0.1006) < UKD (0.1009) < DNN (0.1010) < Joint Learning (0.1017) < CFL (0.1027);<br>(4) LS-Large dataset:<br>CTCVR-AUC: UKD (0.7955) > ESMM (0.7876) > Joint Learning (0.7861) > CFL (0.7849) > ESMM-Division (0.7839) > DNN (0.7833); |
| | | | CTCVR-NLL: UKD (0.0041) < ESMM (0.0041) < CFL (0.0041) < Joint Learning (0.0041) < DNN (0.0041) < ESMM-Division (0.0041);<br>CVR-AUC: UKD (0.7001) > ESMM (0.6924) > Joint Learning (0.6911) > CFL (0.6869) > DNN (0.6835) > ESMM-Division (0.6741);<br>CVR-D-AUC: UKD (0.6872) > CFL (0.6825) > ESMM (0.6791) > Joint Learning (0.6774) > DNN (0.6723) > ESMM-Division (0.6638);<br>CVR-NLL: ESMM-Division (0.0824) < ESMM (0.0842) < UKD (0.0845) < CFL (0.0847) < Joint Learning (0.0849) < DNN (0.0852);<br>CVR-D-NLL: ESMM-Division (0.0999) < UKD (0.1020) < ESMM (0.1021) < DNN (0.1023) < Joint Learning (0.1024) < CFL (0.1027); |
| | Wang et al. (2022) | Multi-task learning models (ESCM$^2$, ESMM), Causal learning models (Doubly Robust) | AUC: ESCM$^2$-IPS (0.7730) > ESCM$^2$-DR (0.7679) > MTL-IPS (0.7586) > MTL-DR (0.7579) > MTL-IMP (0.7563) > ESMM (0.7547) > Doubly Robust (0.7515) > MTL-EIB (0.7272);<br>KS: ESCM$^2$-IPS (0.4144) > ESCM$^2$-DR (0.4119) > MTL-DR (0.4016) > MTL-IMP (0.3974) > MTL-IPS (0.3960) > Doubly Robust (0.3872) > ESMM (0.3856) > MTL-EIB (0.3371);<br>F1: ESCM$^2$-IPS (0.7161) > ESCM$^2$-DR (0.6884) > MTL-IPS (0.6810) > MTL-DR (0.6804) > ESMM (0.6330) > MTL-EIB (0.5808) > Doubly Robust (0.3344) > MTL-IMP (0.1272);<br>Recall: ESCM$^2$-DR (0.5986) > ESCM$^2$-IPS (0.5932) > MTL-IMP (0.5841) > Doubly Robust (0.5789) > ESMM (0.5742) > MTL-IPS (0.5651) > MTL-DR (0.5137) > MTL-EIB (0.5121); |
| | Chan et al. (2023) | Deep learning models (HDR), Causal learning models (ES-DFM, DEFER), Multi-task learning models (DEFUSE) | AUC: HDR (0.0088) > DEFUSE (0.0081) > ES-DFM (0.0080) > DEFER (0.0074);<br>PCOC: HDR (0.0099) > DEFER (0.0047) > DEFUSE (0.0045) > ES-DFM (0.0044);<br>Logloss: HDR (0.0009) < DEFER (0.0014) < DEFUSE (0.0015) < ES-DFM (0.0015);<br>ECE: HDR (0.0000) < DEFUSE (0.0002) < DEFER (0.0002) < ES-DFM (0.0002); |
| | Dai et al. (2023) | Multi-task learning models (DDFM, DEFUSE), Causal learning models (ES-DFM, FSIW) | AUC: DDFM (0.8050) > DEFUSE (0.7987) > ES-DFM (0.7982) > FNW (0.7979) > FNC (0.7977) > DEFER (0.7973) > FSIW (0.7956);<br>RI-AUC: DDFM (0.8301) > DEFUSE (0.6946) > ES-DFM (0.6839) > FNW (0.6774) > FNC (0.6731) > DEFER (0.6645) > FSIW (0.6280);<br>KS: DDFM (0.4591) > DEFUSE (0.4508) > ES-DFM (0.4499) > FNW (0.4496) > DEFER (0.4495) > FNC (0.4491) > FSIW (0.4472);<br>RI-KS: DDFM (0.7528) > DEFUSE (0.5996) > ES-DFM (0.5830) > FNW (0.5775) > DEFER (0.5756) > FNC (0.5683) > FSIW (0.5332);<br>PRAUC: DDFM (0.1036) > DEFUSE (0.1016) > ES-DFM (0.1014) > FNW (0.1013) > FNC (0.1012) > DEER (0.1010) > FSIW (0.1005);<br>RI-PRAUC: DDFM (0.4906) > DEFUSE (0.3019) > ES-DFM (0.2830) > FNW (0.2736) > FNC (0.2642) > DEFER (0.2453) > FSIW (0.1981);<br>NLL: DDFM (0.1035) < DEFUSE (0.1045) < ES-DFM (0.1046) < FNW (0.1047) < FNC (0.1048) < DEFER (0.1055) < FSIW (0.1077);<br>RI-NLL: DDFM (0.9556) < DEFUSE (0.9111) < ES-DFM (0.9067) < FNW (0.9022) < FNC (0.8978) < DEFER (0.8667) < FSIW (0.7689); |
| | Jin et al. (2023) | Deep learning models (AutoFuse), Multi-task learning models (MMOE) | (1) Moment dataset-new ads:<br>GAUC: AutoFuse (0.0083) > AutoInt (0.0082) > AFN (0.0082) > DeepFM (0.0082) > MMOE (0.0082) > MetaEmb (0.0082) > MGQE (0.0082) > DNN (0.0082) > Wide & Deep (0.0079);<br>(2) Moment dataset-old ads:<br>GAUC: AutoFuse (0.0085) > AutoInt (0.0085) > AFN (0.0085) > DeepFM (0.0085) > MMOE (0.0084) = DNN (0.0084) > MetaEmb (0.0084) > MGQE (0.0084) > Wide & Deep (0.0083);<br>(3) Moment dataset-overall:<br>GAUC: AutoFuse (0.0085) > AutoInt (0.0085) > AFN (0.0085) > DeepFM (0.0085) > MMOE (0.0084) > DNN (0.0084) = MetaEmb (0.0084)> MGQE (0.0084) > Wide & Deep (0.0083);<br>(4) Weapp dataset-new ads:<br>GAUC: AutoFuse (0.0091) > AutoInt (0.0091) > AFN (0.0090) > DeepFM (0.0090) > DNN (0.009) > MMOE (0.0090) > MGQE (0.0090) > MetaEmb (0.0090) > Wide & Deep (0.0090);<br>(5) Weapp dataset-old ads:<br>GAUC: AutoFuse(0.0086) > AutoInt (0.0086) > AFN (0.0086) > DeepFM (0.0086) > MGQE (0.0086) > DNN (0.0086) > MMOE (0.0086) > MetaEmb (0.0086) > Wide & Deep (0.0085);<br>(6) Weapp dataset-overall:<br>GAUC: AutoFuse (0.0087) > AutoInt (0.0087) > AFN (0.0087) > DeepFM (0.0087) > MGQE (00087) > DNN (0.0087) > MMOE (0.0086) > MetaEmb (0.0086) > Wide & Deep (0.0086); |
| | Liu et al. (2023b) | Multi-task learning models (TAML, PLE, MMOE, ESMM, AITM), | CVR-AUC: TAML (0.0082) > ESMM (0.0082) > PLE (0.0081) > AITM (0.0081) > MMOE (0.0081) > MLP (0.0076);<br>CTCVR-AUC: TAML (0.0096) > AITM (0.0095) > PLE (0.0095) > MMOE (0.0095) > ESMM (0.0095) > MLP (0.0095); |
| | | Deep learning models (MLP) | |
| | Ouyang et al. (2023a) | Multi-task leaning models (CL4CVR, ESMM) | AUC: CL4CVR (0.0086) > ESMM (0.0086); |
| | Ouyang et al. (2023b) | Deep learning models (MMN), Multi-task learning models (ESMM, MMOE, PLE) | AUC: MMN (0.0083) > STAR (0.0082) > PLE (0.0082) > ESMM (0.0082) > MMOE (type) (0.0082) > DNN (0.0082) > MMOE (scen) (0.0081) > MT-FwFM (0.0079); |
| | Tan et al. (2023) | Deep learning models (USSR) | (1) Dataset 1:<br>AUC: USSR (0.0090) > AutoInt (0.0088);<br>(2) Dataset 2:<br>AUC: USSR (0.0084) > AutoInt (0.0080); |
| | Wang et al. (2023c) | Causal learning models (ULC, FSIW), Multivariate statistical models (DFM) | AUC: ULC > FSIW > DFM;<br>Logloss: ULC < DFM < FSIW; |
| | Yang et al. (2023) | Deep learning models (DCBT) | (1) Warm dataset:<br>AUC: DCN (0.0091) > DCBT (0.0091);<br>PCOC: DCN (0.0068) > DCBT (0.0052);<br>(2) Cold dataset:<br>AUC: DCBT (0.0075) > DCN (0.0074);<br>PCOC: DCN (0.0043) > DCBT (0.0035);<br>(3) Entire dataset:<br>AUC: DCBT (0.0090) > DCN (0.0090);<br>PCOC: DCBT (0.0049) < DCN (0.0063); |
| | Yao et al. (2023) | Multi-task learning models (MVTA) | (1) Neighbors K-10:<br>MSE: MVTA (0.0338) < Doc2vec-DCOW (0.0387) < PCA (0.0391) < Word2Vec (0.0390) = Multi-Hot Encoding (0.0390) < LDA (0.0439) < Doc2vec-DM (0.0444) < Autoencoder (0.0529);<br>MAE: MVTA (0.1321) < Multi-Hot Encoding (0.1392) < Doc2vec-DCOW (0.1395) < Word2Vec (0.1403) < PCA (0.1408) < LDA (0.1483) < Doc2vec-DM (0.1539) < Autoencoder (0.1671);<br>$R^2$: MVTA (0.3902) > Doc2vec-DCOW (0.3034) > Word2Vec (0.2989) > Multi-Hot Encoding (0.2978) > PCA (0.2957) > LDA (0.2102) > Doc2vec-DM (0.2017) > Autoencoder (0.0475);<br>EVS: MVTA (0.3908) > Doc2vec-DCOW (0.3043) > Word2Vec (0.3004) > Multi-Hot Encoding (0.2983) > PCA (0.2958) > LDA (0.2173) > Doc2vec-DM (0.2036) > Autoencoder (0.0565);<br>(2) Neighbors K-20:<br>MSE: MVTA (0.0336) < Doc2vec-DCOW (0.0369) < PCA (0.0375) < Word2Vec (0.0378) < Multi-Hot Encoding (0.0382) < LDA (0.0434) < Doc2vec-DM (0.0447) < Autoencoder (0.0507);<br>MAE: MVTA (0.1319) < Doc2vec-DCOW (0.1367) < Multi-Hot Encoding (0.1370) < Word2Vec (0.1378) < PCA (0.1380) < LDA (0.1472) < Doc2vec-DM (0.1568) < Autoencoder (0.1644);<br>$R^2$: MVTA (0.3944) > Doc2vec-DCOW (0.3361) > PCA (0.3255) > Word2Vec (0.3199) > Multi-Hot Encoding (0.3120) > LDA (0.2185) > Doc2vec-DM (0.1952) > Autoencoder (0.0883);<br>EVS: MVTA (0.3949) > Doc2vec-DCOW (0.3368) > PCA (0.3257) > Word2Vec (0.3215) > Multi-Hot Encoding (0.3134) > LDA (0.2265) > Doc2vec-DM (0.1958) > Autoencoder (0.0942);<br>(3) Neighbors K-30:<br>MSE: MVTA (0.0342) < Doc2vec-DCOW (0.0363) < Word2Vec (0.0373) < PCA (0.0374) < Multi-Hot Encoding (0.0385) < LDA (0.0435) < Doc2vec-DM (0.0452) < Autoencoder (0.0503);<br>MAE: MVTA (0.1339) < Doc2vec-DCOW (0.1364) < Word2Vec (0.1370) < Multi-Hot Encoding (0.1378) < PCA (0.1383) < LDA (0.1483) < Doc2vec-DM (0.1591) < Autoencoder (0.1646);<br>$R^2$: MVTA (0.3843) > Doc2vec-DCOW (0.3468) > Word2Vec (0.3288) > PCA (0.3269) > Multi-Hot Encoding (0.3075) > LDA (0.2166) > Doc2vec-DM (0.1864) > Autoencoder (0.0955);<br>EVS: MVTA (0.3844) > Doc2vec-DCOW (0.3470) > Word2Vec (0.3302) > PCA (0.3272) > Multi-Hot Encoding (0.3098) > LDA (0.2251) > Doc2vec-DM (0.1867) > Autoencoder (0.0989); |
| | Ban et al. (2024a) | Deep learning models (AAGRU) | AUC: AAGRU (0.8689) > LSTM + 1D-CNNs + Interval (0.8564) > LSTM + 1D-CNNs (0.8353) > LSTM (0.8340) > 1D-CNNs (0.8297) > RF (0.7122) > 1D-CNN (0.4996); |
| | | | Accuracy: AAGRU (0.7835) > LSTM + 1D-CNNs + Interval (0.7698) > LSTM + 1D-CNNs (0.7503) > LSTM (0.7476) > 1D-CNNs (0.7457) > RF (0.7158) > 1D-CNN (0.5000); |
| | Ban et al. (2024b) | Deep learning models (DJHAN, DNAMTA, MLP), Multivariate statistical models (LR) | AUC: DJHAN (0.8173) > DTAIN (0.8164) = Global Attention-based RNN (0.8164) > GRU + Self-Attention (0.8127) > LR Attention Mechanism (0.8112) > DIEN (0.8030) > DNAMTA (0.7939) > DeepFM (0.7833) > Behavior Sequence Transformer (0.7139) > MLP (0.7089) > RF (0.6230);<br>RelaImpr: DJHAN (0.5190) > DTAIN (0.5150) = Global Attention-based RNN (0.5150) > GRU + Self-Attention (0.4970) > LR Attention Mechanism (0.4900) > DIEN (0.4500) > DNAMTA (0.4070) > DeepFM (0.3560) > Behavior Sequence Transformer (0.0240) > MLP (0.0000) > RF (-41.4000); |
| | Cui et al. (2024) | Multi-task learning models (LDACP) | (1) Entire dataset:<br>CR: LDACP (0.6232) > CREAD-P (0.5380) > MDME-P (0.5336) > Value Regression-P (0.5321) > MDME-N (0.5299) > TPM-N (0.5157) > ZILN-N (0.5085) > TPM-P (0.4877) > Value Regression-N (0.4212) > ZILN-P (0.4144) > CREAD-N (0.3807) > Ranking Model (0.2888);<br>MAPE: LDACP (0.2228) < CREAD-P (0.2914) < ZILN-N (0.3083) < Value Regression-P (0.3129) < TPM-N (0.3262) < MDME-N (0.3266) < Value Regression-N (0.3436) < TPM-P (0.3453) < MDME-P (0.3554) < CREAD-N (0.4147) < ZILN-P (0.5395) < Ranking Model (0.5604);<br>(2) Life services dataset:<br>CR: LDACP (0.6565) > CREAD-P (0.5891) > MDME-N (0.5691) > ZILN-N (0.5654) > MDME-P (0.5652) > TPM-N (0.5637) > Value Regression-P (0.5577) > CREAD-N (0.4955) > TPM-P (0.4905) > Value Regression-N (0.4856) > ZILN-P (0.3827) > Ranking Model (0.3383);<br>MAPE: LDACP (0.2293) < CREAD-P (0.2998) < Value Regression-P (0.3041) < ZILN-N (0.3079) < MDME-N (0.3260) < Value Regression-N (0.3308) < CREAD-N (0.3463) < MDME-P (0.3519) < TPM-N (0.3536) < TPM-P (0.3899) < Ranking Model (0.5214) < ZILN-P (0.5411);<br>(3) E-commerce dataset:<br>CR: LDACP (0.5906) > CREAD-P (0.5493) > MDME-N (0.5216) > ZILN-N (0.5206) > MDME-P (0.5187) > Value Regression-P (0.4976) > TPM-P (0.4922) > CREAD-N (0.4756) > TPM-N (0.4707) > VR-N (0.4558) > ZILN-P (0.4065) > Ranking Model (0.3322);<br>MAPE: LDACP (0.2433) < CREAD-P (0.2892) < MDME-N (0.3045) < MDME-P (0.3200) < ZILN-N (0.3309) < CREAD-N (0.3405) < VR-P (0.3445) < Value Regression-N (0.3607) < TPM-P (0.4099) < TPM-N (0.4333) < Ranking Model (0.5157) < ZILN-P (0.7023);<br>(4) Media dataset:<br>CR: LDACP (0.6564) > Value Regression-P (0.5679) > MDME-P (0.5643) > CREAD-P (0.5622) > MDME-N (0.5599) > TPM-N (0.5509) > ZILN-N (0.5277) > TPM-P (0.4799) > ZILN-P (0.4258) > Value Regression-N (0.4214) > CREAD-N (0.3541) > Ranking Model (0.2793);<br>MAPE: LDACP (0.2011) < CREAD-P (0.2689) < TPM-N (0.2778) < ZILN-N (0.2815) < Value Regression-P (0.2817) < MDME-N (0.2977) < Value Regression-N (0.3245) < TPM-P (0.3285) < CREAD-N (0.4285) < MDME-P (0.4419) < Ranking Model (0.5065) < ZILN-P (0.5207);<br>(5) Games dataset:<br>CR: LDACP (0.5308) > Value Regression-P (0.4218) > MDME-P (0.4184) > CREAD-P (0.4077) > MDME-N (0.4058) > TPM-N (0.3989) > ZILN-N (0.3854) > TPM-P (0.3592) > Value Regression-N (0.3337) > ZILN-P (0.3116) > CREAD-N (0.2885) > Ranking Model (0.2431);<br>MAPE: LDACP (0.2746) < MDME-P (0.3464) < CREAD-P (0.3694) < TPM-N (0.3812) < ZILN-N (0.3853) < Value Regression-N (0.4101) < Value Regression-P (0.4160) < TPM-P (0.4263) < MDME-N (0.4817) < CREAD-N (0.5032) < Ranking Model (0.9041) < ZILN-P (0.9893); |
| | Feng et al. (2024) | Multi-task learning models (EGEAN, DDPO, ESMM), Multi-task learning models (DCMT, AECM) | CVR-AUC: EGEAN + PVDR (0.7093) > EGEAN + DR (0.6940) > DDPO (0.6802) > AECM (0.6791) > MTL-DR (0.6713) > DCMT (0.6679) > ESMM (0.6507);<br>CTCVR-AUC: EGEAN + PVDR (0.7132) > EGEAN + DR (0.6971) > AECM (0.6825) > DDPO (0.6797) > MTL-DR (0.6789) > DCMT (0.6702) > ESMM (0.6582); |
| | Guo et al. (2024) | Multi-task learning models (MCAC, AITM, ESMM, MMOE, PLE), Tree models (XGBoost) | AUC: MCAC (0.9002) > AITM (0.8649) > PLE (0.8576) > MMOE (0.8513) > ESMM (0.8446) > LightGBM (0.8302) > XGBoost (0.8268);<br>Precision: MCAC (0.1632) > AITM (0.1524) > PLE (0.1473) > MMOE (0.1466) > ESMM (0.1449) > LightGBM (0.1296) > XGBoost (0.1213);<br>Recall: MCAC (0.2217) > AITM (0.1906) > PLE (0.1877) > MMOE (0.1872) > ESMM (0.1831) > LightGBM (0.1719) > XGBoost (0.1680); |
| | Huang et al. (2024) | Multi-task learning models (NISE, $ESCM^2$, | AUC: NISE (0.8172) > ESMM (0.8120) > $ESCM^2$-IPS (0.8107) > DCMT (0.8106) > MMOE (0.7923) > $ESCM^2$-DR (0.7314); |
| | | ESMM, DCMT, MMOE) | Logloss: NISE (0.0861) < ESMM (0.0867) < DCMT (0.0868) < ESCM$^2$-IPS (0.0869) < MMOE (0.0905) < ESCM$^2$-DR (0.4713);<br>KS: NISE (0.5037) > DCMT (0.4972) > ESMM (0.4971) > ESCM$^2$-IPS (0.4951) > MMOE (0.4720) > ESCM$^2$-DR (0.3687); |
| | Li et al. (2024) | Deep learning models (Transformer) | AUC: Fine tune all params (0.0363) > Fine tune LoRA (rank 64) (0.0348) > Fine tune LoRA (rank 32) (0.0344) > Fine tune LoRA (rank 1) (0.0302); |
| | Su et al. (2024) | Multi-task learning models (ESCM$^2$, ESMM, Multi-DR, DCMT) | CVR-AUC: DDPO (0.8357) > MMOE + DDPO (0.8282) > DCMT (0.8199) > ESCM$^2$-IPW (0.8181) > ESCM$^2$-DR (0.8163) > DR-JL (0.8143) > ESMM (0.8139) > MRDR (0.8133) > Multi-DR (0.8132);<br>CTCVR-AUC: DDPO (0.8525) > MMOE + DDPO (0.8444) > DCMT (0.8426) > MRDR (0.8402) > ESCM$^2$-IPW (0.8398) > ESCM$^2$-DR (0.8397) > Multi-DR (0.8390) > ESMM (0.8386) > DR-JL (0.8367); |
| | Yang et al. (2024) | Deep learning models (DESC) | AUC: DESC (0.8641) > Adaptive Calibration (0.8639) > Field-aware Calibration (0.8634) > MBCT (0.8621) > Ensemble Temperature Scaling (0.8613) = Smooth Isotonic Regression (0.8613) = Platt Scaling (0.8613) = Isotonic Regression (0.8613) > Histogram Binning (0.8610);<br>Logloss: DESC (0.5192) < Adaptive Calibration (0.5201) < Field-aware Calibration < MBCT (0.5203) < Smooth Isotonic Regression (0.5211) < Ensemble Temperature Scaling (0.5212) = Platt Scaling (0.5212) = Isotonic Regression (0.5212) < Histogram Binning (0.5218);<br>F-ECE: DESC (0.0132) < Field-aware Calibration (0.0163) < Adaptive Calibration (0.0170) < MBCT (0.0175) < Ensemble Temperature Scaling (0.0199) < Smooth Isotonic Regression (0.0201) = Platt Scaling (0.0201) < Isotonic Regression (0.0202) < Histogram Binning (0.0204);<br>F-RCE: DESC (0.0307) < Field-aware Calibration (0.0402) < MBCT (0.0409) < Adaptive Calibration (0.0413) < Smooth Isotonic Regression (0.0660) = Platt Scaling (0.0660) < Ensemble Temperature Scaling (0.0661) < Isotonic Regression (0.0665) < Histogram Binning (0.0669); |
| | Zhao et al. (2024) | Deep learning models (ConfCalib) | AUC: ConfCalib (0.8303) > Neural Calibration (0.8271) > SIR (0.8226) = Gamma Calibration (0.8226) = Gaussian Calibration (0.8226) = PlattScaling (0.8226) = Isotonic Regression (0.8226) > Histogram Binning (0.8223) > AdaCalib (0.8192);<br>LogLoss: ConfCalib (0.0860) < Neural Calibration (0.0866) < Gamma Calibration (0.0867) = Gaussian Calibration (0.0867) = Isotonic Regression (0.0867) < SIR (0.0868) = PlattScaling (0.0868) = Histogram Binning (0.0868) < AdaCalib (0.0875);<br>Field-RCE: ConfCalib (0.2728) < Neural Calibration (0.2856) < AdaCalib (0.3234) < SIR (0.3555) < Gaussian Calibration (0.3606) < Gamma Calibration (0.3756) < Isotonic Regression (0.3779) < Histogram Binning (0.3838) < PlattScaling (0.3929);<br>Multi-Field-RCE: ConfCalib (0.2243) < AdaCalib (0.2307) < Neural Calibration (0.2330) < SIR (0.2811) < Gaussian Calibration (0.2916) < Gamma Calibration (0.3047) < Isotonic Regression (0.3079) < Histogram Binning (0.3145) < PlattScaling (0.3190);<br>MVCE: ConfCalib (0.0007) < AdaCalib (0.0007 ± 0.0004) = Neural Calibration (0.0007 ± 0.0004) < Gaussian Calibration (0.0015) = Histogram Binning (0.0015) < Gamma Calibration (0.0015) = PlattScaling (0.0015) = Isotonic Regression(0.0015) < SIR (0.0016);<br>ECE: ConfCalib (0.0007) < AdaCalib (0.0010 ±0.0003) < Neural Calibration (0.0012 ±0.0006) < Isotonic Regression (0.0017) = Gaussian Calibration (0.0017) = Histogram Binning (0.0017) < PlattScaling (0.0018) = Gamma Calibration (0.0018) < SIR (0.0020); |
| | Bi ç̇ci (2025) | Deep learning models (DNN) | AUC: Masknet + res.con (0.7564) > Masknet (0.7560) > DeepIM + res.con (0.7552) > DeepIM (0.7551) > AutoInt (0.7544) > DualMLP + res.con (0.7543)> DualMLP (0.7542) > AutoInt + res.con (0.7534);<br>Logloss: Masknet + res.con (0.1940) < Masknet (0.1941) < DeepIM + res.con (0.1943) < DeepIM (0.1944) < AutoInt (0.1945) < DualMLP (0.1945) < DualMLP + res.con (0.1946) < AutoInt + res.con (0.1948);<br>F1: DualMLP + res.con (0.0037) > Masknet + res.con (0.0034) > AutoInt (0.0030) = DualMLP (0.0030) > AutoInt + res.con (0.0024) > Masknet (0.0019) > DeepIM + res.con (0.0017) > DeepIM (0.0007); |
| | Dai et al. (2025) | Deep learning models (MCNet-Field) | AUC: MCNet-Field (0.8500) = NeuCalib (0.8500) = SIR (0.8500) = Gamma Scaling (0.8500) = Gaussian Scaling (0.8500) = Platt Scaling (0.8500) > MCNet-None (0.8499) = AdaCalib (0.8499) = Istonic Regression (0.8499) > Histogram Binning (0.8493);<br>PCOC: Platt Scaling (1.0030) < Istonic Regression (1.0037) = SIR (1.0037) < Histogram Binning (1.0042) < Gaussian Scaling (1.0045) < NeuCalib (1.0050) < MCNet-Field (1.0056) < MCNet-None (1.0059) < Gamma Scaling (1.0065) < AdaCalib (1.0071);<br>F-RCE: MCNet-Field (0. 0061) < Histogram Binning (0. 0080) < Istonic Regression (0. 0081) < SIR (0. 0082) < NeuCalib (0. 0087) < Gaussian Scaling (0. 0093) < Platt Scaling (0. 0095) < MCNet-None (0. 0096) < AdaCalib (0.0104) < Gamma Scaling (0.0113); |
| | Yuan et al. (2025) | Knowledge distillation | (1) Training dataset: |
| | | models (HA-PFD) | AUC: HA-PFD V2 (0.9302) > Review KD (0.9297) > Similarity KD (0.9293) > Adversarial KD (0.9290);<br>Teacher AUC: HA-PFD V2 (0.9708) > Similarity KD (0.9703) > Review KD (0.9700) > Adversarial KD (0.9700);<br>ECE: HA-PFD V2 (0.0378) < Similarity KD (0.0532) < Review KD (0.0547) < Adversarial KD (0.0583);<br>MCE: HA-PFD V2 (0.1896) < Similarity KD (0.3884) < Review KD (0.4002) < Adversarial KD (0.4102);<br>(2) Test dataset:<br>AUC: HA-PFD V2 (0.8877) > Review KD (0.8870) > Similarity KD (0.8852) > Adversarial KD (0.8842);<br>Teacher AUC: HA-PFD V2 (0.9664) > Review KD (0.9663) > Adversarial KD (0.9652) > Similarity KD (0.9651);<br>ECE: HA-PFD V2 (0.0565) < Similarity KD (0.0780) < Review KD (0.0803) < Adversarial KD (0.0816);<br>MCE: HA-PFD V2 (0.2225) < Similarity KD (0.4240) < Review KD (0.4288) < Adversarial KD (0.4333). |

Note: The evaluation metrics reported in this table are rounded to the nearest value at the 0.0001 level. Evaluation metrics without explicitly reported numerical values in the literature were not added to the table.

# 6. Research perspectives

In this section, we discuss current research trends and challenges, and outline potential future directions in advertising CVR prediction.

## 6.1. Current research trends

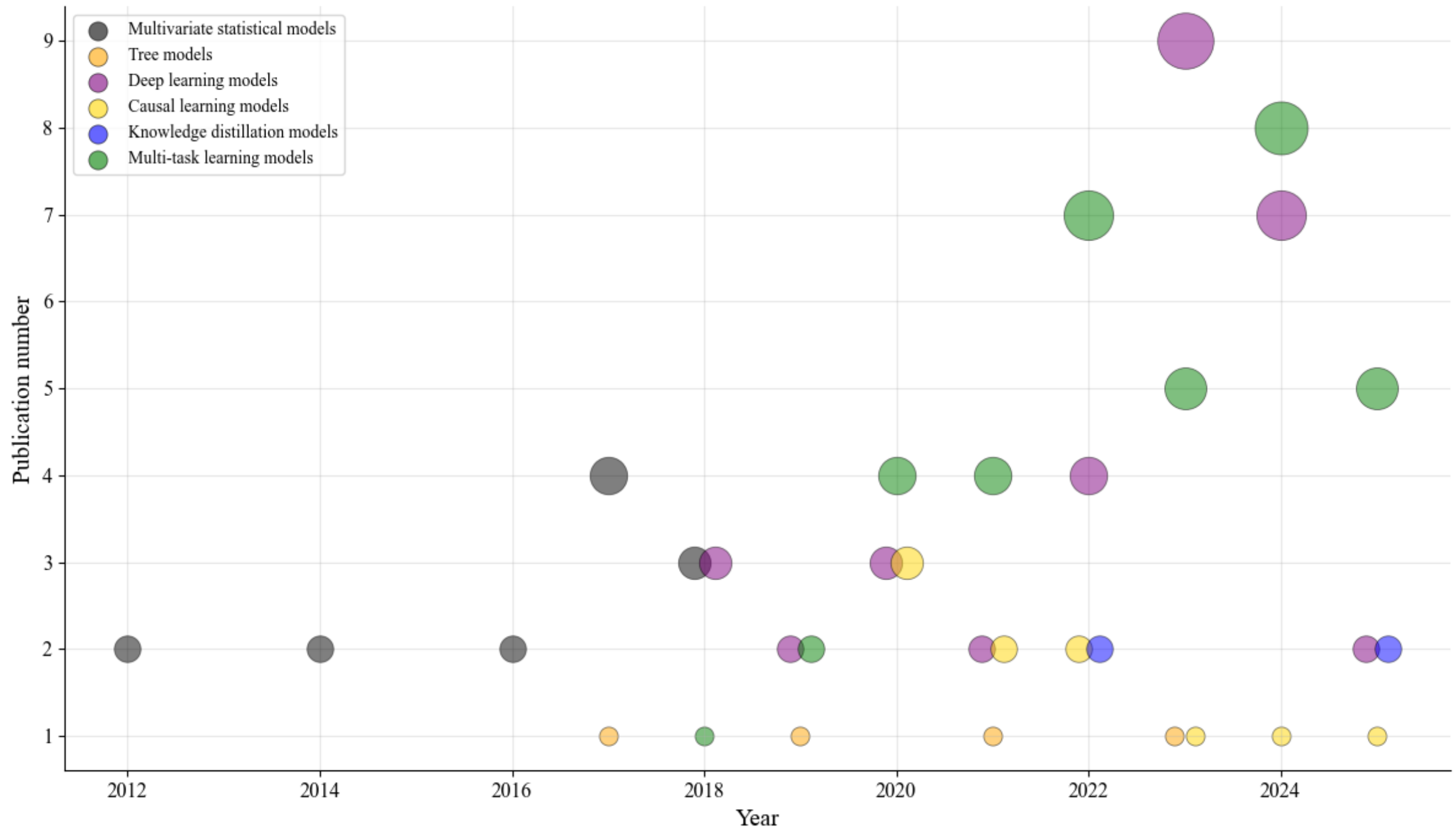

Figure 15. The distribution of publications in categories of CVR prediction models.

Figure 15 illustrates the annual publications in the six categories of advertising CVR prediction models presented in Section 4, where the circle size is in proportion to the number of publications. From Figure 15, we can observe that, in recent years, multi-task learning models and deep learning models have become research hotspots and causal learning models and knowledge distillation models have attracted increasing attention. Based on these observations, we summarize current research trends as follows.

### 6.1.1. Sequential modeling of user behaviors for CVR prediction

User's interactions with ads (e.g., browsing, clicking and purchasing) form a continuous sequence from which high-order collaborative information, dynamic user intents and behavioral patterns can be learned to enhance behavioral predictions (Wan et al., 2025; Yu et al., 2025). Recent studies have developed various deep learning models to capture intricate temporal dynamics, long-range dependencies, and evolving behavior patterns in behavior sequences for CVR prediction (e.g., Agrawal et al., 2022; Min et al., 2023; Ban et al., 2024a). Moreover, it is a promising perspective to capture multiple interests (Li et al., 2021a) and multi-scale periodic patterns (Zhai et al., 2025) from sequential behaviors for CVR prediction.

### 6.1.2. Attribution-enhanced CVR prediction

Conversions are usually driven by a sequence of interactions, each of which makes a certain contribution to the conversion. Attribution analysis quantifies and assigns conversion credits to various interactions, which discerns influential interactions to enhance CVR prediction. In order to achieve accurate attribution credits, time-aware sequential models and the dual-attention mechanism have been employed to learn decay patterns in behavioral sequences for enhancing CVR prediction (Ren et al., 2018; Yang et al., 2020), and the layer-wise relevance propagation technique has been adopted to compute conversion attribution credits by setting relevance back-propagation rules (Agrawal et al., 2022).

### 6.1.3. Reasonable pseudo-labels generation

In online advertising, the number of unclicked samples far exceeds that of clicked samples. The unclicked samples contain rich impression-level information that is essentially beneficial for CVR prediction. However, most existing CVR prediction models treat unclicked samples as negatives, which leads to a non-negligible bias in the predicted CVR. This motivates using pseudo-labels for unclicked samples as informative supervision signals, so that the training and inference of CVR prediction models can explicitly utilize both clicked and unclicked samples in the impression space. A notable research stream for pseudo-labels generation constructs a CVR prediction tower in the click space to produce pseudo-labels through counterfactual mechanisms and semi-supervised learning, and then uses cross-entropy constraints to train a CVR estimator in the unclicked space (Zhu et al., 2023; Huang et al., 2024; Su et al., 2024). Another feasible solution learns click-adaptive representations for impressions via the unsupervised domain adaptation (Xu et al., 2022) and click propensity modeling (Fei et al., 2025), which can be used to produce reliable pseudo-labels.

#### 6.1.4. Knowledge transfer between tasks

In online advertising, behaviors that are closely related to conversions (e.g., browsing and clicks) are helpful to improve CVR prediction. Specifically, knowledge extracted from the related tasks can facilitate the representation learning for CVR prediction, and also serve as regularizers to reduce overfitting risk. However, inter-task knowledge transfers via simple parameter sharing are too rigid to capture the fine-grained relatedness among tasks (Lin et al., 2022). In order to facilitate the knowledge transfer among tasks, one-shot search algorithm has been used to automatically capture the inter-task hierarchical representations and the lay-wise interplay (Wei et al., 2021). Moreover, it needs to align the representation space for CVR prediction and related tasks by using adaptive contrast learning (Ouyang et al., 2023a) and domain adaptation based on adversarial learning (Zhang et al., 2024b).

### 6.2. Main challenges

#### 6.2.1. The delayed feedback from clicks to conversions

In online advertising, clicks usually occur immediately after impressions, whereas conversions may occur after a relatively long time. As a result, conversions that have not been unobserved are easily mislabeled as negatives, which severely affects CVR prediction. For online delayed feedback, a time window is usually pre-determined to ensure that conversion labels are accurate (Yang et al., 2021). However, it is hard to set a proper time window, as short time windows increase fake negatives and long windows may sacrifice data freshness (Yu et al., 2024). The conversion delay can be quantized into multiple time windows with various output heads, which are dynamically aggregated to compute the CVR (Chen et al., 2022; Gao & Yang, 2022; Liu et al., 2024b).

In online environments, it requires to update parameters of delayed feedback models by duplicating fake negative samples from newly observed data and re-ingesting the factual labels during training (Liu et al., 2023a). However, duplicated samples may confuse the model and hinder the relevance of label information. It is suggested to employ influence functions to estimate impacts of newly arrived data and directly update parameters (Ding et al., 2025). Moreover, the conversion process with delayed feedback can be regarded as a stochastic delay differential equation where the diffusion coefficient decreases over the delay time (Zhang et al., 2025a), so as to characterize the conversion probability with the delay feedback.

### 6.2.2. The sample selection bias in CVR prediction

The data distribution of the click space consisting of only clicked samples differs from that of the impression space that contains all samples. Hence, CVR prediction models trained in the click space are inevitably biased by user selection, which raises the issue of sample selection bias.

Although multi-task modeling extends CVR prediction to the entire impression space by introducing co-related auxiliary tasks (Ma et al., 2018; Zhang et al., 2024a) and counterfactual learning models reweight the CVR loss on clicked samples for unbiased CVR prediction (Zhang et al., 2020; Dai et al., 2022; Wang et al., 2022). However, the former may face the gradient conflicts of different prediction tasks caused by inconsistent labels (Cheng et al., 2025), and the latter fails to account for unclicked samples. To this end, knowledge distillation estimates conversion probabilities of unclicked samples which are combined with clicked samples with conversion labels to provide impression-level signals for unbiased CVR estimation (Xu et al., 2022); However, they generally rely on a generalization-capable teacher model and the prediction accuracy of auxiliary tasks (e.g., CTR prediction) (Su et al., 2024). It is promising to explore the augmentation of unobserved samples by using the Markov chain Monte Carlo algorithm, and design Bayesian distributional regression models to correct the sample selection bias for CVR prediction.

### 6.2.3. Spurious correlations between exposures and conversions

In online advertising, ads exposed to users are tailored to their preferences, rather than randomly assigned, which raises the problem of spurious causal relationships in the observed data. That is, user preferences may simultaneously drive both ad impressions and conversions. Thus, it leads to wrong estimations for contributions of exposures to conversion, and in turn the biased CVR prediction (Kumar et al., 2020).

Existing research has employed CRNs and adversarial learning to eliminate the spurious correlation between impressions and conversions and learn invariant balanced representations from behavioral sequences for CVR prediction by treating clicks as pseudo-feedback. However, clicks may also be affected by user preferences, which introduces additional confounding factors (Yao et al., 2022). A potential solution is to employ casual learning and the causal attention mechanism to implement real perturbations in high-dimensional user features and refine causal effects estimation among various types of user behaviors (Tang et al., 2024).

## 6.3. Future directions

### 6.3.1. CVR prediction based on large language models

Large language models (LLMs) have demonstrated strong performance in semantic understanding and contextual reasoning. Especially, LLMs can be utilized to learn semantics-enriched representations of advertising contents (e.g., textual descriptions and visual elements) and contexts and generate complementary textual interpretations for image and video elements in ads (e.g., product highlights, visual summaries, and content abstractions) (Wang et al., 2024), substantially enhancing representation learning for CVR prediction. Moreover, categorical and numerical features can also be transformed into semantics-enriched text embeddings via LLMs (Han et al., 2024). Meanwhile, it is promising to map user attributes and ad textual interpretations generated by prompt engineering into the latent space of LLMs for learning semantic representations and interaction relationships (Lin et al., 2024).

### 6.3.2. Multi-teacher knowledge distillation for CVR prediction

Existing knowledge distillation models employed a single teacher model to generate pseudo-labels for unclicked samples, which places heavy demands on the teacher's capability and is susceptible to noises in the label distribution. It is suggested to construct a multi-teacher knowledge distillation framework to effectively process multi-type features (e.g., categorical features and sequential features) and generate pseudo-labels.

Moreover, most distillation strategies are limited to the logit-level distillation, failing to exploit latent unbiased information (Yuan et al., 2025). It is interesting to incorporate cross-layer distillation approaches and attention mechanisms for efficient multi-level latent information transfer from the teacher model to the student model, thus enhancing the representation learning capability of the student model.

### 6.3.3. Meta-learning for the debiased CVR estimation

Owing to the advantage of "learning to learn", meta-learning enables models to extract valuable knowledge from a small number of samples and quickly adapt to various tasks that have not been encountered before (Pan et al., 2022). On one hand, meta-learning can be integrated with counterfactual learning models to extract the debiased knowledge from small-scale unbiased data and alleviate the high variance in biased data (Li et al., 2025a). On the other hand, meta-learning is also helpful to capture delayed feedback by dynamically adapting modeling parameters through meta-optimization that leverages delayed conversion data across multiple time windows.

### 6.3.4. Multimodal learning for CVR prediction

The proliferation of micro-video sharing platforms has driven a surge in ads with multimodal content (e.g., textual and visual elements) (Li et al., 2020) that reflect advertising creatives and contain rich semantic information. To this end, it calls for multimodal learning models to capture semantic patterns and effects of different modalities for CVR prediction.

In this direction, it is suggested to map multimodal feature embeddings into a unified feature space for alignment through contrast learning (Sheng et al., 2024), handle interaction and fusion by using fine-tuned multimodal large language models, and capture the cooperative effects by constructing multimodal interaction graphs in the framework of GNNs.

### 6.3.5. Data augmentation for CVR prediction

Deep learning models usually require data-rich environments. Data augmentation is an effective technique to enrich training data and ensure the reliable model performance even with limited data. An interesting perspective is to generate behavioral sequences with conversion labels through conditional diffusion models and generative adversarial networks for CVR prediction. Moreover, it is promising to extract semantically consistent behavioral subsequences from users with similar interests that are used to synthesize generated behavioral sequences (Li et al., 2021a).

### 6.3.6. The modeling interpretability of CVR prediction

Deep learning models for CVR prediction have made breakthroughs in the predictive performance, but the learning process is implicit and offers limited interpretability. This hinders managers and advertisers from understanding advertising effects on conversions and identifying influential factors driving advertising decisions.

In this direction, the most important issue is to estimate the importance of each feature to the conversion probability, either by combining probabilistic models and tree models or by integrating GNNs and graph information bottleneck. Moreover, it is desirable to capture causal effects of observed variables on users' conversion decisions by integrating reinforcement learning with Shapley value decomposition methods (Yang et al., 2020) to achieve interpretable CVR predictions.

## 7. Conclusion

This review summarizes the area of advertising CVR prediction based on 99 articles collected from six major academic databases. From the perspective of underlying techniques, we classify CVR prediction models into six categories, introduce the basic framework and its variants, discuss advantages and disadvantages, as well as how they are applied in CVR prediction. Moreover, we summarize performance comparisons among various CVR prediction models on public and proprietary datasets. Lastly, we identify current research trends, imperative challenges, and outline promising prospects for future exploration.

This review consolidates the fragmented research on CVR prediction in the extant literature into a synthetic framework, expecting to provide methodological foundations and valuable insights for conversion modeling innovation and development. This review also offers actionable guidance for practitioners in improving the prediction efficiency of conversion-related performance indexes of advertising campaigns in industrial applications.

# References

[1]. Agrawal, A., Sheoran, N., Suman, S., & Sinha, G. (2022). Multi-touch attribution for complex B2B customer journeys using temporal convolutional networks. In Companion Proceedings of the ACM Web Conference 2022 (WWW’22) (pp. 910-917). Association for Computing Machinery, New York, NY, USA.

[2]. Ban, G., Woo, S. S., & Sung, D. (2024a). Action attention GRU: A data-driven approach for enhancing purchase predictions in digital marketing. In Proceedings of the 39th ACM/SIGAPP Symposium on Applied Computing (SAC’24) (pp. 919-926). Association for Computing Machinery, New York, NY, USA.

[3]. Ban, G., Yun, H., Lee, B., Sung, D., & Woo, S. S. (2024b). Deep journey hierarchical attention networks for conversion predictions in digital marketing. In Proceedings of the 33rd ACM International Conference on Information and Knowledge Management (CIKM’24) (pp. 4358-4365). Association for Computing Machinery, New York, NY, USA.

[4]. Bhamidipati, N., Kant, R., & Mishra, S. (2017). A large scale prediction engine for app install clicks and conversions. In Proceedings of the 26th ACM on Conference on Information and Knowledge Management (CIKM’17) (pp. 167-175). Association for Computing Machinery, New York, NY, USA.

[5]. Biçici, E. (2025). Residual connections improve click-through rate and conversion rate prediction performance. Neural Computing and Applications, 37, 10675-10688.

[6]. Chan, Z., Zhang, Y., Han, S., Bai, Y., Sheng, X. R., Lou, S., Hu, J., Jiang, Y., Xu, J., & Zheng, B. (2023). Capturing conversion rate fluctuation during sales promotions: A novel historical data reuse approach. In Proceedings of the 29th ACM SIGKDD Conference on Knowledge Discovery and Data Mining (KDD’23) (pp. 3774-3784). Association for Computing Machinery, New York, NY, USA.

[7]. Chapelle, O. (2014). Modeling delayed feedback in display advertising. In Proceedings of the 20th ACM SIGKDD International Conference on Knowledge Discovery and Data Mining (KDD’14) (pp. 1097-1105). Association for Computing Machinery, New York, NY, USA.

[8]. Chen, S., Xu, Z., Xu, D., & Gou, X. (2024). Customer purchase prediction in B2C e-business: A systematic review and future research agenda. Expert Systems with Applications, 124261.

[9]. Chen, Y., Jin, J., Zhao, H., Wang, P., Liu, G., Xu, J., & Zheng, B. (2022a). Asymptotically unbiased estimation for delayed feedback modeling via label correction. In Proceedings of the ACM Web Conference 2022 (WWW'22) (pp. 369-379). Association for Computing Machinery, New York, NY, USA.

[10]. Chen, Z., Wu, R., Jiang, C., Li, H., Dong, X., Long, C., He, Y., Cheng, L., & Mo, L. (2022b). CFS-MTL: A causal feature selection mechanism for multi-task learning via pseudo-intervention. In Proceedings of the 31st ACM International Conference on Information and Knowledge Management (CIKM'22) (pp. 3883-3887). Association for Computing Machinery, New York, NY, USA.

[11]. Cheng, W., Lu, Y., Xia, B., Cao, J., Xu, K., Wen, M., Zhang, J., Liu, Z., Hong, L., Gai, K., & Zhou, G. (2025). ChorusCVR: Chorus supervision for entire space post-click conversion rate modeling. arXiv preprint arXiv:2502.08277.

[12]. Cui, P., Yang, Y., Jin, F., Tang, S., Wang, Y., Yang, F., Jia, Y., Cai, Q., Pan, F., Li, C., & Jiang, P. (2024). LDACP: Long-delayed ad conversions prediction model for bidding strategy. arXiv preprint arXiv:2411.16095.

[13]. Cui, Y., Tobossi, R., & Vigouroux, O. (2018). Modelling customer online behaviours with neural networks: Applications to conversion prediction and advertising retargeting. arXiv preprint arXiv:1804.07669.

[14]. Dai, Q., Li, H., Wu, P., Dong, Z., Zhou, X. H., Zhang, R., Zhang, R., & Sun, J. (2022). A generalized doubly robust learning framework for debiasing post-click conversion rate prediction. In Proceedings of the 28th ACM SIGKDD Conference on Knowledge Discovery and Data Mining (KDD'22) (pp. 252-262). Association for Computing Machinery, New York, NY, USA.

[15]. Dai, Q., Xiao, J., Du, Z., Zhu, J., Luo, C., Wu, X. M., & Dong, Z. (2025). MCNet: Monotonic calibration networks for expressive uncertainty calibration in online advertising. arXiv preprint arXiv:2503.00334.

[16]. Dai, S., Zhou, Y., Xu, J., & Wen, J. R. (2023). Dually enhanced delayed feedback modeling for streaming conversion rate prediction. In Proceedings of the 32nd ACM International Conference on Information and Knowledge Management (CIKM'23) (pp. 390-399). Association for Computing Machinery, New York, NY, USA.

[17]. Ding, C., Wu, J., Yuan, Y., Fang, J., Li, C., Wang, X., & He, X. (2025). Addressing delayed feedback in conversion rate prediction via influence functions. arXiv preprint arXiv:2502.01669.

[18]. Ding, K., He, Y., Dong, X., Yang, J., Zhang, L., Li, A., Zhang, X., & Mo, L. (2022). GFlow-FT: Pick a child network via gradient flow for efficient fine-tuning in recommendation systems. In Proceedings of the 31st ACM International Conference on Information and

Knowledge Management (CIKM'22) (pp. 3918-3922). Association for Computing Machinery, New York, NY, USA.

[19]. Dishi, Y., Friedler, O., Karni, Y., Silberstein, N., & Stolin, Y. (2025). Practical multi-task learning for rare conversions in Ad Tech. arXiv preprint arXiv:2507.20161.

[20]. Du, G., Chen, Y., & Wang, X. (2019a). Boosted trees model with feature engineering: An approach for post-click conversion rate prediction. In Proceedings of the 2nd International Conference on Big Data Technologies (ICBDT'19) (pp. 136-141). Association for Computing Machinery, New York, NY, USA.

[21]. Du, R., Zhong, Y., Nair, H., Cui, B., & Shou, R. (2019b). Causally driven incremental multi touch attribution using a recurrent neural network. arXiv preprint arXiv:1902.00215.

[22]. Fei, K., Zhang, X., & Li, J. (2025). Entire-Space variational information exploitation for post-click conversion rate prediction. In Proceedings of the 39th AAAI Conference on Artificial Intelligence (AAAI'25) (pp. 11654-11662). AAAI Press, Palo Alto, California, USA.

[23]. Feng, H., Zhang, G., Zhang, Y., We, Y., & Liu, Q. (2024). EGEAN: An exposure-guided embedding alignment network for post-click conversion estimation. arXiv preprint arXiv:2412.06852.

[24]. Gandhudi, M., Gangadharan, G. R., Alphonse, P. J. A., Velayudham, V., & Nagineni, L. (2023). Causal aware parameterized quantum stochastic gradient descent for analyzing marketing advertisements and sales forecasting. Information Processing & Management, 60(5), 103473.

[25]. Gao, H., & Yang, Y. (2022). Multi-head online learning for delayed feedback modeling. arXiv preprint arXiv:2205.12406.

[26]. Gharibshah, Z., & Zhu, X. (2021). User response prediction in online advertising. ACM Computing Surveys (CSUR), 54(3), 1-43.

[27]. Gligorijevic, D., Gligorijevic, J., & Flores, A. (2020). Prospective modeling of users for online display advertising via deep time-aware model. In Proceedings of the 29th ACM International Conference on Information and Knowledge Management (CIKM'20) (pp. 2461-2468). Association for Computing Machinery, New York, NY, USA.

[28]. Gu, S., Sheng, X. R., Fan, Y., Zhou, G., & Zhu, X. (2021). Real negatives matter: Continuous training with real negatives for delayed feedback modeling. In Proceedings of the 27th ACM SIGKDD Conference on Knowledge Discovery and Data Mining (KDD'21) (pp. 2890-2898). Association for Computing Machinery, New York, NY, USA.

[29]. Guo, B., Song, X., Wang, S., Gong, W., He, T., & Liu, X. (2024). Multi-task conditional attention network for conversion prediction in logistics advertising. In Proceedings of the 30th ACM SIGKDD Conference on Knowledge Discovery and Data Mining (KDD'24) (pp. 5028-5037). Association for Computing Machinery, New York, NY, USA.

[30]. Guo, Y., Ao, X., Liu, Q., & He, Q. (2023). Leveraging post-click user behaviors for calibrated conversion rate prediction under delayed feedback in online advertising. In Proceedings of the 32nd ACM International Conference on Information and Knowledge

Management (CIKM'23) (pp. 3918-3922). Association for Computing Machinery, New York, NY, USA.

[31]. Guo, Y., Li, H., Ao, X., Lu, M., Liu, D., Xiao, L., Jiang, J., & He, Q. (2022). Calibrated conversion rate prediction via knowledge distillation under delayed feedback in online advertising. In Proceedings of the 31st ACM International Conference on Information and Knowledge Management (CIKM'22) (pp. 3983-3987). Association for Computing Machinery, New York, NY, USA.

[32]. Han, R., Li, Q., Jiang, H., Li, R., Zhao, Y., Li, X., & Lin, W. (2024). Enhancing CTR prediction through sequential recommendation pre-training: Introducing the SRP4CTR framework. In Proceedings of the 33rd ACM International Conference on Information and Knowledge Management (CIKM'24) (pp. 3777-3781). Association for Computing Machinery, New York, NY, USA.

[33]. Hao, H., Wang, Y., Xue, S., Xia, Y., Zhao, J., & Shen, F. (2020). Temporal convolutional attention-based network for sequence modeling. arXiv preprint arXiv:2002.12530.

[34]. Hou, Y., Zhao, G., Liu, C., Zu, Z., & Zhu, X. (2021). Conversion prediction with delayed feedback: A multi-task learning approach. In 2021 IEEE International Conference on Data Mining (ICDM'21) (pp. 191-199). IEEE Computer Society, Los Alamitos, CA, USA.

[35]. Huang, J., Zhang, L., Wang, J., Jiang, S., Huang, D., Ding, C., & Xu, L. (2024). Utilizing non-click samples via semi-supervised learning for conversion rate prediction. In Proceedings of the 18th ACM Conference on Recommender Systems (RecSys'24) (pp. 350-359). Association for Computing Machinery, New York, NY, USA.

[36]. Ji, W., & Wang, X. (2017). Additional multi-touch attribution for online advertising. In Proceedings of the 31st AAAI Conference on Artificial Intelligence (AAAI'17) (pp. 1360-1366). AAAI Press, Palo Alto, California, USA.

[37]. Ji, W., Wang, X., & Zhang, D. (2016). A probabilistic multi-touch attribution model for online advertising. In Proceedings of the 25th ACM International on Conference on Information and Knowledge Management (CIKM'16) (pp. 1373-1382). Association for Computing Machinery, New York, NY, USA.

[38]. Jiang, J., & Jiang, H. (2017). Conversion rate estimation in online advertising via exploring potential impact of creative. In International Conference on Database and Expert Systems Applications (DEXA'17) (pp. 299-309). Lecture Notes in Computer Science, Springer, Cham.

[39]. Jin, J., Lu, G., Li, S., Gao, X., Tan, A., & Wang, L. (2023). Automatic fusion network for cold-start CVR prediction with explicit multi-level representation. In 2023 IEEE 39th International Conference on Data Engineering (ICDE'23) (pp. 3440-3452). IEEE, Piscataway, NJ, USA.

[40]. Jin, Y., & Yang, Y. (2025). A survey on knowledge graph-based click-through rate prediction. Expert Systems with Applications, 281, 127501.

[41]. Jin, Y., Yang, Y., & Ma, B. (2025). Dynamic heterogeneous graph convolutional networks for click-through rate prediction in recommender systems. Electronic Commerce Research, 1-30.

[42]. Kitada, S., Iyatomi, H., & Seki, Y. (2019). Conversion prediction using multi-task conditional attention networks to support the creation of effective ad creatives. In Proceedings of the 25th ACM SIGKDD International Conference on Knowledge Discovery and Data Mining (KDD'19) (pp. 2069-2077). Association for Computing Machinery, New York, NY, USA.

[43]. Kuhn, M., & Johnson, K. (2019). Feature engineering and selection: A practical approach for predictive models. CRC Press.

[44]. Kumar, S., Gupta, G., Prasad, R., Chatterjee, A., Vig, L., & Shroff, G. (2020). CAMTA: Causal attention model for multi-touch attribution. In 2020 International Conference on Data Mining Workshops (ICDMW'20) (pp. 79-86). IEEE Computer Society, Los Alamitos, CA, USA.

[45]. Lane, R. O. (2025). A comprehensive review of classifier probability calibration metrics. arXiv preprint arXiv:2504.18278.

[46]. Lea, C., Flynn, M. D., Vidal, R., Reiter, A., & Hager, G. D. (2017). Temporal convolutional networks for action segmentation and detection. In Proceedings of the IEEE Conference on Computer Vision and Pattern Recognition (CVPR'17) (pp. 156-165). IEEE, Piscataway, NJ, USA.

[47]. Lee, K. C., Orten, B., Dasdan, A., & Li, W. (2012). Estimating conversion rate in display advertising from past performance data. In Proceedings of the 18th ACM SIGKDD International Conference on Knowledge Discovery and Data Mining (KDD'12) (pp. 768-776). Association for Computing Machinery, New York, NY, USA.

[48]. Li, D., Hu, B., Chen, Q., Wang, X., Qi, Q., Wang, L., & Liu, H. (2021a). Attentive capsule network for click-through rate and conversion rate prediction in online advertising. Knowledge-based Systems, 211, 106522.

[49]. Li, H., Pan, F., Ao, X., Yang, Z., Lu, M., Pan, J., Liu, D., Xiao, L., & He, Q. (2021b). Follow the prophet: Accurate online conversion rate prediction in the face of delayed feedback. In Proceedings of the 44th International ACM SIGIR Conference on Research and Development in Information Retrieval (SIGIR'21) (pp. 1915-1919). Association for Computing Machinery, New York, NY, USA.

[50]. Li, K., Chen, X., Leng, L., Xu, J., Sun, J., & Rezaei, B. (2024). Privacy preserving conversion modeling in data clean room. In Proceedings of the 18th ACM Conference on Recommender Systems (RecSys'24) (pp. 819-822). Association for Computing Machinery, New York, NY, USA.

[51]. Li, N., Arava, S. K., Dong, C., Yan, W., & Pani, A. (2018). Deep neural net with attention for multi-channel multi-touch attribution. In Proceedings of the 24th ACM SIGKDD Conference on Knowledge Discovery and Data Mining (KDD'18) (pp. 1-6). Association for Computing Machinery, New York, NY, USA.

[52]. Li, P., Tong, X., Wang, Y., & Zhang, Q. (2025a). Meta doubly robust: Debiasing CVR prediction via meta-learning with a small amount of unbiased data. Knowledge-Based Systems, 310, 112898.

[53]. Li, W., Wang, D., Ding, Z., Sohrabizadeh, A., Qin, Z., Cong, J., & Sun, Y. (2025b). Hierarchical mixture of experts: Generalizable learning for high-level synthesis. In Proceedings of the 39th AAAI Conference on Artificial Intelligence (AAAI'25) (pp. 18476-18484). AAAI Press, Palo Alto, California, USA.

[54]. Li, X., Wang, C., Tan, J., Zeng, X., Ou, D., Ou, D., & Zheng, B. (2020). Adversarial multimodal representation learning for click-through rate prediction. In Proceedings of the Web Conference 2020 (WWW'20) (pp. 827-836). Association for Computing Machinery, New York, NY, USA.

[55]. Lin, J., Chen, B., Wang, H., Xi, Y., Qu, Y., Dai, X., Zhang, K., Tang, R., Yu, Y., & Zhang, W. (2024). ClickPrompt: CTR models are strong prompt generators for adapting language models to CTR prediction. In Proceedings of the ACM Web Conference 2024 (WWW'24) (pp. 3319-3330). Association for Computing Machinery, New York, NY, USA.

[56]. Lin, Z., Yang, X., Peng, X., Zhao, W. X., Liu, S., Wang, L., & Zheng, B. (2022). Modeling adaptive fine-grained task relatedness for joint CTR-CVR estimation. arXiv preprint arXiv:2208.13442.

[57]. Ling, X., Deng, W., Gu, C., Zhou, H., Li, C., & Sun, F. (2017). Model ensemble for click prediction in bing search ads. In Proceedings of the 26th International Conference on World Wide Web Companion (WWW'17 Companion) (pp. 689-698). Association for Computing Machinery, New York, NY, USA.

[58]. Liu, E., & Liu, C. (2021). XGBoost-based method for internet advertising conversion rate computer intelligent prediction model. In 2021 3rd International Conference on Artificial Intelligence and Advanced Manufacture (AIAM'21) (pp. 2432-2436). Association for Computing Machinery, New York, NY, USA.

[59]. Liu, Q., Ao, X., Guo, Y., & He, Q. (2024a). Online conversion rate prediction via multi-interval screening and synthesizing under delayed feedback. In Proceedings of the 38th AAAI Conference on Artificial Intelligence (AAAI'24) (pp. 8796-8804). AAAI Press, Palo Alto, California, USA.

[60]. Liu, Q., Li, H., Ao, X., Guo, Y., Dong, Z., Zhang, R., Chen, Q., Tong, J., & He, Q. (2023a). Online conversion rate prediction via neural satellite networks in delayed feedback advertising. In Proceedings of the 46th International ACM SIGIR Conference on Research and Development in Information Retrieval (SIGIR'23) (pp. 1406-1415). Association for Computing Machinery, New York, NY, USA.

[61]. Liu, X., Jia, Q., Wu, C., Li, J., Quanyu, D., Bo, L., Zhang, R., & Tang, R. (2023b). Task adaptive multi-learner network for joint CTR and CVR estimation. In Companion Proceedings of the ACM Web Conference 2023 (WWW'23) (pp. 490-494). Association for Computing Machinery, New York, NY, USA.

[62]. Liu, Y., Jia, Q., Shi, S., Wu, C., Du, Z., Xie, Z., Tang, R., Zhang, M., & Li, M. (2024b). Ranking-aware unbiased post-click conversion rate estimation via AUC optimization on entire exposure space. In Proceedings of the 18th ACM Conference on Recommender Systems (RecSys'24) (pp. 360-369). Association for Computing Machinery, New York, NY, USA.

[63]. Lu, Q., Pan, S., Wang, L., Pan, J., Wan, F., & Yang, H. (2017). A practical framework of conversion rate prediction for online display advertising. In Proceedings of the 23rd ACM SIGKDD International Conference on Knowledge Discovery and Data Mining (KDD'17) (pp. 1-9). Association for Computing Machinery, New York, NY, USA.

[64]. Mansourian, A. M., Ahmadi, R., Ghafouri, M., Babaei, A. M., Golezani, E. B., Ghamchi, Z. Y., Ramezanian, V., Taherian, A., Dinashi, K., Miri, A., & Kasaei, S. (2025). A Comprehensive survey on knowledge distillation. arXiv preprint arXiv:2503.12067.

[65]. Ma, X., Zhao, L., Huang, G., Wang, Z., Hu, Z., Zhu, X., & Gai, K. (2018). Entire space multi-task model: An effective approach for estimating post-click conversion rate. In the 41st International ACM SIGIR Conference on Research and Development in Information Retrieval (SIGIR'18) (pp. 1137-1140). Association for Computing Machinery, New York, NY, USA.

[66]. Mao, K., Zhu, J., Su, L., Cai, G., Li, Y., & Dong, Z. (2023). FinalMLP: An enhanced two-stream MLP model for CTR prediction. In Proceedings of the 37th AAAI Conference on Artificial Intelligence (AAAI'23) (pp. 4552-4560). AAAI Press, Palo Alto, California, USA, 37(4).

[67]. Min, X., Zhang, X., Shen, B., Wang, S., He, Y., Li, C., & Zhou, J. (2023). SeqGen: A sequence generator via user side information for behavior sparsity in recommendation. In Proceedings of the 32nd ACM International Conference on Information and Knowledge Management (CIKM'23) (pp. 4205-4209). Association for Computing Machinery, New York, NY, USA.

[68]. Oldfield, J., Georgopoulos, M., Chrysos, G., Tzelepis, C., Panagakis, Y., Nicolaou, M., Deng, J., & Patras, I. (2024). Multilinear mixture of experts: Scalable expert specialization through factorization. Advances in Neural Information Processing Systems (NeurIPS'24) (pp. 53022-53063). Curran Associates, New York, NY, USA.

[69]. Ouyang, W., Dong, R., Zhang, X., Guo, C., Luo, J., Liu, X., & Du, Y. (2023a). Contrastive learning for conversion rate prediction. In Proceedings of the 46th International ACM SIGIR Conference on Research and Development in Information Retrieval (SIGIR'23) (pp. 1909-1913). Association for Computing Machinery, New York, NY, USA.

[70]. Ouyang, W., Zhang, X., Guo, C., Ren, S., Sui, Y., Zhang, K., Luo, J., Chen, Y., Xu, D., Liu, X., & Du, Y. (2023b). Masked multi-domain network: Multi-type and multi-scenario conversion rate prediction with a single model. In Proceedings of the 32nd ACM International Conference on Information and Knowledge Management (CIKM'23) (pp. 4773-4779). Association for Computing Machinery, New York, NY, USA.

[71]. Pan, J., Mao, Y., Ruiz, A. L., Sun, Y., & Flores, A. (2019). Predicting different types of conversions with multi-task learning in online advertising. In Proceedings of the 25th ACM

SIGKDD International Conference on Knowledge Discovery and Data Mining (KDD’19) (pp. 2689-2697). Association for Computing Machinery, New York, NY, USA.

[72]. Pan, X., Li, M., Zhang, J., Yu, K., Wen, H., Wang, L., Mao, C., & Cao, B. (2022). MetaCVR: Conversion rate prediction via meta learning in small-scale recommendation scenarios. In Proceedings of the 45th International ACM SIGIR Conference on Research and Development in Information Retrieval (SIGIR’22) (pp. 2110-2114). Association for Computing Machinery, New York, NY, USA.

[73]. Qiu, Y., Tziortziotis, N., Hue, M., & Vazirgiannis, M. (2020). Predicting conversions in display advertising based on URL embeddings. In Proceedings of the 14th International Workshop on Data Mining for Online Advertising (ADKDD’20) (pp. 1-6). Association for Computing Machinery, New York, NY, USA.

[74]. Ren, K., Fang, Y., Zhang, W., Liu, S., Li, J., Zhang, Y., Yu, Y., & Wang, J. (2018). Learning multi-touch conversion attribution with dual-attention mechanisms for online advertising. In Proceedings of the 27th ACM International Conference on Information and Knowledge Management (CIKM’18) (pp. 1433-1442). Association for Computing Machinery, New York, NY, USA.

[75]. Rosales, R., Cheng, H., & Manavoglu, E. (2012). Post-click conversion modeling and analysis for non-guaranteed delivery display advertising. In Proceedings of the 5th ACM International Conference on Web Search and Data Mining (WSDM’12) (pp. 293-302). Association for Computing Machinery, New York, NY, USA.

[76]. Saito, Y. (2020). Doubly robust estimator for ranking metrics with post-click conversions. In Proceedings of the 14th ACM Conference on Recommender Systems (RecSys’20) (pp. 92-100). Association for Computing Machinery, New York, NY, USA.

[77]. Saito, Y., Morisihta, G., & Yasui, S. (2020). Dual learning algorithm for delayed conversions. In Proceedings of the 43rd International ACM SIGIR Conference on Research and Development in Information Retrieval (SIGIR’20) (pp. 1849-1852). Association for Computing Machinery, New York, NY, USA.

[78]. Shan, L., Lin, L., & Sun, C. (2018). Combined regression and tripletwise learning for conversion rate prediction in real-time bidding advertising. In Proceedings of the 41st International ACM SIGIR Conference on Research & Development in Information Retrieval (SIGIR’18) (pp. 115-123). Association for Computing Machinery, New York, NY, USA.

[79]. Sheng, X. R., Yang, F., Gong, L., Wang, B., Chan, Z., Zhang, Y., Cheng, Y., Zhu, Y-N., Ge, T., Zhu, H., Jiang, Y., Xu, J., & Zheng, B. (2024). Enhancing Taobao display advertising with multimodal representations: Challenges, approaches and insights. In Proceedings of the 33rd ACM International Conference on Information and Knowledge Management (CIKM’24) (pp. 4858-4865). Association for Computing Machinery, New York, NY, USA.

[80]. Shtoff, A., Kaplan, Y., & Raviv, A. (2023). Improving conversion rate prediction via self-supervised pre-training in online advertising. In 2023 IEEE International Conference on Big Data (BigData’23) (pp. 1835-1842). IEEE Computer Society, Los Alamitos, CA, USA.

[81]. Statista, 2025, Online advertising revenue in the U.S. 2000-2024, Access on May 13, 2025, https://www.statista.com/statistics/183816/us-online-advertising-revenue-since-2000/.

[82]. Su, H., Meng, L., Zhu, L., Lu, K., & Li, J. (2024). DDPO: Direct dual propensity optimization for post-click conversion rate estimation. In Proceedings of the 47th International ACM SIGIR Conference on Research and Development in Information Retrieval (SIGIR'24) (pp. 1179-1188). Association for Computing Machinery, New York, NY, USA.

[83]. Su, Y., Zhang, L., Dai, Q., Zhang, B., Yan, J., Wang, D., Bao, Y., Xu, S., He, Y., & Yan, W. (2021). An attention-based model for conversion rate prediction with delayed feedback via post-click calibration. In Proceedings of the 29th International Conference on International Joint Conferences on Artificial Intelligence (IJCAI'21) (pp. 3522-3528). Morgan Kaufmann, San Francisco, CA, USA.

[84]. Tallis, M., & Yadav, P. (2018). Reacting to variations in product demand: An application for conversion rate (CR) prediction in sponsored search. In 2018 IEEE International Conference on Big Data (Big Data'18) (pp. 1856-1864). IEEE, Piscataway, NJ, USA.

[85]. Tan, X., Deng, Y., Qu, C., Xue, S., Shi, X., Zhang, J., & Qiu, X. (2023). Adaptive learning on user segmentation: Universal to specific representation via bipartite neural interaction. In Proceedings of the Annual International ACM SIGIR Conference on Research and Development in Information Retrieval in the Asia Pacific Region (SIGIR-AP'23) (pp. 205-211). Association for Computing Machinery, New York, NY, USA.

[86]. Tang, H., Liu, J., Zhao, M., & Gong, X. (2020). Progressive layered extraction (PLE): A novel multi-task learning (MTL) model for personalized recommendations. In Proceedings of the 14th ACM Conference on Recommender Systems (RecSys'20) (pp. 269-278). Association for Computing Machinery, New York, NY, USA.

[87]. Tang, J., Wen, J., & Jing, L. (2024). DCRMTA: Unbiased causal representation for multi-touch attribution. arXiv preprint arXiv:2401.08875.

[88]. Terven, J., Cordova-Esparza, D. M., Romero-González, J. A., Ramírez-Pedraza, A., & Chávez-Urbiola, E. A. (2025). A comprehensive survey of loss functions and metrics in deep learning. Artificial Intelligence Review, 58(7), 195.

[89]. Vasile, F., Lefortier, D., & Chapelle, O. (2017). Cost-sensitive learning for utility optimization in online advertising auctions. In Proceedings of the 23rd ACM SIGKDD International Conference on Knowledge Discovery and Data Mining (KDD'17) (pp. 1-6). Association for Computing Machinery, New York, NY, USA.

[90]. Wan, S., Yang, S., & Fu, Z. (2025). Focus on user micro multi-behavioral states: Time-sensitive user behavior conversion prediction and multi-view reinforcement learning based recommendation approach. Information Processing & Management, 62(2), 103967.

[91]. Wang, H., Chang, T. W., Liu, T., Huang, J., Chen, Z., Yu, C., Li, R., & Chu, W. (2022). $ESCM^2$: Entire space counterfactual multi-task model for post-click conversion rate estimation. In Proceedings of the 45th International ACM SIGIR Conference on Research and Development in Information Retrieval (SIGIR'22) (pp. 363-372). Association for Computing Machinery, New York, NY, USA.

[92]. Wang, H., Du, Y., Jin, C., Li, Y., Wang, Y., Sun, T., Qin, P., & Fan, C. (2023a). GACE: Learning graph-based cross-page ads embedding for click-through rate prediction. In International Conference on Neural Information Processing (ICONIP'23) (pp. 429-443). Springer, Singapore.

[93]. Wang, R., Fu, B., Fu, G., & Wang, M. (2017). Deep & cross network for ad click predictions. In Proceedings of the ADKDD'17 (pp. 1-7). Association for Computing Machinery, New York, NY, USA.

[94]. Wang, Y., Lam, H. T., Wong, Y., Liu, Z., Zhao, X., Wang, Y., Chen, B., Guo, H., & Tang, R. (2023b). Multi-task deep recommender systems: A survey. arXiv preprint arXiv:2302.03525.

[95]. Wang, Y., Sun, P., Zhang, M., Jia, Q., Li, J., & Ma, S. (2023c). Unbiased delayed feedback label correction for conversion rate prediction. In Proceedings of the 29th ACM SIGKDD Conference on Knowledge Discovery and Data Mining (KDD'23) (pp. 2456-2466). Association for Computing Machinery, New York, NY, USA.

[96]. Wang, Y., Sha, Z., Lin, K., Feng, C., Zhu, K., Wang, L., Jiao, X., Huang, F., Ye, C., He, D., Guo, Z., Li, S., & Liu, L. (2024). One-step reach: LLM-based keyword generation for sponsored search advertising. In Companion Proceedings of the ACM Web Conference 2024 (WWW'24) (pp. 1604-1608). Association for Computing Machinery, New York, NY, USA.

[97]. Wang, Y., Zhang, J., Da, Q., & Zeng, A. (2020). Delayed feedback modeling for the entire space conversion rate prediction. arXiv preprint arXiv:2011.11826.

[98]. Wang, Z., She, Q., & Zhang, J. (2021). MaskNet: Introducing feature-wise multiplication to CTR ranking models by instance-guided mask. arXiv preprint arXiv:2102.07619.

[99]. Wei, P., Zhang, W., Xu, Z., Liu, S., Lee, K. C., & Zheng, B. (2021). AutoHERI: Automated hierarchical representation integration for post-click conversion rate estimation. In Proceedings of the 30th ACM International Conference on Information and Knowledge Management (CIKM'21) (pp. 3528-3532). Association for Computing Machinery, New York, NY, USA.

[100]. Wen, H., Zhang, J., Wang, Y., Lv, F., Bao, W., Lin, Q., & Yang, K. (2020). Entire space multi-task modeling via post-click behavior decomposition for conversion rate prediction. In Proceedings of the 43rd International ACM SIGIR Conference on Research and Development in Information Retrieval (SIGIR'20) (pp. 2377-2386). Association for Computing Machinery, New York, NY, USA.

[101]. Xi, D., Chen, Z., Yan, P., Zhang, Y., Zhu, Y., Zhuang, F., & Chen, Y. (2021). Modeling the sequential dependence among audience multi-step conversions with multi-task learning in targeted display advertising. In Proceedings of the 27th ACM SIGKDD Conference on Knowledge Discovery and Data Mining (KDD'21) (pp. 3745-3755). Association for Computing Machinery, New York, NY, USA.

[102]. Xie, B., Sun, Y., Lin, Z., & Ma, B. (2022). Att-TCN: An advertising attribution method based on temporal convolutional network with attention mechanism. In 2022 IEEE 8th

International Conference on Computer and Communications (ICCC'22) (pp. 1460-1465). IEEE, Piscataway, NJ, USA.

[103]. Xu, Z., Wei, P., Zhang, W., Liu, S., Wang, L., & Zheng, B. (2022). UKD: Debiasing conversion rate estimation via uncertainty-regularized knowledge distillation. In Proceedings of the ACM Web Conference 2022 (WWW'22) (pp. 2078-2087). Association for Computing Machinery, New York, NY, USA.

[104]. Yang, D., Dyer, K., & Wang, S. (2020). Interpretable deep learning model for online multi-touch attribution. arXiv preprint arXiv:2004.00384.

[105]. Yang, H., Lu, Q., Qiu, A. X., & Han, C. (2016). Large scale CVR prediction through dynamic transfer learning of global and local features. In Proceedings of the 5th International Workshop on Big Data, Streams and Heterogeneous Source Mining: Algorithms, Systems, Programming Models and Applications at KDD 2016 (pp. 103-119). PMLR, New York, NY, USA.

[106]. Yang, J. Q., Li, X., Han, S., Zhuang, T., Zhan, D. C., Zeng, X., & Tong, B. (2021). Capturing delayed feedback in conversion rate prediction via elapsed-time sampling. In Proceedings of the 35th AAAI Conference on Artificial Intelligence (AAAI'21) (pp. 4582-4589). AAAI Press, Palo Alto, California, USA.

[107]. Yang, J., & Zhan, D. C. (2022). Generalized delayed feedback model with post-click information in recommender systems. Advances in Neural Information Processing Systems (NeurIPS'22) (pp. 26192-26203). Curran Associates, New York, NY, USA.

[108]. Yang, J., Zhang, L., He, Y., Ding, K., Huan, Z., Zhang, X., & Mo, L. (2023). DCBT: A simple but effective way for unified warm and cold recommendation. In Proceedings of the 46th International ACM SIGIR Conference on Research and Development in Information Retrieval (SIGIR'23) (pp. 3369-3373). Association for Computing Machinery, New York, NY, USA.

[109]. Yang, S., Yang, H., Zou, Z., Xu, L., Yuan, S., & Zeng, Y. (2024). Deep ensemble shape calibration: Multi-field post-hoc calibration in online advertising. In Proceedings of the 30th ACM SIGKDD Conference on Knowledge Discovery and Data Mining (KDD'24) (pp. 6117-6126). Association for Computing Machinery, New York, NY, USA.

[110]. Yang, Y., & Zhai, P. (2022). Click-through rate prediction in online advertising: A literature review. Information Processing & Management, 59(2), 102853.

[111]. Yang, Y., Zhao, K., Zeng, D. D., & Jansen, B. J. (2022). Time-varying effects of search engine advertising on sales-An empirical investigation in E-commerce. Decision Support Systems, 163, 113843.

[112]. Yao, D., Gong, C., Zhang, L., Chen, S., & Bi, J. (2022). CausalMTA: Eliminating the user confounding bias for causal multi-touch attribution. In Proceedings of the 28th ACM SIGKDD Conference on Knowledge Discovery and Data Mining (KDD'22) (pp. 4342-4352). Association for Computing Machinery, New York, NY, USA.

[113]. Yao, Z., Kong, D., Lu, M., Bai, X., Yang, J., & Xiong, H. (2023). Multi-view multi-task campaign embedding for cold-start conversion rate forecasting. IEEE Transactions on Big Data, 9(1), 280-293.

[114]. Yasui, S., & Kato, M. (2022). Learning classifiers under delayed feedback with a time window assumption. In Proceedings of the 28th ACM SIGKDD Conference on Knowledge Discovery and Data Mining (CIKM'22) (pp. 2286-2295). Association for Computing Machinery, New York, NY, USA.

[115]. Yasui, S., Morishita, G., Komei, F., & Shibata, M. (2020). A feedback shift correction in predicting conversion rates under delayed feedback. In Proceedings of the Web Conference 2020 (WWW'20) (pp. 2740-2746). Association for Computing Machinery, New York, NY, USA.

[116]. Yi, Q., Tang, J., Zhao, X., Zeng, Y., Song, Z., & Wu, J. (2025). An adaptive entire-space multi-scenario multi-task transfer learning model for recommendations. IEEE Transactions on Knowledge and Data Engineering, 37(4), 1585-1598.

[117]. Yoshikawa, Y., & Imai, Y. (2018). A nonparametric delayed feedback model for conversion rate prediction. arXiv preprint arXiv:1802.00255.

[118]. Yu, B., Zhang, J., Fu, Y., & Xu, Z. (2025). A weighted heterogeneous graph attention network method for purchase prediction of potential consumers with multibehaviors. Information Processing & Management, 62(5), 104175.

[119]. Yu, L., Cai, Y., Chen, L., Zhang, M., Day, W. Y., Li, L., Chen, R., Choi, S., & Hu, X. (2024). Addressing delayed feedback in conversion rate prediction: A domain adaptation approach. In 2024 IEEE International Conference on Data Mining (ICDM'24) (pp. 917-922). IEEE Computer Society, Los Alamitos, CA, USA.

[120]. Yuan, H., Zhang, W., Hao, Z., & Deng, Z. (2025). Hardness-aware privileged features distillation with latent alignment for CVR prediction. In Proceedings of the 31st ACM SIGKDD Conference on Knowledge Discovery and Data Mining (KDD'25) (pp. 5182-5193). Association for Computing Machinery, New York, NY, USA.

[121]. Zhai, P., Yang, Y., & Zhang, C. (2023). Causality-based CTR prediction using graph neural networks. Information Processing & Management, 60(1), 103137.

[122]. Zhai, P., Yang, Y., & Zhang, C. (2025). Periodic graph neural networks for click-through rate prediction in online advertising. ACM Transactions on Information Systems. Forthcoming. https://doi.org/10.1145/3769103.

[123]. Zhang, D., Wu, H., Zeng, G., Yang, Y., Qiu, W., Chen, Y., & Hu, H. (2022). CTnoCVR: A novelty auxiliary task making the lower-CTR-higher-CVR upper. In Proceedings of the 45th International ACM SIGIR Conference on Research and Development in Information Retrieval (SIGIR'22) (pp. 2272-2276). Association for Computing Machinery, New York, NY, USA.

[124]. Zhang, Q., Zhang, Z., Lu, B., He, B., Li, L., & Dong, Z. (2023). Large scale CVR prediction through hierarchical history modeling. In Proceedings of the Recommender Systems Challenge 2023 (RecSysChallenge'23) (pp. 18-22). Association for Computing Machinery, New York, NY, USA.

[125]. Zhang, W., Bao, W., Liu, X. Y., Yang, K., Lin, Q., Wen, H., & Ramezani, R. (2020). Large-scale causal approaches to debiasing post-click conversion rate estimation with multi-task learning. In Proceedings of the Web Conference 2020 (WWW'20) (pp. 2775-2781). Association for Computing Machinery, New York, NY, USA.

[126]. Zhang, X., Guo, Y., & Ao, X. (2024a). Predicting calibrated conversion rate of online advertising using a multi-task mixture-of-experts calibration model. In Proceedings of the 12nd CCF Conference on Big Data (BigData'24) (pp. 189-201). Springer Nature Singapore, Singapore.

[127]. Zhang, X., Huang, C., Zheng, K., Su, H., Ji, T., Wang, W., Qi, H., & Li, J. (2024b). Adversarial-enhanced causal multi-task framework for debiasing post-click conversion rate estimation. In Proceedings of the ACM Web Conference 2024 (WWW'24) (pp. 3287-3296). Association for Computing Machinery, New York, NY, USA.

[128]. Zhang, X., Li, W., Li, R., Fu, Z., Tang, T., Zhang, Z., Chen, W. Y., Noorshams, N., Jasapara, N., Ding, X., Wen, E., & Feng, X. (2025a). Personalized interpolation: An efficient method to tame flexible optimization window estimation. arXiv preprint arXiv:2501.14103.

[129]. Zhang, Y., Wei, Y., & Ren, J. (2014). Multi-touch attribution in online advertising with survival theory. In 2014 IEEE International Conference on Data Mining (ICDM'14) (pp. 687-696). IEEE, Piscataway, NJ, USA.

[130]. Zhang, Z., Yang, B., & Lu, Y. (2025b). A local context enhanced consistency-aware mamba-based sequential recommendation model. Information Processing & Management, 62(3), 104076.

[131]. Zhao, Y., Wu, C., Jia, Q., Zhu, H., Yan, J., Zong, L., Zhang, L., Dong, Z., & Zhang, M. (2024). Confidence-aware multi-field model calibration. In Proceedings of the 33rd ACM International Conference on Information and Knowledge Management (CIKM'24) (pp. 5111-5118). Association for Computing Machinery, New York, NY, USA.

[132]. Zheng, J., Chen, S., Du, Y., & Song, P. (2022). A multiview graph collaborative filtering by incorporating homogeneous and heterogeneous signals. Information Processing & Management, 59(6), 103072.

[133]. Zhou, Y., Mishra, S., Gligorijevic, J., Bhatia, T., & Bhamidipati, N. (2019). Understanding consumer journey using attention based recurrent neural networks. In Proceedings of the 25th ACM SIGKDD International Conference on Knowledge Discovery and Data Mining (KDD'19) (pp. 3102-3111). Association for Computing Machinery, New York, NY, USA.

[134]. Zhu, F., Zhong, M., Yang, X., Li, L., Yu, L., Zhang, T., Zhou, J., Chen, C., Wu, F., Liu, G., & Wang, Y. (2023). DCMT: A direct entire-space causal multi-task framework for post-click conversion estimation. In 2023 IEEE 39th International Conference on Data Engineering (ICDE'23) (pp. 3113-3125). IEEE, Piscataway, NJ, USA.

[135]. Zhuang, J., Li, Y., Su, R., Xu, K., Shao, Z., Li, K., Long, L., Sun, H., Qi, M., Meng, Y., Tang, Y., Liu, Z., Shen, Q., & Mudgal, A. (2025). On the practice of deep hierarchical ensemble network for ad conversion rate prediction. In Companion Proceedings of the ACM on Web

Conference 2025 (WWW’25) (pp. 671-680). Association for Computing Machinery, New York, NY, USA.

# A. Research articles included in this review

Table A1. The list of reviewed articles.

| No. | Title | Year | Authors | Publication Outlet |
|---|---|---|---|---|
| 1 | Estimating conversion rate in display advertising from past performance data | 2012 | Lee et al. | *The 18th ACM SIGKDD International Conference on Knowledge Discovery and Data Mining (KDD'12)* |
| 2 | Post-click conversion modeling and analysis for non-guaranteed delivery display advertising | 2012 | Rosales et al. | *The 5th ACM International Conference on Web Search and Data Mining (WSDM'12)* |
| 3 | Modeling delayed feedback in display advertising | 2014 | Chapelle. | *The 20th ACM SIGKDD International Conference on Knowledge Discovery and Data Mining (KDD'14)* |
| 4 | Multi-touch attribution in online advertising with survival theory | 2014 | Zhang et al. | *The 14th IEEE International Conference on Data Mining (ICDM'14)* |
| 5 | A probabilistic multi-touch attribution model for online advertising | 2016 | Ji et al. | *The 25th ACM International on Conference on Information and Knowledge Management (CIKM'16)* |
| 6 | Large scale CVR prediction through dynamic transfer learning of global and local features | 2016 | Yang et al. | *The 5th International Workshop on Big Data, Streams and Heterogeneous Source Mining: Algorithms, Systems, Programming Models and Applications* |
| 7 | A large scale prediction engine for app install clicks and conversions | 2017 | Bhamidipati et al. | *The 26th ACM on Conference on Information and Knowledge Management (CIKM'17)* |
| 8 | Additional multi-touch attribution for online advertising | 2017 | Ji & Wang. | *The 31st AAAI Conference on Artificial Intelligence (AAAI'17)* |
| 9 | Conversion rate estimation in online advertising via exploring potential impact of creative | 2017 | Jiang & Jiang. | *The 28th International Conference on Database and Expert Systems Applications (DEXA'17)* |
| 10 | A practical framework of conversion rate prediction for online display advertising | 2017 | Lu et al. | *The 23rd ACM SIGKDD International Conference on Knowledge Discovery and Data Mining (KDD'17)* |
| 11 | Cost-sensitive learning for utility optimization in online advertising auctions | 2017 | Vasile et al. | *The 23rd ACM SIGKDD International Conference on Knowledge Discovery and Data Mining (KDD'17)* |
| 12 | Deep neural net with attention for multi-channel multi-touch attribution | 2018 | Li et al. | *The 24th ACM SIGKDD Conference on Knowledge Discovery and Data Mining (KDD'18)* |
| 13 | Modelling customer online behaviours with neural networks: applications to conversion prediction and advertising retargeting | 2018 | Cui et al. | *Pre-print (arXiv)* |
| 14 | Entire space multi-task model: An effective approach for estimating post-click conversion rate | 2018 | Ma et al. | *The 41st International ACM SIGIR Conference on Research and Development in Information Retrieval (SIGIR'18)* |
| 15 | Learning multi-touch conversion attribution with dual-attention mechanisms for online advertising | 2018 | Ren et al. | *The 27th ACM International Conference on Information and Knowledge Management (CIKM'18)* |
| 16 | Combined regression and tripletwise learning for conversion rate prediction in real-time bidding advertising | 2018 | Shan et al. | *The 41st International ACM SIGIR Conference on Research and Development in Information Retrieval (SIGIR'18)* |
| 17 | Reacting to variations in product demand: An application for conversion rate (CR) prediction in sponsored search | 2018 | Tallis & Yadav. | *The 6th IEEE International Conference on Big Data (BigData'18)* |
| 18 | A nonparametric delayed feedback model for conversion rate prediction | 2018 | Yoshikawa & Imai. | *Pre-print (arXiv)* |
| 19 | Boosted trees model with feature engineering: An approach for post-click conversion rate prediction | 2019 | Du et al. | *The 2nd International Conference on Big Data Technologies (ICBDT'19)* |
| 20 | Causally driven incremental multi touch attribution using a recurrent neural network | 2019 | Du et al. | *Pre-print (arXiv)* |
| 21 | Conversion prediction using multi-task conditional attention networks to support the creation of effective ad creatives | 2019 | Kitada et al. | *The 25th ACM SIGKDD International Conference on Knowledge Discovery and Data Mining (KDD'19)* |
| 22 | Predicting different types of conversions with multi-task learning in online advertising | 2019 | Pan et al. | *The 25th ACM SIGKDD International Conference on Knowledge Discovery and Data Mining (KDD'19)* |
| 23 | Understanding consumer journey using attention based recurrent neural networks | 2019 | Zhou et al. | *The 25th ACM SIGKDD International Conference on Knowledge Discovery and Data Mining (KDD'19)* |
| 24 | Delayed feedback modeling for the entire space conversion rate prediction | 2020 | Wang et al. | *Pre-print (arXiv)* |
| 25 | Prospective modeling of users for online display advertising via deep time-aware model | 2020 | Gligorijevic et al. | *The 29th ACM International Conference on Information and Knowledge Management (CIKM'20)* |
| 26 | CAMTA: Causal attention model for multi-touch attribution | 2020 | Kumar et al. | *The 20th International Conference on Data Mining Workshops (ICDMW'20)* |
| 27 | Predicting conversions in display advertising based on URL embeddings | 2020 | Qiu et al. | *Pre-print (arXiv)* |
| 28 | Dual learning algorithm for delayed conversions | 2020 | Satio et al. | *The 43rd International ACM SIGIR Conference on Research and Development in Information Retrieval (SIGIR'20)* |
| 29 | Progressive layered extraction (PLE): A novel multi-task learning (MTL) model for personalized recommendations | 2020 | Tang et al. | *The 14th ACM Conference on Recommender Systems (RecSys'20)* |
| 30 | Entire space multi-task modeling via post-click behavior decomposition for conversion rate prediction | 2020 | Wen et al. | *The 43rd International ACM SIGIR Conference on Research and Development in Information Retrieval (SIGIR'20)* |
| 31 | Interpretable deep learning model for online multi-touch attribution | 2020 | Yang et al. | *Pre-print (arXiv)* |
| 32 | A feedback shift correction in predicting conversion rates under delayed feedback | 2020 | Yasui et al. | *The 29th ACM Web Conference (WWW'20)* |
| 33 | Large-scale causal approaches to debiasing post-click conversion rate estimation with multi-task learning | 2020 | Zhang et al. | *The 29th ACM Web Conference (WWW'20)* |
| 34 | Real negatives matter: Continuous training with real negatives for delayed feedback modeling | 2021 | Gu et al. | *The 27th ACM SIGKDD Conference on Knowledge Discovery and Data Mining (KDD'21)* |
| 35 | Conversion prediction with delayed feedback: A multi-task learning approach | 2021 | Hou et al. | *The 21st International Conference on Data Mining (ICDM'21)* |
| 36 | XGBoost-based method for internet advertising conversion rate computer intelligent prediction model | 2021 | Liu & Liu. | *The 3rd International Conference on Artificial Intelligence and Advanced Manufacture (AIAM'21)* |
| 37 | Attentive capsule network for click-through rate and conversion rate prediction in online advertising | 2021 | Li et al. | *Knowledge-based Systems* |
| 38 | Follow the prophet: Accurate online conversion rate prediction in the face of delayed feedback | 2021 | Li et al. | *The 44th International ACM SIGIR Conference on Research and Development in Information Retrieval (SIGIR'21)* |
| 39 | Modeling the sequential dependence among audience multi-step conversions with multi-task learning in targeted display advertising | 2021 | Xi et al. | *The 27th ACM SIGKDD Conference on Knowledge Discovery and Data Mining (KDD'21)* |
| 40 | An attention-based model for conversion rate prediction with delayed feedback via post-click calibration | 2021 | Su et al. | *The 29th International Conference on International Joint Conferences on Artificial Intelligence (IJCAI'21)* |
| 41 | AutoHERI: Automated hierarchical representation integration for post-click conversion rate estimation | 2021 | Wei et al. | *The 30th ACM International Conference on Information and Knowledge Management (CIKM'21)* |
| 42 | Capturing delayed feedback in conversion rate prediction via elapsed-time sampling | 2021 | Yang et al. | *The 35th AAAI Conference on Artificial Intelligence (AAAI'21)* |
| 43 | Multi-touch attribution for complex B2B customer journeys using temporal convolutional networks | 2022 | Agrawal et al. | *The 31st ACM Web Conference (WWW'22)* |
| 44 | CFS-MTL: A causal feature selection mechanism for multi-task learning via pseudo-intervention | 2022 | Chen et al. | *The 31st ACM International Conference on Information and Knowledge Management (CIKM'22)* |
| 45 | Asymptotically unbiased estimation for delayed feedback modeling via label correction | 2022 | Chen et al. | *The 31st ACM Web Conference (WWW'22)* |
| 46 | A generalized doubly robust learning framework for debiasing post-click conversion rate prediction | 2022 | Dai et al. | *The 28th ACM SIGKDD Conference on Knowledge Discovery and Data Mining (KDD'22)* |
| 47 | GFlow-FT: Pick a child network via gradient flow for efficient fine-tuning in recommendation systems | 2022 | Ding et al. | *The 31st ACM International Conference on Information and Knowledge Management (CKIM'22)* |
| 48 | Multi-head online learning for delayed feedback modeling | 2022 | Gao & Yang. | *Pre-print (arXiv)* |
| 49 | Calibrated conversion rate prediction via knowledge distillation under delayed feedback in online advertising | 2022 | Guo et al. | *The 31st ACM International Conference on Information & Knowledge Management (CIKM'22)* |
| 50 | Modeling adaptive fine-grained task relatedness for joint CTR-CVR estimation | 2022 | Lin et al. | *Pre-print (arXiv)* |
| 51 | ESCM$^2$: Entire space counterfactual multi-task model for post-click conversion rate estimation | 2022 | Wang et al. | *The 45th International ACM SIGIR Conference on Research and Development in Information Retrieval (SIGIR'22)* |
| 52 | Att-TCN: An advertising attribution method based on temporal convolutional network with attention mechanism | 2022 | Xie et al. | *The 8th International Conference on Computer and Communications (ICCC'22)* |
| 53 | UKD: Debiasing conversion rate estimation via uncertainty-regularized knowledge distillation | 2022 | Xu et al. | *The 31st ACM Web Conference (WWW'22)* |
| 54 | Generalized delayed feedback model with post-click information in recommender systems | 2022 | Yang & Zhan. | *The 36th Annual Conference on Neural Information Processing Systems (NeurIPS'22)* |
| 55 | CausalMTA: Eliminating the user confounding bias for causal multi-touch attribution | 2022 | Yao et al. | *The 28th ACM SIGKDD Conference on Knowledge Discovery and Data Mining (KDD'22)* |
| 56 | Learning classifiers under delayed feedback with a time window assumption | 2022 | Yasui & Kato. | *The 28th ACM SIGKDD Conference on Knowledge Discovery and Data Mining (CIKM'22)* |
| 57 | CTnoCVR: A novelty auxiliary task making the lower-CTR-higher-CVR upper | 2022 | Zhang et al. | *The 45th International ACM SIGIR Conference on Research and Development in Information Retrieval (SIGIR'22)* |
| 58 | Capturing conversion rate fluctuation during sales promotions: A novel historical data reuse approach | 2023 | Chan et al. | *The 29th ACM SIGKDD Conference on Knowledge Discovery and Data Mining (KDD'23)* |
| 59 | Dually enhanced delayed feedback modeling for streaming conversion rate prediction | 2023 | Dai et al. | *The 32nd ACM International Conference on Information and Knowledge Management (CIKM'23)* |
| 60 | Leveraging post-click user behaviors for calibrated conversion rate prediction under delayed feedback in online advertising | 2023 | Guo et al. | *The 32nd ACM International Conference on Information and Knowledge Management (CIKM'23)* |
| 61 | Automatic fusion network for cold-start CVR prediction with explicit multi-level representation | 2023 | Jin et al. | *The 39th IEEE International Conference on Data Engineering (ICDE'23)* |
| 62 | Online conversion rate prediction via neural satellite networks in delayed feedback advertising | 2023 | Liu et al. | *The 46th International ACM SIGIR Conference on Research and Development in Information Retrieval (SIGIR'23)* |
| 63 | Task adaptive multi-learner network for joint CTR and CVR estimation | 2023 | Liu et al. | *The 32nd ACM Web Conference (WWW'23)* |
| 64 | SeqGen: A sequence generator via user side information for behavior sparsity in recommendation | 2023 | Min et al. | *The 32nd ACM International Conference on Information and Knowledge Management (CIKM'23)* |
| 65 | Contrastive learning for conversion rate prediction | 2023 | Ouyang et al. | *The 46th International ACM SIGIR Conference on Research and Development in Information Retrieval (SIGIR'23)* |
| 66 | Masked multi-domain network: multi-type and multi-scenario conversion rate prediction with a single model | 2023 | Ouyang et al. | *The 32nd ACM International Conference on Information and Knowledge Management (CIKM'23)* |
| 67 | Improving conversion rate prediction via self-supervised pre-training in online advertising | 2023 | Shtoff et al. | *The 11st IEEE International Conference on Big Data (BigData'23)* |
| 68 | Adaptive learning on user segmentation: Universal to specific representation via bipartite neural interaction | 2023 | Tan et al. | *The 1st Annual International ACM SIGIR Conference on Research and Development in Information Retrieval in the Asia Pacific Region (SIGIR-AP'23)* |
| 69 | Unbiased delayed feedback label correction for conversion rate prediction | 2023 | Wang et al. | *The 29th ACM SIGKDD Conference on Knowledge Discovery and Data Mining (KDD'23)* |
| 70 | DCBT: A simple but effective way for unified warm and cold recommendation | 2023 | Yang et al. | *The 46th International ACM SIGIR Conference on Research and Development in Information Retrieval (SIGIR'23)* |
| 71 | Multi-view multi-task campaign embedding for cold-start conversion rate forecasting | 2023 | Yao et al. | *IEEE Transactions on Big Data* |
| 72 | Large scale CVR prediction through hierarchical history modeling | 2023 | Zhang et al. | *The 17th Recommender Systems Challenge (RecSysChallenge'23)* |
| 73 | DCMT: A direct entire-space causal multi-task framework for post-click conversion estimation | 2023 | Zhu et al. | *The 39th International Conference on Data Engineering (ICDE'23)* |
| 74 | Action attention GRU: A data-driven approach for enhancing purchase predictions in digital marketing | 2024 | Ban et al. | *The 39th ACM/SIGAPP Symposium on Applied Computing (SAC'24)* |
| 75 | Deep journey hierarchical attention networks for conversion predictions in digital marketing | 2024 | Ban et al. | *The 33rd ACM International Conference on Information and Knowledge Management (CIKM'24)* |
| 76 | LDACP: Long-delayed ad conversions prediction model for bidding strategy | 2024 | Cui et al. | *Pre-print (arXiv)* |
| 77 | EGEAN: An exposure-guided embedding alignment network for post-click conversion estimation | 2024 | Feng et al. | *Pre-print (arXiv)* |
| 78 | Multi-task conditional attention network for conversion prediction in logistics advertising | 2024 | Guo et al. | *The 30th ACM SIGKDD Conference on Knowledge Discovery and Data Mining (KDD'24)* |
| 79 | Utilizing non-click samples via semi-supervised learning for conversion rate prediction | 2024 | Huang et al. | *The 18th ACM Conference on Recommender Systems (RecSys'24)* |
| 80 | Privacy preserving conversion modeling in data clean room | 2024 | Li et al. | *The 18th ACM Conference on Recommender Systems (RecSys'24)* |
| 81 | Online conversion rate prediction via multi-interval screening and synthesizing under delayed feedback | 2024 | Liu et al. | *The 38th AAAI Conference on Artificial Intelligence (AAAI'24)* |
| 82 | Ranking-aware unbiased post-click conversion rate estimation via AUC optimization on entire exposure space. | 2024 | Liu et al. | *The 18th ACM Conference on Recommender Systems (RecSys'24)* |
| 83 | DDPO: Direct dual propensity optimization for post-click conversion rate estimation | 2024 | Su et al. | *The 47th International ACM SIGIR Conference on Research and Development in Information Retrieval (SIGIR'24)* |
| 84 | DCRMTA: Unbiased causal representation for multi-touch attribution | 2024 | Tang et al. | *Pre-print (arXiv)* |
| 85 | Deep ensemble shape calibration: Multi-field post-hoc calibration in online advertising | 2024 | Yang et al. | *The 30th ACM SIGKDD Conference on Knowledge Discovery and Data Mining (KDD'24)* |
| 86 | Addressing delayed feedback in conversion rate prediction: A domain adaptation approach | 2024 | Yu et al. | *The 24th IEEE International Conference on Data Mining (ICDM'24)* |
| 87 | Predicting calibrated conversion rate of online advertising using a multi-task mixture-of-experts calibration model | 2024 | Zhang et al. | *The 12nd CCF Conference on Big Data (BigData'24)* |
| 88 | Adversarial-enhanced causal multi-task framework for debiasing post-click conversion rate estimation | 2024 | Zhang et al. | *The 33rd ACM Web Conference (WWW'24)* |
| 89 | Confidence-aware multi-field model calibration | 2024 | Zhao et al. | *The 33rd ACM International Conference on Information and Knowledge Management (CIKM'24)* |
| 90 | Residual connections improve click-through rate and conversion rate prediction performance | 2025 | Bicici. | *Neural Computing and Applications* |
| 91 | ChorusCVR: Chorus supervision for entire space post-click conversion rate modeling | 2025 | Cheng et al. | *Pre-print (arXiv)* |
| 92 | MCNet: monotonic calibration networks for expressive uncertainty calibration in online advertising | 2025 | Dai et al. | *Pre-print (arXiv)* |
| 93 | Addressing delayed feedback in conversion rate prediction via influence functions | 2025 | Ding et al. | *Pre-print (arXiv)* |
| 94 | Practical multi-task learning for rare conversions in Ad Tech | 2025 | Dishi et al. | *Pre-print (arXiv)* |
| 95 | Entire-Space variational information exploitation for post-click conversion rate prediction | 2025 | Fei et al. | *The 39th AAAI Conference on Artificial Intelligence (AAAI'25)* |
| 96 | An adaptive entire-space multi-scenario multi-task transfer learning model for recommendations | 2025 | Yi et al. | *IEEE Transactions on Knowledge and Data Engineering* |
| 97 | Hardness-aware privileged features distillation with latent alignment for CVR prediction | 2025 | Yuan et al. | *The 31st ACM SIGKDD Conference on Knowledge Discovery and Data Mining (KDD'25)* |
| 98 | Personalized interpolation: An efficient method to tame flexible optimization window estimation | 2025 | Zhang et al. | *Pre-print (arXiv)* |
| 99 | On the practice of deep hierarchical ensemble network for ad conversion rate prediction | 2025 | Zhuang et al. | *The 34th ACM Web Conference (WWW'25)* |

## B. Fundamental structures of CVR prediction models

### B.1. Gradient boosting decision tree (GBDT)

Gradient boosting decision tree (GBDT) combines a set of weak learners with the additive model to fit the negative gradient of the loss function (Ren et al., 2020). GBDT makes predictions by iteratively training weak trees in $M$ rounds to minimize prediction errors (Du et al., 2019). For each sample $i$, the negative gradient of the loss function ($r_{im}$) can be given as:

$$r_{im} = -[\frac{\partial L(y_i,F(x_i))}{\partial F(x_i)}]_{F(x_i)=F_{m-1}(x_i)}, \quad (1)$$

where $F_{m-1}(x_i)$ is the output in the $m-th$ iteration and $x_i$ denotes characteristics of the $i-th$ sample.

A regression tree is fitted to obtain the leaf node region $R_{im}$, and values of leaf nodes are estimated to minimize the loss function and obtain the best fit value $c_{mj} = \arg\min_{c} \sum_{x_i \in R_{im}} L(y_i, F_{m-1}(x_i) + c)$, where $y_i$ denotes the true label of the $i-th$ sample and $c$ is a constant value. Then the regression tree is updated, i.e., $F_m(x) = F_{m-1}(x) + \sum_{j=1}^{J} c_{mj} I(x \in R_{im})$, where $I(\cdot)$ denotes the indicator function. The final learner is given as $\hat{F}(x) = F_m(x) + F_0(x) + \sum_{m=1}^{M}\sum_{j=1}^{J} c_{mj} I(x \in R_{im})$, where $F_0(x)$ denotes the initialized weak learner.

### B.2. Multilayer perceptron (MLP)

The structure of multilayer perceptron (MLP) is inspired by the biological neuron model, consisting of multiple layers of processing units called neurons, along with biases and weights that mimic brain functions. Typically, MLP comprises three components: an input layer, a hidden layer and an output layer. Specifically, the input layer receives multi-dimensional feature embeddings, the output layer and the hidden layer learns complex relationships among features through nonlinear transformation (Mao et al., 2023). In the training phase, MLP utilizes the backpropagation algorithm to optimize its parameters (i.e., weights and biases). The computation process of MLP is given as

$$\begin{cases} \boldsymbol{h_0} = \sigma_0(\boldsymbol{W}_0\boldsymbol{x} + b_0) \\ \boldsymbol{h_1} = \sigma_1(\boldsymbol{W}_1\boldsymbol{h}_0 + b_1), \\ \boldsymbol{O} = \sigma_2(\boldsymbol{W}_2\boldsymbol{h}_1 + b_2) \end{cases} \quad (2)$$

where $\boldsymbol{h}_0$, $\boldsymbol{h}_1$, and $\boldsymbol{O}$ are the output of the input layer, hidden layer, and output layer, respectively; $\boldsymbol{x}$ is the input feature embedding; $\sigma_1$ and $\sigma_2$ are activation functions; $\boldsymbol{W}_0$, $\boldsymbol{W}_1$, and $\boldsymbol{W}_2$ represent the learned weight matrices; $b_0$, $b_1$, and $b_2$ are bias terms.

### B.3. Deep neural networks (DNNs)

The structure of deep neural networks (DNNs) is comprised of an input layer, an output layer, and more than two hidden layers (Montesinos et al., 2022). In the process of feed-forward propagation, the input features are processed through hidden layers, where each layer computes a weighted sum of representation from the preceding hidden layer, followed by a nonlinear activation function. The mathematical formulation of DNNs is given as:

$$\begin{cases} \boldsymbol{h}_0 = \sigma_0(\boldsymbol{W}_0\boldsymbol{x} + b_0) \\ \boldsymbol{h}_l = \sigma_l(\boldsymbol{W}_l\boldsymbol{h}_{l-1} + b_l) \\ \boldsymbol{O} = \sigma_{L+1}(\boldsymbol{W}_{L+1}\boldsymbol{h}_L + b_{L+1}) \end{cases}, \quad (3)$$

where $\boldsymbol{h}_0$ is output of the input layer, $\boldsymbol{h}_l$ denotes the output of the $l-th$ hidden layer, $l = 1,2,\dots,L$, $\boldsymbol{h}_{l-1}$ is the output of the preceding hidden layer, $\boldsymbol{O}$ is the output of the output layer; $\sigma_0$, $\sigma_l$, and $\sigma_{L+1}$ are activation functions; $\boldsymbol{W}_0$, $\boldsymbol{W}_l$, and $\boldsymbol{W}_{L+1}$ represent the learned weight matrices; $b_0$, $b_l$, and $b_{L+1}$ are bias terms.

In the training process, DNNs employ the backpropagation algorithm to minimize the loss function by iteratively updating the model parameters, bringing the predicted values closer to the ground truth.

### B.4. Recurrent neural networks (RNNs)

In the architecture of RNNs, neurons receive feedback from their previous states through recurrent connections, forming a cyclic network structure. The updated process of hidden state $\boldsymbol{h}_t$ at time $t$ is given as:

$$\boldsymbol{h}_t = \sigma_t(\boldsymbol{W}_{xh}x_t + \boldsymbol{W}_{hh}\boldsymbol{h}_{t-1} + b_h), \quad (4)$$

where $x_t$ is the input vector at time $t$; $\sigma_t$ denotes the activation functions of the hidden; $\boldsymbol{W}_{xh}$ and $\boldsymbol{W}_{hh}$ are the corresponding weight matrices; $b_h$ represents the bias term.

**(1) LSTM** incorporates a gate mechanism to regulate the flow of information, enabling the network to retain or discard information selectively (Van Houdt et al., 2020). Core components of LSTM include the input gate $\boldsymbol{i}_t$, the forget gate $\boldsymbol{f}_t$, the output gate $\boldsymbol{o}_t$, and the memory cell $\boldsymbol{c}_t$. The learning process of LSTM involves the following steps: (1) based on the previous hidden state $\boldsymbol{h}_{t-1}$ and the current input $\boldsymbol{x}_t$, LSTM calculates the three gates and the candidate memory cell $\tilde{\boldsymbol{c}}_t$; (2) the memory cell $\boldsymbol{c}_t$ is updated by combining the forget gate $\boldsymbol{f}_t$ and the input gate $\boldsymbol{i}_t$; (3) the updated memory cell is then filtered by the output gate $\boldsymbol{o}_t$ to generate the new hidden state $\boldsymbol{h}_t$. The computation process of LSTM is given as

$$\begin{cases} \boldsymbol{f}_t = \sigma\left(\boldsymbol{W}_f[\boldsymbol{h}_{t-1}, \boldsymbol{x}_t] + b_f\right) \\ \boldsymbol{i}_t = \sigma(\boldsymbol{W}_i[\boldsymbol{h}_{t-1}, \boldsymbol{x}_t] + b_i) \\ \boldsymbol{o}_t = \sigma(\boldsymbol{W}_o[\boldsymbol{h}_{t-1}, \boldsymbol{x}_t] + b_o) \\ \tilde{\boldsymbol{c}}_t = tanh(\boldsymbol{W}_c[\boldsymbol{h}_{t-1}, \boldsymbol{x}_t] + b_c) \\ \boldsymbol{c}_t = \boldsymbol{f}_t \odot \boldsymbol{c}_{t-1} + \boldsymbol{i}_t \odot \tilde{\boldsymbol{c}}_t \\ \boldsymbol{h}_t = \boldsymbol{o}_t \odot tan\,h(\boldsymbol{c}_t) \end{cases}, \quad (5)$$

where $\boldsymbol{W}_f$, $\boldsymbol{W}_i$, $\boldsymbol{W}_o$, and $\boldsymbol{W}_c$ are weight matrices in the three gates and the cell state, respectively; and $b_f$, $b_i$, $b_o$ and $b_c$ are the corresponding bias terms.

**(2) GRU** is a simplified variant of the LSTM, which omits a separate memory cell (Xiao et al., 2022). GRU relies on two gates (i.e., the update gate $\boldsymbol{z}_t$ and the reset gate $\boldsymbol{r}_t$) to control the flow of information, the former determines how much of the past information to forget, and the latter controls how much of the past information should be retained and how much of the new input should be incorporated. Formally, GRU can be formulated as

$$\begin{cases} \boldsymbol{z}_t = \sigma(\boldsymbol{W}_z[\boldsymbol{h}_{t-1}, \boldsymbol{x}_t] + b_z) \\ \boldsymbol{r}_t = \sigma(\boldsymbol{W}_r[\boldsymbol{h}_{t-1}, \boldsymbol{x}_t] + b_r) \\ \widetilde{\boldsymbol{h}}_t = tanh(\boldsymbol{W}_h[\boldsymbol{r}_t \odot \boldsymbol{h}_{t-1}, \boldsymbol{x}_t] + b_h) \\ \boldsymbol{h}_t = (1 - \boldsymbol{z}_t) \odot \boldsymbol{h}_{t-1} + \boldsymbol{z}_t \odot \widetilde{\boldsymbol{h}}_t \end{cases}, \quad (6)$$

where $\boldsymbol{W}_z$, $\boldsymbol{W}_r$ and $\boldsymbol{W}_h$ are weight matrices in the two gates and the cell state, respectively; and $b_z$, $b_r$, and $b_h$ are the corresponding bias terms.

### B.5. Transformer

Transformer is a sequence-to-sequence model, comprised of the multi-head self-attention mechanism, the positional encoding, feed-forward neural networks, residual connections and layer normalization (Vaswani et al., 2017). For behavioral sequences, Transformer employs the positional encoding to provide information about the relative or absolute positions of input tokens within the sequence, which is given as

$$\begin{cases} PE(pos, 2i) = sin\left(\frac{pos}{10000^{\frac{2i}{d_{model}}}}\right) \\ PE(pos, 2i+1) = cos\left(\frac{pos}{10000^{\frac{2i}{d_{model}}}}\right) \end{cases}, \quad (7)$$

where $pos$ is the position, $i$ is the dimension index, and $d_{model}$ is the model dimensionality.

The positional embeddings $\boldsymbol{P} = \{\boldsymbol{p}_1, \boldsymbol{p}_2, \dots, \boldsymbol{p}_N\}$ are added into sequence embeddings $\boldsymbol{E} = \{\boldsymbol{e}_1, \boldsymbol{e}_2, \dots, \boldsymbol{e}_N\}$ to obtain the positional sequence embeddings $\boldsymbol{X} = \boldsymbol{P} + \boldsymbol{E}$.

The three learnable matrices $\boldsymbol{W}_q$, $\boldsymbol{W}_k$, and $\boldsymbol{W}_v$ are used to project $\boldsymbol{X}$ into the query ($\boldsymbol{Q}$), key ($\boldsymbol{Q}$), and value ($\boldsymbol{V}$) matrices. The scaled dot-product attention is calculated as

$$Attention(\boldsymbol{Q}, \boldsymbol{K}, \boldsymbol{V}) = softmax(\frac{\boldsymbol{QK}^T}{\sqrt{d_k}})\boldsymbol{V}, \quad (8)$$

where $d_k$ is the dimension of the key matrix.

The multi-head self-attention consists of multiple self-attention, which is given as

$$\begin{cases} \boldsymbol{head}_m = Attention(\boldsymbol{Q}_m, \boldsymbol{K}_m, \boldsymbol{V}_m) \\ MultiHead(\boldsymbol{Q}, \boldsymbol{K}, \boldsymbol{V}) = \boldsymbol{W}_0[\boldsymbol{head}_1; \dots; \boldsymbol{head}_n] \end{cases}, \quad (9)$$

where $\boldsymbol{head}_m$ denotes the output of the $m - th$ attention head and $\boldsymbol{W}_0$ is the output projection matrix.

The output representation of the multi-head attention (i.e., $MultiHead(\boldsymbol{Q}, \boldsymbol{K}, \boldsymbol{V})$) will be processed by a feed-forward network (FFN), which is given as

$$FFN\big(MultiHead(\boldsymbol{Q}, \boldsymbol{K}, \boldsymbol{V})\big) = \boldsymbol{W}_2\, max(0, \boldsymbol{W}_1 MultiHead(\boldsymbol{Q}, \boldsymbol{K}, \boldsymbol{V}) + b_1) + b_2, \quad (10)$$

where $\boldsymbol{W}_1$ and $\boldsymbol{W}_2$ are the weight matrices; $b_1$ and $b_2$ are the bias items.

To facilitate gradient flow and prevent degradation, each sub-layer (e.g., the multi-head self-attention and the feed-forward network) is followed by a residual connection and layer normalization, which is given as

$$LayerNorm(FFN\big(MultiHead(\boldsymbol{Q}, \boldsymbol{K}, \boldsymbol{V})\big) + sublayer\Big(FFN\big(MultiHead(\boldsymbol{Q}, \boldsymbol{K}, \boldsymbol{V})\big)\Big)),$$

(11)

where $sublayer(\cdot)$ denotes the output of the corresponding sublayer.

### B.6. Counterfactual recurrent networks (CRNs)

The structure of CRNs is comprised of two components: an encoder based on RNNs learning balanced temporal representations from user behaviors, and a decoder leveraging the representations to estimate the potential outcomes of future treatment strategies (Bica et al., 2020).

CRNs first generate an equilibrium representation by maximizing the treatment classifier loss and minimizing the outcome predictor loss. Specifically, the equilibrium representation $\Phi(\boldsymbol{H}_t)$ is constructed such that it satisfies $P(\Phi(\boldsymbol{H}_t)|A_t = A_1) = P(\Phi(\boldsymbol{H}_t)|A_t = A_2) \ldots = P(\Phi(\boldsymbol{H}_t)|A_t = A_k)$, where $A_t \in \{A_1, \ldots, A_K\}$ is the possible treatment in time $t$ and $\boldsymbol{H}_t$ is the historical representation that is invariant across treatments. The loss functions of CRNs are given as

$$\begin{cases} \mathcal{L}_t^{(i)}(\theta_r, \theta_y, \theta_a) = \sum_{i=1}^{N} \mathcal{L}_{t,y}^{(i)}(\theta_r, \theta_y) - \lambda \mathcal{L}_{t,a}^{(i)}(\theta_r, \theta_a) \\ \mathcal{L}_{t,y}^{(i)}(\theta_r, \theta_y) = ||y_{t+1}^{(i)} - G_y(\Phi(\boldsymbol{H}_t; \theta_r); \theta_y)||^2 \\ \mathcal{L}_{t,a}^{(i)}(\theta_r, \theta_a) = -\sum_{j=1}^{K} \mathbb{I}\left\{a_t^{(i)} = A_j\right\} log\, G_j^a(\Phi(\boldsymbol{H}_t; \theta_r); \theta_a) \end{cases}, \quad (12)$$

where $\mathcal{L}_t^{(i)}$ is the overall loss at time $t$, $\mathcal{L}_{t,y}^{(i)}$ is the loss of the output predictor, $\mathcal{L}_{t,a}^{(i)}$ is the loss of the treatment classifier, $\theta_r$ is the parameter of the representation function $\Phi$, $\theta_y$ is the parameter of the output predictor $G_y$, $\theta_a$ is the parameter of the treatment classifier $G_j^a$, $\lambda$ is the trade-off parameter, $y_{t+1}^{(i)}$ denotes the outcome of the $i - th$ user behavior at time $t + 1$, $a_t^{(i)}$ is the treatment of the $i - th$ user behavior at time $t$, and $\mathbb{I}\{\cdot\}$ is the indicator function.

### B.7. Knowledge distillation models

The framework of knowledge distillation consists of a large and complex teacher model pretrained on a large-scale dataset, and a compact and efficient student model trained by mimicking the output distribution of teacher model (Mansourian et al., 2025). Specifically, the teacher model is used to generate soft labels for the training data. The soft labels are the predicted probabilities of all classes, and can be estimated by using a softmax function

$$q_i = \frac{exp(\boldsymbol{z}_j/T)}{\sum_j exp(\boldsymbol{z}_j/T)}, \quad (13)$$

where $q_i$ is the soft prediction probability of the $i$-th class, $\boldsymbol{z}_j$ denotes the logits of the teacher model, and $T$ is the distillation temperature parameter, higher values of $T$ yield flatter output distributions.

The student model is trained on the soft labels generated by the teacher to mimic the teacher's behavior and enhance its predictive performance. In the training process, the student model inherits knowledge from the teacher by jointly minimizing the cross-entropy loss with respect to the true labels, and the Kullback-Leibler divergence with respect to the teacher's soft outputs (For details, refer to Section 4.5).

## References

Bica, I., Alaa, A. M., Jordon, J., & van der Schaar, M. (2020). Estimating counterfactual treatment outcomes over time through adversarially balanced representations. Proceedings of the 8th International Conference on Learning Representations (ICLR'20) (pp. 1-28), virtually, April 26th – May 1st, 2020. https://openreview.net/group?id=ICLR.cc/2020/Conference.

Du, G., Chen, Y., & Wang, X. (2019). Boosted trees model with feature engineering: An approach for post-click conversion rate prediction. Proceedings of the 2nd International Conference on Big Data Technologies (ICBDT'19) (pp. 136-141). New York, NY, USA: Association for Computing Machinery.

Mansourian, A. M., Ahmadi, R., Ghafouri, M., Babaei, A. M., Golezani, E. B., Ghamchi, Z. Y., Ramezanian, V., Taherian, A., Dinashi, K., Miri, A., & Kasaei, S. (2025). A Comprehensive survey on knowledge distillation. arXiv preprint arXiv:2503.12067.

Mao, K., Zhu, J., Su, L., Cai, G., Li, Y., & Dong, Z. (2023). FinalMLP: An enhanced two-stream MLP model for CTR prediction. Proceedings of the 37th AAAI Conference on Artificial Intelligence (AAAI'23) (pp. 4552-4560). Palo Alto, California, USA: AAAI Press.

Montesinos López, O. A., Montesinos López, A., & Crossa, J. (2022). Fundamentals of artificial neural networks and deep learning. Multivariate Statistical Machine Learning Methods for Genomic Prediction (pp. 379-425). Cham: Springer.

Ren, J., Zhang, J., & Liang, J. (2020). Feature engineering of click-through-rate prediction for advertising. International Conference in Communications, Signal Processing, and Systems (CSPS'18) (pp. 204-211). Singapore: Springer.

Van Houdt, G., Mosquera, C., & Nápoles, G. (2020). A review on the long short-term memory model. Artificial Intelligence Review, 53(8), 5929-5955.

Vaswani, A., Shazeer, N., Parmar, N., Uszkoreit, J., Jones, L., Gomez, A. N., Kaiser, L., & Polosukhin, I. (2017). Attention is all you need. Advances in Neural Information Processing Systems (NIPS'17) (pp. 1-11). New York, NY, USA: Curran Associates.

Xiao, Y., He, W., Zhu, Y., & Zhu, J. (2022). A click-through rate model of e-commerce based on user interest and temporal behavior. Expert Systems with Applications, 207, 117896.

# C. Datasets for CVR prediction models

Table C1. The summary of datasets for advertising CVR prediction.

| Dataset | URL | Description | References |
|---|---|---|---|
| Criteo conversion | https://ailab.criteo.com/criteo-sponsored-search-conversion-log-dataset/ | The Criteo conversion dataset contains users' clicks and conversion behaviors, covering key fields such as impression, clicks, and conversions. The 2020 Criteo conversion dataset is the most recent dataset and contains Criteo predictive search logs from August 3, 2020 to October 16, 2020. The 2013 Criteo conversion dataset is no longer publicly available. | Chapelle (2014); Vasile et al. (2017); Tallis & Yadav (2018); Yoshikawa & Imai (2018); Yasui et al. (2020); Gu et al. (2021); Hou et al. (2021); Li et al. (2021b); Yang et al. (2021); Chen et al. (2022a); Gao & Yang (2022); Guo et al. (2022); Yasui & Kato (2022); Yang & Zhan (2022); Dai et al. (2023); Liu et al. (2023a); Ouyang et al. (2023b); Wang et al. (2023c); Liu et al. (2024a); Yu et al. (2024); Ding et al. (2025) |
| Ali-CCP | https://tianchi.aliyun.com/dataset/dataDetail?dataId=408 | The Ali-CCP dataset is collected from Taobao advertising traffic logs, which contains contains 84 exposures, 3.4M clicks, and 18K conversions. | Ma et al. (2018); Tang et al. (2020); Zhang et al. (2020); Wen et al. (2020); Wei et al. (2021); Xi et al. (2021); Lin et al. (2022); Wang et al. (2022); Xu et al. (2022); Zhang et al. (2022); Liu et al. (2023b); Ouyang et al. (2023a); Zhu et al. (2023); Feng et al. (2024); Guo et al. (2024); Huang et al. (2024); Liu et al. (2024b); Su et al. (2024); Zhang et al. (2024b); Cheng et al. (2025); Yi et al. (2025); Yuan et al. (2025) |
| Criteo attribution | https://ailab.criteo.com/criteo-attribution-modeling-bidding-dataset/ | The Criteo attribution dataset is a widely used dataset in online advertising attribution research. It contains 30 days of real-time traffic data covering 675 advertising channels, 16.46 million advertising impressions, and 6.14 million users. Each impression includes a timestamp, user information, advertising information, and behavioral labels indicating whether the user clicked or converted. | Ren et al. (2018); Kumar et al. (2020); Xie et al. (2022); Yao et al. (2022); Tang et al. (2024) |
| Tencent advertising algorithm competition 2017 | https://github.com/guicunbin/Tencent | The Tencent advertising algorithm competition 2017 dataset records 14 days of advertising log data. | Li et al. (2021b); Guo et al. (2022); Liu et al. (2023a); Liu et al. (2024a); Yu et al. (2024) |
| iPinYou | https://contest.ipinyou.com/ | The iPinYou dataset contains log data on advertising bidding, display, clicks, and conversions. This dataset is divided into three seasons covering multiple advertisers and users. | Shan et al. (2018) |
| Miaozhen | http://www.miaozhen.com/en/index.html | The Miaozhen dataset is provided by a marketing company in China, containing about 1.24 billion advertising logs from May to June 2013. It contains detailed information on advertising impressions, user behaviors, channels, and conversions. | Zhang et al. (2014); Ji et al. (2016); Ji & Wang (2017); Ren et al. (2018) |
| Tenc_UnionAds | https://drive.weixin.qq.com/s?k=AJEAIQdfAAo57eHKjr#/ | The Tenc_UnionAds dataset is an industrial-scale dataset comprising millions of samples collected from Tencent's advertising platform over five consecutive days. It includes over 300 features, encompassing user, advertising, and contextual information. Moreover, it contains three types of samples, i.e., click and conversion, click without conversion, and non-click. | Zhang et al. (2024b) |
| Alimama | https://tianchi.aliyun.com/dataset/147588 | The Alimama dataset is a real e-commerce advertising dataset released by Alimama advertising platform in 2018 to support the prediction research of search advertising CVR. It provides five types of data, i.e., basic data, advertising product information, user information, contextual information, and store information. | Yi et al. (2025) |
| ShareChat | https://www.recsyschallenge.com/2023/ | The ShareChat dataset includes approximately 10 million randomly selected users who accessed the ShareChat + Moj app over a three-month period. It provides user features (e.g., demographic attributes and content preference embeddings) and ad features (e.g., advertising categorical features and embeddings), with the objective of predicting user installation probabilities. | Zhang et al. (2023) |
| Synthetic dataset | https://algo.qq.com/?lang=en | The Synthetic dataset refers to artificial data generated through algorithmic or model-based simulations, designed to replicate various behaviors, including advertising exposures, user clicks, and conversions. | Saito et al. (2020); Yao et al. (2022); Yasui & Kato (2022); Tang et al. (2024) |
| Proprietary datasets | - | | Lee et al. (2012); Rosales et al. (2012); Yang et al. (2016); Bhamidipati et al. (2017); Lu et al. (2017); Li et al. (2018); Cui et al. (2018); Ma et al. (2018); Du et al. (2019a); Du et al. (2019b); Kitada et al. (2019); Pan et al. (2019); Zhou et al. (2019); Gligorijevic et al. (2020); Qiu et al. (2020); Yang et al. (2020); Yasui et al. (2020); Zhang et al. (2020); Hou et al. (2021); Liu & Liu (2021); Li et al. (2021a); Li et al. (2021b); Gu et al. (2021); Su et al. (2021); Wei et al. (2021); Agrawal et al. (2022); Chen et al. (2022b); Dai et al. (2022); Ding et al. (2022); Lin et al. (2022); Wang et al. (2022); Xu et al. (2022); Chan et al. (2023); Dai et al. (2023); Jin et al. (2023); Liu et al. (2023b); Min et al. (2023); Ouyang et al. (2023a); Ouyang et al. (2023b); Shtoff et al. (2023); Tan et al. (2023); Wang et al. (2023c); Yang et al. (2023); Yao et al. (2023); Ban et al. (2024a); Ban et al. (2024b); Cui et al. (2024); Feng et al. (2024); Guo et al. (2024); Huang et al. (2024); Li et al. (2024); Su et al. (2024); Yang et al. (2024); Zhao et al. (2024); Bi ç̇ci (2025); Dai et al. (2025); Dishi et al. (2025); Yuan et al. (2025); Zhang et al. (2025a); Zhuang et al. (2025) |